%% file: b2.tex
\documentclass[12pt]{article}
\pdfoutput=1
\usepackage[
top = 2.5cm, 
bottom = 2.5cm, 
left = 2.5cm, 
right = 2.5cm]{geometry}

\usepackage{jheppub}
\usepackage[sort&compress]{natbib}
\usepackage[utf8]{inputenc}
\usepackage{ifpdf}
\usepackage{braket}
\usepackage{graphicx}
\usepackage{caption}
\usepackage{subcaption}
\usepackage{slashed}
\usepackage{feynmf}
\usepackage{epstopdf}
\usepackage{amsmath, amssymb,url,bm,textgreek,upgreek}
\usepackage[svgnames]{xcolor}
\usepackage{hyperref}
\usepackage{dsfont}
\hypersetup{linkcolor=DarkBlue,citecolor=magenta}

\pdfoutput=1
\makeatletter
\def\@fpheader{\relax}
\makeatother

\def\be{\begin{eqnarray}}
\def\ee{\end{eqnarray}}

\def\tr{\operatorname{tr}}

\def\Vol{\operatorname{Vol}}

\def\tr{\operatorname{tr}}

\def\d{{\rm d}}

\def\K{\kappa}

\def\id{\mathds{1}}
\def \bea {\begin{eqnarray}}
\def \eea {\end{eqnarray}}
\def \nn {\nonumber}

\newcommand{\oE}{\mathring{E}}
\newcommand{\oD}{\mathring{{\rm \nabla}}}

\newcommand{\oR}{\mathring{R}}
\newcommand{\op}{\mathring{{\partial}}}
\newcommand{\oO}{\mathring{{O}}}
\newcommand{\oB}{\mathring{{B}}}
\def \la {\langle}
\def \ra {\rangle}
\def \a {\alpha}
\def \b {\beta}
\def \g {\gamma}

\def \de{\delta}

\def \m {\mu}
\def \n {\nu}

\def\si{\sigma}
\def \del {\partial}
\def \ts{ \textstyle}

\title{Boundary Conformal Field Theory \\
and a Boundary Central Charge}

\author{Christopher P. Herzog}
\author{and Kuo-Wei Huang}

\affiliation{C. N. Yang Institute for Theoretical Physics, Department of Physics and Astronomy,\\
Stony Brook University, Stony Brook, NY 11794, USA 
}

\abstract{
We consider the structure of current  and stress tensor two-point functions in conformal field theory with a boundary.
The main result of this paper is a relation between a boundary central charge and the coefficient of 
a displacement operator correlation function in the boundary limit.
The boundary central charge under consideration is the coefficient of the
product of the extrinsic curvature and the Weyl curvature in the conformal anomaly.  
Along the way, we describe several auxiliary results.  
Three of the more notable are as follows: 
(1) we give the bulk and boundary conformal blocks for the current two-point function; 
(2) we show that the structure of these current and stress tensor two-point functions 
is essentially universal for all free theories; 
(3) we introduce a class of interacting conformal field theories with boundary degrees of freedom, where the interactions
are confined to the boundary.  
The most interesting example we consider can be thought of as the infrared fixed point of graphene. 
This particular 
interacting conformal model in four dimensions provides a counterexample of a previously conjectured relation between a boundary central charge and a bulk central charge.   
The model also demonstrates that the boundary central charge can change in response to marginal deformations.
}

\preprint{%UT-16-30,
YITP-17-24}

\begin{document}
\maketitle
\setcounter{page}{2}

\section{Introduction}

\input{introduction}

\section{Boundary Conformal Anomalies}
\label{sec:boundarycharge}
\input{boundarycharge}

\section{Boundary Conformal Field Theory and Two-Point Functions}
\label{sec:twopoint}
\input{twopoint}

\section{A Boundary Central Charge}
\label{sec:proof}
\input{proof}

\section{Free Fields and Universality}
\label{sec:free}
\input{free}

\section{Models with Boundary Interactions}
\label{sec:interaction}
\input{interaction}

\section{Conclusions and Open Problems}

Motivated by recent classification of the boundary trace anomalies for bCFTs \cite{Herzog:2015ioa,Fursaev:2015wpa,Solodukhin:2015eca}, we 
studied the structure of two-point functions in bCFTs.
Our main result \eqref{mainresult} states a relation between the $b_2$ boundary central charge in $d=4$ bCFTs and
the spin-zero displacement operator correlation function near the boundary.
Since $\alpha(1)=2 \alpha(0)$ in free theories, we can explain the $b_2=8c$ relation observed in \cite{Fursaev:2015wpa}.
Indeed, from our study of free theories,
we find that two-point functions of free bCFTs have a simple universal structure. 

Going beyond free theory, we defined a class of interacting models with the interactions restricted to the boundary.
We computed their beta functions and pointed out the locations of the fixed points.
In particular, the mixed dimensional QED is expected to be exactly conformal in $d=4$.
We have provided evidence that this model can be a counterexample of the $b_2=8c$ relation in 4d bCFTs.
As we summarized in the introduction,
this mixed QED theory is interesting for at least three other reasons as well: its connection with graphene,
its connection with three dimensional QED, and its behavior under electric-magnetic duality.
It doubtless deserves further exploration.

%For the graphene-like theory, the near boundary limit of the stress tensor two-point function, 
%characterized by $\alpha(1)$, depends on the exactly marginal coupling $g$.
%We were not quite able to demonstrate this dependence here because of an order of limits issue.
A feature of this graphene-like theory is that the near boundary
 limit of the stress tensor two-point function, characterized by $\alpha(1)$, depends on the exactly marginal coupling $g$.  
Given the claimed relationship between $b_2$ and $\alpha(1)$ \eqref{mainresult}, it follows that $b_2$ also depends on the exactly marginal coupling $g$.  This dependence stands in contrast to the situation for the bulk charges $a$ and $c$.  Wess-Zumino consistency rules out the possibility of any such dependence for $a$ \cite{Osborn1991weyl}.  
The idea is to let $a(g(x))$ depend on the coupling $g$ which we in turn promote to a coordinate dependent external field.  Varying the Euler density must produce a total derivative.  Any spatial dependence of $a$ spoils this feature. 

The situation is different for $c$ (and hence also $\alpha(0)$).  While the Euler density varies to produce a total derivative, the integrated $W^2$ term has zero Weyl variation.  Thus in principle, one might be able to find examples of field theories where $c$ depends on marginal couplings.
In \cite{Nakayama:2017oye}, an AdS/CFT model without 
supersymmetry is constructed suggesting the possibility that the $c$-charge can change under exactly marginal deformations. In practice, guaranteeing an exactly marginal direction in four dimensions is difficult and usually requires supersymmetry.  Supersymmetry in turn fixes $c$ to be a constant.  

For $b_2$, the situation is similar to the situation for $c$.  The integrated $KW$ boundary term also has a zero Weyl variation, and $b_2$ could in principle depend on marginal couplings.  In contrast to the situation without a boundary, the presence of a boundary has allowed us to construct a non-supersymmetric theory with an exactly marginal direction in the moduli space -- this mixed dimensional QED.  Correspondingly, we are finding that $\alpha(1)$ and $b_2$ can depend on the position in this flat direction. 
A similar situation is that the boundary entropy $g$ in two dimensional conformal field theories is known to depend on marginal directions in the moduli space \cite{Elitzur:1998va}.\footnote{%
 We would like to thank T.~Dumitrescu for this remark.
}
There is a potential downside to this dependence. 
If we are looking for a quantity that orders quantum field theories under RG flow, it is inconvenient for that quantity to depend on marginal directions.  
We normally would like such a quantity to stay constant on the space of exactly marginal couplings 
and only change when we change the energy scale.
It is nevertheless interesting to understand better how these 4d boundary central charges 
behave under (boundary) RG flow.

Many interesting open problems are not answered in the present paper.
Let us conclude by listing them:
\begin{itemize}

\item
Find extensions to other dimensional bCFTs, in particular three dimensional bCFTs.  In the three dimensional case, it would
be interesting if the central charge $b$ could be related to a stress tensor correlation function.
Some incomplete speculations about the 3d case can be found in appendix \ref{sec:3d}.
%\footnote{%
%Soon after finishing this paper, we realized a straightforward way to relate $b$ and $b_1$ to correlation functions of the 
%displacement operator 
%\cite{HHJ2017}.
%}

\item
Classify the structure of three-point functions in bCFTs.
We believe such a classification will provide a way to better
understand the $b_1$ boundary central charge in 4d bCFTs that we have not considered in this paper.

\item
Compute directly the $b_2$ boundary central charge in the mixed dimensional QED in curved space and verify the relation \eqref{mainresult}.

\item
In order to check our proposal, it would be interesting to compute stress tensor two-point functions in bottom-up holographic models of bCFTs and compare the result with central charges computed in refs.\ \cite{Miao:2017gyt,Chu:2017aab,Astaneh:2017ghi}.  More ambitiously, one could consider top-down holographic Janus solutions as well.

\item
While some higher loop results exist already for mixed QED (see e.g.\ \cite{Teber:2012de,Teber:2014hna}), one can further extend the perturbative analysis in this and other boundary interacting models studied here.

\item
Search for new boundary interacting models in four and other dimensions and find new fixed points.

\item
Search for stronger bounds on these boundary central charges, 
building perhaps on the reflection positivity results in this paper.

\item
We are sympathetic to the idea discussed in ref.\ \cite{Astaneh:2017ghi} 
that, like for $c$, supersymmetry restricts the possible behaviors of $b_2$.    
Here we say nothing about the role of supersymmetry beyond some observations given in footnote \ref{susynote}.
It would be interesting to investigate supersymmetric constraints on the boundary charges, both from a purely field theoretic
standpoint and in holographic models of bCFT.  For example, what can one say about the relative values of $b_2$ and $c$
for maximally supersymmetric Yang-Mills theory in four dimensions in the presence of supersymmetry preserving boundary conditions?

\item 
It will be interesting to consider the stress tensor two-point function with a codimension-2 surface. 
Such geometry has an important relationship to quantum entanglement.

\end{itemize}

\section*{Acknowledgments}

We would like to thank Alexander Abanov, Christopher Beem, Nikolay Bobev, Francesco Bonetti, Michele Del Zotto, 
Thomas Dumitrescu, Dmitri Fursaev, Diego Hofman, Matthijs Hogervorst, Kristan Jensen, 
Igor Klebanov, Zohar Komargodski, Pedro Liendo, Robert Myers, Hugh Osborn, Silviu Pufu, 
Eliezer Rabinovici, Leonardo Rastelli, Nathan Seiberg, Robert Shrock, 
Sergey Solodukhin, Dam T. Son, Sofian Teber 
and Julio Virrueta for discussion.  
Special thanks go to Davide Gaiotto for catching an error in an earlier version of this manuscript.
This work was supported in part by the National Science Foundation 
under Grant No.\ PHY-1620628.

\newpage

\appendix
\section{Null Cone Formalism}
\label{sec:nullcone}
\input{nullcone}

\section{Variation Rules}
\label{sec:Weyl}

Here we give a brief review on the definitions of the Weyl tensor and extrinsic curvature. 
We list relevant metric perturbation formulae.

Under the metric perturbation $g_{\mu\nu}\to g_{\mu\nu}+ \delta g_{\mu\nu}$,
the transformed Christoffel connection is given by
\be
\delta^{(n)}\Gamma^\lambda_{\mu\nu}
&=&{n\over 2} \delta^{(n-1)} (g^{\lambda \rho}) \Big(\nabla_\mu\delta
g_{\rho\nu}+\nabla_\nu \delta g_{\rho\mu}-\nabla_\rho \delta 
g_{\mu\nu}\Big) \ .
\ee
The Riemann and Ricci curvature tensors transform as
\be 
 \delta {R^\lambda_{~\mu\sigma\nu}}&=&\nabla_\sigma
\delta\Gamma^\lambda_{\mu\nu}-\nabla_\nu
\delta\Gamma^\lambda_{\mu\sigma} \ ,\\
\delta {R_{\mu\nu}}&=& {1\over 2} \Big(
\nabla^\lambda \nabla_\mu \delta g_{\lambda \nu}+\nabla^\lambda
\nabla_\nu \delta g_{\mu\lambda}-g^{\lambda\rho}\nabla_\mu \nabla_\nu
\delta g_{\lambda\rho}-\Box \delta g_{\mu\nu} \Big) \ , \\ 
\delta {R}
&=& - R^{\mu\nu} \delta g_{\mu\nu} + \nabla^\mu \Big(\nabla^\nu \delta
g_{\mu\nu}-g^{\lambda \rho}\nabla_\mu \delta g_{\lambda \rho} \Big)\ .
\ee 
The Weyl tensor in $d$-dimensions (for $d>3$) is defined as
\be 
\label{weyl}
W^{(d)}_{\mu\si\rho\nu} = R_{\mu\si\rho\nu} - {2\over d-2} \Big(
g_{\mu[\rho}R_{\nu]\si} - g_{\si[\rho}R_{\nu]\mu}  -
{g_{\mu[\rho}g_{\nu]\si}\over (d-1)}R\Big) \ .
\ee 
Note $W_{\mu\si\rho\nu}=W_{[\mu\si][\rho\nu]}$,
$W_{\mu[\si\rho\nu]}=0$ and $W^{\mu}_{~\si\rho\mu}=0$.
One can write the transformation of the Weyl tensor as    
\be
\delta W_{\mu\si\rho\nu} =- 2 P_{\mu\si\rho\nu,\alpha\gamma\delta\beta} \del^\gamma \del^\delta \delta
g^{\alpha\beta} \ ,
\ee 
where  $P_{\mu\si\rho\nu,\alpha\gamma\delta\beta}$ is a projector given by
\be 
\label{P}
P_{\mu\sigma\rho\nu,\alpha\gamma\delta\beta}&=&
{\ts {1\over 12}}\big(
\de_{\mu\alpha}\de_{\nu\beta}\de_{\si\gamma}\de_{\rho\de} +
\de_{\mu\de}\de_{\si\beta}\de_{\rho\alpha}\de_{\nu\gamma} 
- \mu \leftrightarrow \sigma, \n \leftrightarrow \rho\big ) \nn\\ 
&& +
{\ts{1\over 24}}\big(
\de_{\mu\alpha}\de_{\nu\gamma}\de_{\rho\de}\de_{\si\beta} -\mu \leftrightarrow \sigma, \n \leftrightarrow \rho, \a \leftrightarrow \gamma, \delta \leftrightarrow \beta  \big ) \nn\\ 
&& - {\ts {1\over
8 (d-2)}}\big( \de_{\mu\rho}\de_{\alpha\delta}\de_{\si\gamma}\de_{\nu\beta} +
\de_{\mu\rho}\de_{\alpha\delta}\de_{\si\beta}\de_{\nu\gamma} -\mu \leftrightarrow \sigma, \n \leftrightarrow \rho, \a \leftrightarrow \gamma, \delta \leftrightarrow \beta \big ) \nn\\ 
&& + {\ts {1\over
2(d-1)(d-2)}}\big( \de_{\mu\rho} \de_{\nu\si} -  \de_{\mu\n } \de_{\rho \si} \big )
\big ( \de_{\alpha\de}\de_{\beta\gamma} -
\de_{\alpha\beta}\de_{\delta\gamma}\big) \ .  
\ee   
For a symmetric tensor or operator $t^{\g\delta}$ one has the following symmetric property:
\be
\label{symt}
P_{\mu\sigma\rho\nu,\alpha\gamma\delta\beta} t^{\gamma\delta}=P_{\mu\sigma\rho\nu,\b\gamma\delta\a} t^{\gamma\delta} \ ,
\ee 
while in general $P_{\mu\sigma\rho\nu,\alpha\gamma\delta\beta}\neq P_{\mu\sigma\rho\nu,\b\gamma\delta\a}$.

Defining the induced metric by $h_{\mu\nu}= g_{\mu\nu}-n_\mu n_\nu$, where $n_{\mu}$ is the 
outward-pointing normal vactor, the extrinsic curvature is
\bea
K_{\mu\nu}= h_\m^\lambda h_\n^\sigma \nabla_{\lambda} n_{\sigma}= \nabla_\mu n_\nu- n_\mu a_\nu \ ,
\eea  
where $a^\mu= n^\lambda \nabla_\lambda n^{\mu}$. 
On the boundary 
we have the following variations in general coordinates:
\bea
\delta n_\mu &=& {1\over 2}n_\mu \delta g_{nn} \ , \\
\delta n^\mu &=& -{1\over 2}n^\mu  \delta g_{nn}-  h^{\mu\nu}\delta g_{n\nu} \ , \\
\delta {K_{\mu\nu}}
 &=&  {K_{\mu\nu}\over 2}\delta g_{nn} + \Big(n_\mu K^\lambda_\nu+n_\nu K^\lambda_\mu \Big)\delta g_{\lambda n}- {
h^\lambda_\mu h_\nu^\rho n^\alpha\over 2} \Big(\nabla_\lambda \delta
g_{\alpha\rho}+ \nabla_\rho \delta g_{\lambda\alpha}-\nabla_\alpha
\delta g_{\lambda \rho}\Big) \ , \nn\\
\delta K&=& -{1\over 2} K^{\mu\nu} \delta g_{\mu\nu}-{1\over 2} n^\mu \Big(\nabla^\nu \delta g_{\m\n}
-g^{\n\lambda}\nabla_\mu \delta g_{\n\lambda}\Big)-{1\over 2} \oD_A (h^{AB} \delta g_{B n})\ , 
\eea 
where $\oD^\mu$ denotes the covariant derivative compatible with the boundary metric. 

We can foliate the spacetime with hypersurfaces 
labelled by $y\equiv n^\mu x_\mu$ and adopt the Gaussian normal coordinates. 
The metric reads
\bea
\label{gnc}
ds^2= dy^2+ h_{AB}(y,x_A) dx^A dx^B \ .
\eea
In the Gaussian normal coordinate $a^\mu=0$, and one has
\bea
K_{AB}={1\over 2} \partial_n h_{AB} \ ,
\eea 
and $\Gamma^y_{AB}= - K_{AB}$, $\Gamma^A_{yB}=  K^A_{B}$, 
$\Gamma^A_{yy}=  \Gamma^y_{yA}= \Gamma^y_{yy}=0$.
The transformation rules of the extrinsic curvature become
\be
\label{varyKab}
\delta K_{AB}&=& {1\over 2} \nabla_n \delta g_{AB}+{1\over 2} K^C_A \delta g_{BC}+{1\over 2} K^C_B \delta g_{AC}
-{1\over 2}K_{AB}\delta g_{nn} - \oD_{(A} \delta g_{B) n} \ , \\
\delta K&=&  {1\over 2}h^{AB}  \nabla_n \delta g_{AB}- {1\over 2}K \delta g_{nn}
- \oD^{A} \delta g_{A n}\ , \\
 \nabla_n \delta g_{AB}&=&\partial_{n}\delta g_{AB}-K^C_A \delta g_{BC}-K^C_B \delta g_{AC} \ .
\ee  

\section{Gauge Fixing Mixed Dimensional QED}
\label{sec:gaugefixing}

In the presence of a planar boundary, which already breaks the full Lorentz invariance of the theory, it can be more convenient
to consider a more general type of gauge fixing, characterized by two constants $\eta$ and $\zeta$ instead of just the usual $\xi$:
%Let $y = x^n$ be the normal coordinate to the boundary:
\be
I= \int_{{\cal M}} \d^4 x \left( - \frac{1}{4} F_{\mu\nu} F^{\mu\nu} - \frac{1}{2}(\eta \partial_n A^n - \zeta \partial_A A^A)^2 \right)+  \int_{\partial {\cal M}} \d^3 x  \left( i\bar \psi \slashed{D} \psi  \right) \ ,
\ee where the boundary fermions do not affect the discussion of the gauge field Green's function in what follows.
%The equations of motion that follow from this action are
%\be
%\partial_A (\partial^A A_n - \partial_n A^A) + \eta \partial_n (\eta \partial_n A^n + \zeta \partial_A A^A) &=& 0 \ , \\
%\partial_A (\partial^A A_B - \partial_B A^A) + \zeta \partial_B (\eta \partial_n A^n + \zeta \partial_A A^A) &=& 0 \ .
%\ee
Standard Feynman gauge is achieved by setting $\zeta = \eta = 1$.  We will kill the off-diagonal terms in the equations of motion
by setting $\eta = 1/\zeta$.  

Our strategy will be to first proceed by ignoring the presence of a boundary and then to take it into account at a later stage using the method of images.
The (Euclidean) Green's function is defined by the equation:
\be
\left(
\begin{array}{cc}
\partial^2 \delta_{AB} + (\zeta^2 - 1) \partial_A \partial_B & 0 \\
0 & \partial^2 + (\zeta^{-2} - 1) \partial_n^2 
\end{array}
\right) G^{\mu\nu}(x,x') = \delta^{(4)} (x-x') \ .
\ee
Fourier transforming, we obtain
\be
\left(
\begin{array}{cc}
k^2 \delta_{AB}+ (\zeta^2 - 1) k_A k_B & 0 \\
0 & k^2 + ( \zeta^{-2} - 1) k_n^2 
\end{array}
\right) \tilde G^{\mu\nu}(k) = -1 \ .
\ee
Inverting this matrix, we don't quite get the usual result because $k^2 \neq k_A k^A$.  
The full result is a bit messy.  Instead, let us take $\eta^2 = 1 + \delta \eta$ and expand to linear order in $\delta \eta$.
We find 
\be
\tilde G_{\mu\nu}(k) = 
\delta_{\mu\nu} \frac{1}{k^2} + 
\frac{\delta \eta}{k^4} \left(
\begin{array}{cc}
-k_A k_B & 0 \\
0 & k_n^2
\end{array}
\right)  + {\cal O}(\delta \eta^2)\ .
\ee

The next step is to undo the Fourier transform in the normal direction.  We have a handful of contour integrals to perform:
\be
I_0 &=& \int \frac{\d q}{2 \pi} \frac{e^{i q \, \delta y}}{{\bf k}^2 + q^2} = \frac{e^{- |{\bf k}| |\delta y|}}{2 | {\bf k}|} \ , \\
I_1 &=& \int \frac{\d q}{2 \pi} \frac{e^{i q \, \delta y}}{({\bf k}^2 + q^2)^2}  = \frac{e^{-| {\bf k}| |\delta y|} (1 + |{\bf k}| |\delta y|)}{4 |{\bf k}|^3} \ , \\
I_2 &=& \int \frac{\d q}{2 \pi} \frac{q^2 e^{i q \, \delta y}}{({\bf k}^2 + q^2)^2} =\frac{e^{-| {\bf k}| |\delta y|} (1 - |{\bf k}| |\delta y|)}{4 |{\bf k}|}\ ,
\ee where we denote $q= k_n$.
In the absence of a boundary, we can then write the partially Fourier transformed Green's function in the form
\be
\tilde G_{\mu\nu}({\bf k}, \delta y) &=&
\delta_{\mu\nu} \frac{e^{- |{\bf k}| |\delta y|}}{2|{\bf k}|} + \frac{\delta \eta \, e^{- |{\bf k}| |\delta y|} }{4 |{\bf k}|^3}
\left(
\begin{array}{cc}
- k_A k_B (1 + |{\bf k}| | \delta y|) & 0 \\
0 & |{\bf k}|^2 (1 - |{\bf k}| | \delta y| )
\end{array}
\right)
\ .
\ee
Recall that $\delta y = y - y'$.
In the presence of a boundary, depending on our choice of absolute or relative boundary conditions, we can add or subtract the reflected
Green's function $\tilde G_{\mu\nu}({\bf k}, y + y')$.  Let the resulting Green's function be $\tilde G_{\mu\nu}^{(B)}({\bf k}, y, y')$.    
To make contact with the mixed QED theory considered in the text, we would like absolute boundary conditions, i.e.\ Dirichlet for $A_n$ and Neumann for $A_B$.  In this case, the partially transformed Green's function restricted to the boundary is
\be
\tilde G_{\mu\nu}^{(B)}({\bf k},0,0) &=& \delta_{\mu\nu} \frac{1}{|{\bf k}|} + \frac{\delta \eta \, }{2 |{\bf k}|^3}
\left(
\begin{array}{cc}
- k_A k_B  & 0 \\
0 & 0
\end{array}
\right)
\ .
\ee
We can thus adopt a small gauge transformation to compensate for the additional ${\cal O}(g^2) ~ k_A k_B$ dependence in the photon self-energy (\ref{photonselfenergy}) when performing the Fourier transform \eqref{pigammaf}.

\section{Remarks on Boundary Charge in 3D}
\label{sec:3d}

Based on $d=3$ free theories \cite{Jensen:2015swa, Fursaev:2016inw}, it is tempting to conjecture the following relation between the boundary charge $b$ and $\alpha(1)$:
\be
\label{3dbrel}
b= {\pi^2 \over 8} \alpha(1) \ .
\ee  

However, it turns out there are additional subtleties in odd dimensional bCFTs so the derivation would be different from that in $d=4$ bCFTs we have considered.
The anomaly effective action for $d=3$ CFTs with a boundary is given by
\be
\label{3dacion}
\widetilde W={\mu^{\epsilon}\over \epsilon} {1\over 4 \pi}\Big( a \int_{\partial \cal{M}} \oR + b \int_{\partial \cal{M}} \tr \hat K^2 \Big) \ .
\ee 
The $a$-anomaly is topological, so only the $b$-anomaly would contribute to the $\mu$-dependent part of the stress tensor two-point function.
Denote
\be
\hat \delta^{(d)}_{AB,CD}
= {1\over 2}(\delta_{AD}\delta_{BC}+\delta_{AC}\delta_{BD})-{1\over (d-1)} \delta_{AB}\delta_{CD} \ .
\ee 
(Note $d$ is the bulk dimensionality and here $\delta^A_{A}=d-1$.) 
We can  then write the variation of the traceless part of the extrinsic curvature as
\be
\lim_{g^{CD} \to \delta^{CD}}{\delta \hat K_{AB}(x)\over \delta g^{CD}(x')}
= -{1\over 2} \hat \delta^{(d)}_{AB,CD} \partial_n \delta^d(x-x') \ .
\ee
We obtain
\be
\label{3d2}
\mu {\del \over \del \mu}\la T_{AB}(x') T_{CD}(x'') \ra^{(b)}= {b\over 2 \pi} \hat \delta^{(d=3)}_{AB,CD}\partial_{y} \delta (y-y') \partial_{y} \delta^3 (x'-x'')|_{y=0} \ .
\ee 
Note there is no normal-normal component and it is traceless only using the tangential indices.  
%It does not vanish if acting \eqref{3d2} by $\partial_A$, indicating that the conservation of the stress tensor 
%is modified and a displacement operator will contribute. By turning on  $\delta g_{nA}$, one can also have the spin-%one displacement operator correlation function:
%\be
%\label{3d3}
%\mu {\del \over \del \mu}\la T_{nA}(x') T_{nB}(x'') \ra^{(b)}&=& -{b\over \pi} \delta_{AB} \delta (y-y')  \op^2 \delta^3 %(x'-x'')|_{y=0} \  .
%\label{3d4}
%\mu {\del \over \del \mu}\la T_{nB}(x') T_{CD}(x'') \ra^{(b)}&=& -{b\over 2 \pi}  \delta (y-y') \hat \delta^{(d=3)}_{AB,CD} \partial^A \partial_{y} \delta^3 (x'-x'')|_{y=0} \ .
%\ee
%We see that the additional issue in $d=3$ bCFTs 
%is that there can be additional $purely$ boundary contributions to the two-point function that produce the log divergence, unlike
%in $d=4$  where there is no log term produced from the two-point function on the boundary, based on the power counting as we discussed before. 
%Indeed, we have used this argument to require a cancellation between $b_2$ and $c_{\rm{bry}}$ boundary contributions to the two-point function in $d=4$ bCFTs.
%In $d=3$, however, one should first classify all possible purely boundary contributions to the correlation functions (including terms that correspond to displacement
%operators which violate the conservation condition) in order to relate the $b$-charge with $\alpha(1)$.
We anticipate a key difference between the 3d and 4d cases is that the 3d stress tensor two-point function may have anomalous scale dependence associated with purely boundary terms.  
At this point, we are unsure how to derive \eqref{3dbrel} based on 
the anomalous boundary behavior of the stress tensor two-point function.
The classification of purely boundary terms of the two-point function in odd dimensions, however, goes beyond the scope of the present work so we wish to discuss them elsewhere \cite{HHJ2017}.\footnote{%
Soon after finishing this paper, we realized a straightforward way to 
relate $b$ in 3d bCFTs and $b_1$ in 4d bCFTs to correlation functions of the displacement operator.
In \cite{HHJ2017} we will provide a derivation of the relation \eqref{3dbrel}, and also show that $b_1={2\pi^6\over 35} C_{nnn}$, where $C_{nnn}$ is the coefficient of the displacement operator three-point function.}

\end{document}

%% file: introduction.tex
Quantum field theory on manifolds with a boundary finds important applications ranging from 
condensed matter physics to particle physics, from cosmology to string theory. 
In this paper, we continue an investigation into the structure of boundary conformal field theory (bCFT) begun over thirty years 
ago \cite{JC,McAvity:1993ue,McAvity:1995zd}.  We are motivated by recent progress in classifying boundary terms in the 
trace anomaly of the stress tensor for bCFTs \cite{Herzog:2015ioa,Fursaev:2015wpa,Solodukhin:2015eca}.
The trace anomaly plays a central role in our understanding of conformal field theory (CFT) and arguably, thinking of CFTs 
as fixed points of renormalization group (RG) flow, of quantum field theory more generally. In particular, the coefficient $a$ 
of the Euler-density term in the trace anomaly for even dimensional CFTs is conjectured to be monotonic under RG 
flow, $a_{\rm UV} > a_{\rm IR}$.  In two \cite{Zamolodchikov:1986gt} and four \cite{Komargodski:2011vj} dimensions, the 
conjecture is in fact proven, providing a primitive tool to map out the space of quantum field theories.

Another term in the trace anomaly of four dimensional CFTs is the square of the Weyl 
curvature with a coefficient conventionally called $c$.  In flat space, the form 
of the two-point function of the stress tensor is fixed up to an overall normalization 
constant, a constant determined by $c$ as well \cite{Osborn:1993cr}.  
Less well known is what happens when there is a boundary. In curved space, one of the additional 
boundary localized terms in the trace of the stress tensor can be schematically written $KW$ 
where $W$ is the bulk Weyl curvature and $K$ the extrinsic curvature of the boundary 
\cite{Herzog:2015ioa,Fursaev:2015wpa,Solodukhin:2015eca}.  Let us call the coefficient of this term $b_2$.  
  Ref.\ \cite{Fursaev:2015wpa} observed that for free theories, $b_2$ and $c$ were 
linearly related: $b_2=8c$ with our choice of normalization.  A bottom up holographic 
approach to the problem suggests that for interacting theories, this relation may not 
always hold \cite{Chu:2017aab,Miao:2017gyt,Astaneh:2017ghi}.    
In this paper, generalizing a method of ref.\ \cite{Osborn:1993cr} (see also \cite{Erdmenger:1996yc}), 
we argue that $b_2$ is fixed instead by the near boundary limit of the stress tensor 
two-point function in the case where the two-point function is computed in a flat half space.  
For free theories, the bulk and boundary limits of the 
two-point function are related by a factor of two, and our proposal is then 
consistent with the $b_2 = 8c$ observation.  
More generally, we find that interactions modify the relation between these limits.

To cross the logical chasm between $b_2$ and the stress tensor two-point function, our 
approach is to try to fill the chasm rather than just to build a bridge.  
With a view toward understanding the $b_2$ charge, we 
investigate bCFT more generally, in dimension $d>2$, employing 
a variety of techniques from conformal block decompositions to Feynman diagrams.  
As a result, we find a number of auxiliary results which may have interest in their own right.  

One such result is the observation that current and stress tensor two-point functions of 
free bCFTs have a universal structure.  We consider stress tensor two-point functions for the scalar, fermion, and 
$p$-form in $2p+2$ dimensions 
as well as the current two-point functions for the scalar and fermion.  Additionally, we describe their 
conformal block decompositions in detail.  
These calculations  
follow and generalize earlier work \cite{McAvity:1993ue,McAvity:1995zd, Liendo:2012hy}.  
By conformal block decomposition, we are referring to a representation of the 
two-point functions as a sum over primary operators.  
In bCFT, there are two distinct such decompositions.  Taking an operator in the bulk close to the 
boundary, we can re-express it as a sum over boundary primary fields, allowing for the 
boundary conformal block decomposition.  Alternately, bringing two operators close 
together, we have the standard operator product expansion (OPE) where we can 
express the two operators as a sum over primary fields in the 
bulk, leading to the bulk conformal block decomposition.  
Our discussion of conformal blocks is in section \ref{sec:conformalblockdecomp} and appendix \ref{sec:nullcone}.   
Figure \ref{fig:crossing} represents the two types of conformal block decomposition in pictorial form.

We find generically for free theories that the two-point 
correlators can be described by a function of an invariant cross ratio $v$ of the form 
$f(v) \sim 1 \pm v^{2 \Delta}$, where 
$\Delta$ is a scaling dimension.  
Here, $v \to 1$ is the limit that the points get close to the boundary and $v \to 0$ is the coincident limit.  
(The behavior for free scalars is in general more complicated, but the limits 
$v \to 0$ and $v \to 1$ of $f(v)$ are the same as for the functions $1 \pm v^{2 \Delta}$.)  
The 1 in $1 \pm v^{2 \Delta}$  then corresponds to the two-point function 
in the absence of a boundary, and morally at least, we can think of the $v^{2 \Delta}$ 
as the contribution of an image point on the other side of the boundary. 
 
In the context of the $b_2$-charge, let us call the relevant cross-ratio function 
for the stress tensor $\alpha(v) \sim 1 + v^{2d}$.  (Again, the function $\alpha(v)$ for a scalar is 
more complicated, but the limits $v \to 0$ and $v \to 1$ are the same.)  
In this case, we have the relation $\alpha(1)  = 2 \alpha(0)$.  
As we will see, $c$ is proportional to the bulk limit $\alpha(0)$. It follows that there will be 
a corresponding relation between $c$ and $\alpha(1)$ for 
free theories, which can be understood, given our proposed general relation \eqref{mainresult} 
between $\alpha(1)$ and $b_2$, as the equality $b_2 = 8c$ in free theories. 

What then happens for interacting theories?  
A canonical example of an interacting bCFT is the Wilson-Fisher fixed point, analyzed in either the $\epsilon$ \cite{Ei1993,McAvity:1993ue} or large $N$ \cite{McAvity:1995zd} expansion or more recently using boot strap ideas
\cite{Liendo:2012hy,Gliozzi:2015qsa}.  
(Also see \cite{Diehl:1996kd,Diehl2}.)
Two choices of boundary conditions at the planar boundary are  Dirichlet (ordinary) or Neumann (special).   Indeed, one finds generically, in both the $\epsilon$ expansion and in the large $N$ expansion, that $\alpha(1) \neq 2\alpha(0)$. 
In precisely the limit $d=4$, the Wilson-Fisher theory however becomes free and the relation $\alpha(1) = 2 \alpha(0)$ or equivalently $b_2 = 8c$ is recovered. 

We would then like to search for an interacting bCFT in $d=4$ dimensions that is tractable.  
Our strategy  
is to consider a free field in four dimensions coupled to a free field on the planar boundary in three dimensions through a classically marginal interaction that lives purely on the boundary.  We consider in fact three different examples.  Two of our examples turn out to be cousins of the Wilson-Fisher theory with a boundary, 
in the sense that, with appropriate fine tuning, they have a perturbative
IR fixed point in the $\epsilon$ expansion, $d = 4 - \epsilon$. The first example is a mixed dimensional Yukawa theory with a four dimensional scalar coupled to a three dimensional fermion. The second is a mixed dimensional scalar theory with coupled three and four dimensional scalar fields.
At their perturbative interacting fixed points,
%, which appear to be different from each other and different from the Wilson-Fisher theory, 
%naively commuting the small coupling and near boundary limits,
both interacting theories give $\alpha(1) \neq 2 \alpha(0)$ at leading order in perturbation theory.  
To our knowledge, neither theory has been examined in the literature.  Given the 
interest in the Wilson-Fisher theory with a boundary, we suspect these cousins may deserve a more in depth analysis.
Our calculations stop at one loop corrections to the propagators and interaction vertex.  While these theories are free in the IR in $d=4$ dimensions with $\epsilon =0$, if we set $\epsilon = 1$ we may be able to learn some interesting data about fixed point theories in $d=3$ with a two dimensional boundary.\footnote{%
 We considered also a theory with a five dimensional scalar $\phi$ coupled to a four dimensional boundary scalar $\eta$ in the $\epsilon$ expansion, with a classically marginal boundary coupling, $\phi^2 \eta$, in the hopes that
 there might be an IR fixed point at $\epsilon = 1$ associated with a four dimensional bulk theory.  At one loop, however, we find only a UV fixed point in $d > 5$ dimensions.  For the sake of brevity, we do not include those calculations here.
}
Unfortunately, neither of these interacting theories gives us an example of $b_2 \neq 8 c$.

The third and perhaps most interesting example consists of a four dimensional photon coupled to a three
dimensional massless fermion of charge $g$.  
The photon wave-function is not renormalized at one or two loops \cite{Teber:2012de,Teber:2014hna}.  Indeed, a simple power counting argument suggests it is not perturbatively renormalized at all. 
A Ward identity then guarantees that the $\beta$ function for the coupling $g$ vanishes.  Perturbatively, it follows that this mixed dimensional QED is exactly conformal for all values of $g$.  
The theory provides a controllable example where $\alpha(1) \neq 2 \alpha(0)$ in exactly 
four dimensions.  
A leading order calculation in perturbation theory 
indeed demonstrates that $\alpha(1) \neq 2 \alpha(0)$.

While we do not demonstrate the relation between $\alpha(1)$ and $b_2$ for mixed QED in particular, we do provide a general argument based on an effective anomaly action.  The argument is similar in spirit to Osborn and Petkou's argument \cite{Osborn:1993cr} relating $c$ and $\alpha(0)$.  The basic idea is the following.  On the one hand, an effective anomaly action for the stress tensor will produce delta-function distributions that contribute to the stress tensor two-point function in the coincident and near boundary limits.  As the effective anomaly action is constructed from the $W^2$ and $KW$ curvature terms with coefficients $c$ and $b_2$, these delta-function distributions will also have $c$ and $b_2$ dependent coefficients.  At the same time, the coincident limit of the stress tensor two-point function has UV divergences associated with similar delta-function distributions. 
Keeping track of boundary contributions, by matching the coefficients of these distributions, we obtain a constraint \eqref{mainresult} relating $b_2$ and $\alpha(1)$.

 The quantity $\alpha(1)$ is related to the coefficient of the two-point function of the displacement operator.  
 In the presence of a boundary, the Ward identity for stress tensor conservation is modified to
\be
\label{displacementdef}
\partial_\mu T^{\mu n} &=& D^n \delta(x_\perp) \ , \\
\partial_\mu T^{\mu A}  &=&  - \partial_B \hat T^{AB} \delta(x_\perp) \ ,
\ee where $\delta(x_\perp)$ is the Dirac delta function with support on the boundary, $\mu, \nu$ 
are $d$ dimensional indices, $A, B$ are tangential indices and $n$ is the normal direction.
We can identify a scalar $D^n$ displacement operator, sourced by perturbing the location of the boundary. 
Through a Gauss law pill box type argument, the operator $D^n$ is equal to the boundary limit of $T^{nn}$.
Moreover,
the value $\alpha(1)$ is proportional to the contribution of $D^n$ to the stress tensor two-point function in the boundary limit. 
A novel feature of all three boundary interacting theories, which distinguishes them from the Wilson-Fisher theory, is that in the perturbative limit, they have degrees of freedom that propagate on the boundary and an associated boundary stress tensor $\hat T^{AB} \delta(x_\perp)$.  
We expect that a classical non-zero $ \hat T^{AB}$ generally exists in theories with boundary degrees of freedom that are coupled to bulk degrees of freedom.
This boundary stress tensor is not conserved on its own, $\partial_B \hat T^{AB} \neq 0$, and 
 conservation of energy and momentum in the full theory is guaranteed through an inflow mechanism involving the boundary limit of the normal-tangential component of the full stress tensor. We have 
\be
{\rm{Classical}}:~~~T^{nn}|_{\rm bry}= D^n \ , ~~ T^{nA}|_{\rm bry} = - \partial_B \hat T^{AB} \ .
\ee
 
 While this story makes sense classically, renormalization effects alter the story non-perturbatively.  
 Because $\hat T^{AB}$ is not conserved, its scaling dimension will shift upward from the unitarity bound at $\Delta = d-1$.  
 It then no longer makes sense to separate out $\hat T^{AB}$ as a delta function-localized stress tensor; renormalization has ``thickened'' the degrees of freedom living on the boundary.  Instead, one has just the bulk stress tensor $T^{\mu \nu}$, which is conserved, and whose conservation implies 
\be
{\rm{Operator}}:~~~T^{nn}|_{\rm bry}= D^n \ , ~~ T^{nA}|_{\rm bry} = 0 \ ,
\ee  understood as an operator statement (at quantum level). Any insertion of $T^{nA}|_{\rm bry}$ in a correlation function sets that correlation function to zero.  In other words, there can be a localized, nonzero $T^{nA}|_{\rm bry}$ classically, but quantum effects smear it out.
 This renormalization effect leads to subtleties with commuting the small coupling and near boundary limits in our perturbative calculations.
For recent discussions of displacement operators, see 
\cite{Gaiotto:2013sma, Dias:2013bwa, Billo:2016cpy, Bianchi:2015liz, Balakrishnan:2016ttg, Drukker:2017xrb}.\footnote{%
	As an application of the boundary conformal anomaly, in \cite{Herzog:2016kno} we introduced 
	a notation of reduction entropy (RE). We observed that the RE intriguingly reproduces the universal entanglement entropy 
	upon a dimensional reduction, provided that $b_2=8c$ and a term $\la T^{nn} \ra$ is added in the RE.
	Interestingly, from the present work, we realize that $\la T^{nn} \ra$ in RE is the displacement operator.
	Moreover, since we find more generally that $b_2\sim \alpha(1)$, the RE 
	encodes the information about boundary conditions for interacting CFTs.  
	The entanglement entropy, when computed by introducing a conical singularity, 
	to our knowledge, however, does not seem to depend on boundary conditions. 
	It would be interesting to revisit the calculations in ref.\ \cite{Herzog:2016kno} in view
	of the results presented in this paper.
} 

 Before moving to the details,  it is worth remarking several features of this mixed dimensional QED theory.
While its bCFT aspects have not to our knowledge been emphasized, 
the theory is closely related to models of graphene and has been studied over the years 
\cite{Gorbar:2001qt,Teber:2012de,Teber:2014hna,Kotikov:2013eha, Kotikov:2016yrn} in various contexts.  
Son's model \cite{Son:2007ja} of graphene starts with charged, relativistic fermions that propagate in 2+1 dimensions 
with a speed $v_f<1$ and their electric interactions with 4d photons.  There is a $\beta$ function for $v_f$ with an IR fixed point at $v_f=1$.  Restoring the magnetic field and interactions at this IR fixed point, one finds precisely this mixed dimensional
QED  \cite{Teber:2012de}.  
Similar statements about the non-renormalization of the coupling $g$ 
can be found in the graphene literature (see e.g.\ \cite{Vozmediano:2010fz}).\footnote{This mixed QED was recently considered as a relativistic theory exhibiting fractional quantum Hall effect \cite{Son:2015xqa}.}

In the large $N$ limit where one has many fermions, this QED-like theory can be mapped to three dimensional QED in a similar large $N$ limit, with $g \sim 1/N$ \cite{Kotikov:2013eha}.  Indeed, three dimensional QED is expected to flow to a conformal fixed point in the IR for sufficiently large $N$ \cite{ANW1998,Nash1989}.  This map thus replaces a discrete family of CFTs, indexed by  $N$, with a continuous family of bCFTs, indexed by $g$.
Such a map is reminiscent of AdS/CFT, with $g$ playing the role of Newton's constant $G_N$.
More recently, Hsiao and Son \cite{Hsiao:2017lch} conjectured that this mixed QED theory should have an exact S-duality.  
Such an S-duality has interesting phenomenological consequences.  Using it, they calculate the conductivity at the self-dual point.  Their calculation is in spirit quite similar to a calculation in an AdS/CFT context for the M2-brane theory \cite{Herzog:2007ij}.

\subsection*{Outline}
An outline of the paper is as follows.  
In section \ref{sec:boundarycharge}, we review the various boundary terms that appear in the trace anomaly of bCFTs.  
In section \ref{sec:twopoint}, we first review the general structure of the two-point functions in bCFT. Then, we discuss constraints on these two-point functions. We also give the boundary and bulk conformal block decompositions.  
Our decompositions for the current two-point function (\ref{Q2formula}, \ref{currentzeropi}--\ref{currentonefirstrho}, \ref{currentonepi}) have not yet been discussed in the literature to our knowledge.  Nor have certain symmetry properties of the boundary blocks (\ref{reflectionrule}) and positivity properties of the current and stress tensor correlators (\ref{currentreflection}, \ref{stresstensorreflection}).  
In section \ref{sec:proof}, we give our argument relating $\alpha(1)$ to $b_2$-charge in 4d bCFTs. We also review how $\alpha(0)$ is related to the standard bulk $c$-charge. 
In section \ref{sec:free}, we discuss two-point functions for free fields, including a conformal scalar, a Dirac fermion and gauge fields.  
In particular, the discussion of $p$-forms in $2p+2$-dimensions is to our knowledge new. 
Lastly, in section \ref{sec:interaction}, we introduce our theories with classically marginal boundary interactions.
A discussion section describes some projects for the future.
In Appendix \ref{sec:nullcone} we review how to derive the conformal blocks for scalar, vector, and tensor operators in the null cone formulation. 
Appendix \ref{sec:Weyl} describes some curvature tensors and variation rules relevant to the discussion of the trace anomaly 
in sections \ref{sec:boundarycharge} and \ref{sec:proof}.
We discuss gauge fixing of the mixed QED in Appendix \ref{sec:gaugefixing}.  Some remarks on the 3d boundary anomalies are given in Appendix \ref{sec:3d}.

%% file: boundarycharge.tex
Considering a classically Weyl invariant theory embedded in a curved spacetime background, the counterterms added to regularize divergences give rise to the conformal (Weyl) anomaly, which is defined as a non-vanishing expectation of the trace of the stress tensor. 
The conformal anomaly in the absence of a boundary is well-known, in particular in $d=2$ and $d=4$ dimensions; see for instance \cite{Deser:1993yx, Duff:1993wm} for reviews. There is no conformal anomaly in odd dimensions in a compact spacetime.
In the presence of a boundary, there are new Weyl anomalies localized on the boundary and their structure turns out to be rather rich. There are also new central charges defined as the coefficients of these boundary invariants. 
One expects that these boundary central charges can be used to characterize CFTs with a boundary or a defect, in a similar way that one characterizes CFTs without a boundary using the bulk central charges.

For an even dimensional CFT$_d$ with $d=2n+2; n=0,1,2,...$, the Weyl anomaly can be written as
\be
\label{traceanomalyeven}
\langle {T^\mu}_\mu \rangle^{d=2n+2} &=& \frac{4}{d! \Vol(S^d)} \Big[ \sum_i c_i {\cal I}_i+\delta(x_\perp) \sum_j b_j I_j- (-1)^{{d\over2}} a_{d}\Big(E_d+\delta(x_\perp) E^{\rm{(bry)}}\Big) \Big]  \ . \nn\\
\ee 
We normalize the Euler density $E_d$ such that integrating $E_d$ over an $S^d$ yields $d! \rm{Vol}(S^d)$. 
We denote $E^{\rm{(bry)}}$ as the boundary term of the Euler characteristic, which has a Chern-Simons-like structure \cite{EGH, Myers}. See \cite{Herzog:2015ioa} for an extensive discussion. 
Notice that $E^{\rm{(bry)}}$ is used to preserve the conformal invariance of the bulk Euler density when a boundary is present, so its coefficient is fixed by the bulk $a$-charge. In this paper we are interested in a smooth and compact codimension-one boundary so we do not include any corner terms. 
The normalizations of local Weyl covariant terms, ${\cal I}_i$ and $I_j$, are defined such that they simply have the same overall factor of the Euler anomaly. 
One can certainly adopt a different convention and rescale central charges $a, c_i$ and $b_j$. 
The numbers of the local Weyl covariant terms vary depending on the dimensions. We emphasize that, since $ {\cal I}_i$ and $I_j$ are independently Weyl covariant, there are no constraints relating bulk charges $c_i$ to $b_j$ from an argument based solely on Weyl invariance of the integrated anomaly.

For an odd dimensional CFT$_d$ with $d=2n+1$, $n=1,2,3,\ldots$ there is no bulk Weyl anomaly. 
In the presence of a boundary, however, there can be boundary contributions. We write
\be
\label{traceanomalyodd}
\langle {T^\mu}_\mu \rangle^{d=2n+1} &=& \frac{2}{(d-1)! \Vol(S^{d-1})} \delta(x_\perp) \Big(\sum_i b_i I_i + (-1)^{(d+1)\over 2} a_{d} \oE_{d-1} \Big) \ ,
\ee 
where $\oE_{d-1}$ is the boundary Euler density defined on the $d-1$ dimensional boundary. 
The coefficient $a_d$ with odd $d$ is an $a$-type  boundary charge. 
Similarly, $I_i$ represents independent local Weyl covariant terms on the boundary. 

An important boundary object is the traceless part of the extrinsic curvature defined as
\be
\hat K_{AB}= K_{AB}- {h_{AB}\over d-1} K \ ,
\ee where $h_{AB}$ is the induced metric on the boundary. (See appendix \ref{sec:Weyl} for notation.) 
$\hat K_{AB}$ transforms covariantly under the Weyl transformation. 

Note that we have dropped terms that depend on the regularization
scheme in \eqref{traceanomalyeven} and \eqref{traceanomalyodd}. For instance, the $\Box R$ anomaly in $d=4$ CFTs can be removed by adding a finite counterterm $R^2$.
It is worth mentioning that Wess-Zumino consistency rules out the possibility of a boundary total derivative anomaly in $d=4$ CFTs \cite{Herzog:2015ioa}.

Let us consider explicit examples. In $d=2$ one has
\be
\langle T^{\mu}{}_{\mu}\rangle^{d=2} = \frac{a}{2 \pi}\left( R + 2 K \delta(x_{\perp})\right)\ .
\ee 
One can replace the anomaly coefficient $a$ with the more common $d=2$ central charge $c = 12 a$. Note $c=1$ for a free conformal scalar or a Dirac fermion. The $d=2$ bCFTs have been a rich subject but since there is no new central charge, in this paper we will not discuss $d=2$ bCFTs. Interested readers may refer to \cite{Cardy:2004hm} for relevant discussion of $d=2$ bCFTs and their applications.

In $d=3$ the anomaly contributes purely on the boundary. One has \cite{Graham:1999pm}
\be
\langle T^{\mu}{}_{\mu}\rangle^{d=3} = \frac{\delta(x_{\perp})}{4 \pi}\left(a \oR + b \tr \hat K^2\right)\ ,
\ee 
where $\tr \hat K^2= \tr K^2 - {1\over 2} K^2$ and $\oR $ is the boundary Ricci scalar. 
Restricting to free fields of different spin $s$, the values of these charges are
\be
a^{s=0}&=&-{1\over 96} ~~ {\rm{(D)}}\ ,~~~ a^{s=0}={1\over 96} ~~ {\rm{(R)}} \ , ~~~a^{s={1\over 2}}=0 \ ,
\ee
and
\be
b^{s=0}&=&{1\over 64} ~~ {\rm{(D~or~R)}}\ ,~~~b^{s={1\over 2}}= {1\over 32} \ ,
\ee 
where  $(D)/(R)$ stands for Dirichlet/Robin boundary conditions.  
Neumann boundary conditions in general do not preserve conformal symmetry, but there is a particular choice of
Robin boundary condition involving the extrinsic curvature which does.
The quantity $b$ for the scalar with Dirichlet and Robin boundary conditions was first computed to our knowledge by refs.\ 
 \cite{Nozaki:2012qd} and \cite{Jensen:2015swa}, respectively.  The complete table can be found in \cite{Fursaev:2016inw}.

In $d=4$ CFT, the conformal anomaly reads
\be
\label{4dTrace}
\langle {T^\mu}_\mu \rangle^{d=4}&=&
{1\over 16 \pi^2} \Big( c W_{\mu\nu\lambda\rho}^2- a E_4\Big)\nn\\
&&~~~~~~~~~~~~+{\delta(x_{\perp})\over 16 \pi^2} \Big(a E^{\rm{(bry)}}_4-b_1 \tr\hat{K}^3-b_2
h^{\alpha\gamma}\hat{K}^{\beta\delta}W_{\alpha\beta\gamma\delta} \Big) \ ,
\ee where
\be
E_4
&=&{1\over 4} \delta^{\mu \nu \lambda\rho}_{\sigma\omega\eta\delta} R^{\sigma\omega}{}_{\mu \nu} R^{\eta\delta}{}_{\lambda\rho} \ , \\
E^{\rm{(bry)}}_4
&=& -4 \delta^{ABC}_{DEF}~
K^{D}_{A}\left({1\over 2} R^{EF}{}_{BC} +
{2\over 3} K^{E}_{B} K^{F}_{C} \right)\ , \\
\tr\hat{K}^3&=&\tr K^3-K \tr K^2+{2\over 9} K^3 \ , \\
h^{\alpha\gamma}\hat{K}^{\beta\delta}W_{\alpha\beta\gamma\delta} &=&
R^{\mu}_{~\nu \lambda \rho} K_{\mu}^{\lambda} n^\nu n^\rho -{1\over 2}
R_{\mu\nu} (n^\mu n^\nu K + K^{\mu\nu}) +{1\over 6} KR \ ,
\ee with $\delta^{\mu \nu \lambda\rho}_{\sigma\omega\eta\delta}/\delta^{ABC}_{DEF}$ being the bulk/boundary generalized Dirac delta function, which evaluates to $\pm 1$ or 0. 
%
%In our notation, the Greek indices are bulk indices while the Latin indices are boundary indices. 
%(When one uses Greek indices for boundary variables one has to keep in mind that the normal component is empty.) 
%
Because of the tracelessness and symmetry of the Weyl tensor, one can write $h^{\alpha\gamma}\hat{K}^{\beta\delta}W_{\alpha\beta\gamma\delta} = -{K}^{AB}W_{nAnB}$. The coefficients $b_1$ and $b_2$ are new central charges. The values of these charges were computed for free theories. %Let us list the results. 
The bulk charges are independent of boundary conditions and are given by
\be
a^{s=0}&=&{1\over 360}\ ,
~a^{s={1\over 2}}={11\over 360} \ ,
~a^{s={1}}={31\over 180}\ , \\
c^{s=0}&=&{1\over 120}\ ,
~c^{s={1\over 2}}={1\over 20} \ ,
\label{cs1}
~c^{s={1}}={1\over 10}\ ,
\ee
(see e.g.\ \cite{BirrellDavies}).
The boundary charge $b_1$ of a scalar field depends on boundary conditions. One has 
\be
b_1^{s=0}&=&{2\over 35}~{\rm{(D)}}\ ,
~ b_1^{s=0}={2\over 45}~{\rm{(R)}}\ ,
~b_1^{s={1\over 2}}={2\over 7}~{\rm{(D ~or~ R)}} \ ,
~b_1^{s={1}}={16\over 35}~{\rm{(D~or~R)}}\ .
\ee
For scalar fields, these results were first obtained for Dirichlet boundary conditions by \cite{JM} and for Robin conditions by \cite{Moss}.  This list is duplicated from the more recent 
ref.\ \cite{Fursaev:2015wpa} where standard heat kernel methods are employed. 
Finally, from free theories one finds
\be
\label{8c}
b_2=8c \ ,
\ee 
independent of boundary condition \cite{Fursaev:2015wpa,Solodukhin:2015eca}.  (The result for $b_2$ for scalar fields with Dirichlet boundary conditions was computed first to our knowledge in \cite{DS}.)
It is one of the main motivations of this work to understand how general the relation \eqref{8c} is.

The complete classification of conformal anomaly with boundary terms in five and six dimensions, to our knowledge, has not been given; see \cite{Solodukhin:2015eca} for recent progress. 
Certainly, it is expected that the numbers of boundary Weyl invariants increase as one considers higher dimensional bCFTs.

%% file: twopoint.tex
We would like to first review the general construction of conformal field theory two-point functions involving a scalar operator $O$, a conserved current $J^\mu$, and a stress tensor $T^{\mu\nu}$ in the presence of a planar boundary. Much of our construction can be found in the literature, for example in refs.\ \cite{McAvity:1993ue,McAvity:1995zd, Liendo:2012hy}.  However, some details are to our knowledge new. %\CH{Changed.}  
We provide the conformal blocks for the current-current 
two-point functions
 (\ref{Q2formula}, \ref{currentzeropi}--
 \ref{currentonefirstrho}, 
 \ref{currentonepi}).  
We also remark on order of limits, positivity (\ref{currentreflection}, \ref{stresstensorreflection}) and some symmetry (\ref{reflectionrule}) properties more generally.

\subsection{General Structure of Two-Point Functions}

A conformal transformation $g$ is a combination of a diffeomorphism $x^\mu \to x_g^\mu(x)$ and a local scale transformation $\delta_{\mu\nu} \to \Omega_g(x)^{-2} \delta_{\mu\nu}$ that preserves the usual flat metric $\delta_{\mu\nu}$ on ${\mathbb R}^d$.  The group is isomorphic to $O(d+1,1)$ and is generated by rotations and translations, for which $\Omega_g = 1$, and spatial inversion $x^\mu \to x^\mu / x^2$, for which $\Omega_g = x^2$.  
In analogy to the rule for transforming the metric, given a tensor operator $O^{\mu_1 \cdots \mu_s}$ of weight $\Delta$, we can define an action of the conformal group
\be
O^{\mu_1 \cdots \mu_s}(x) \to \Omega_g^{\Delta + s} \left( \prod_{j=1}^s \frac{\partial {x_g}^{\mu_j}}{\partial x^{\nu_j}} \right)
O^{\nu_1 \cdots \nu_s}(x)  \ .
\ee
In this language, $J^\mu$ and $T^{\mu\nu}$ have their usual engineering weights of $\Delta = d-1$ and $d$, respectively.  
Notationally, it is useful to define the combination ${(R_g)^\mu}_\nu \equiv \Omega_g \frac{\partial {x_g}^\mu}{\partial x^\nu}$.
Given the action of $R_g$ on the metric, it is clearly an element of $O(d)$.  In a coordinate system $x = (y, {\bf x})$, a planar boundary at $y=0$ is kept invariant by only a $O(d,1)$ subgroup of the full conformal group, in particular, the subgroup generated by rotations and translations in the plane $y=0$ along with inversion $x^\mu \to x^\mu / x^2$.  

While in the absence of a boundary, one-point functions of quasi-primary operators vanish and two-point functions have a form fixed by conformal symmetry, the story is more complicated with a boundary.
A quasi-primary scalar field $O_\Delta$ of dimension $\Delta$ can have an expectation value:
\be
\label{O}
\la O_\Delta (x) \ra = {a_\Delta \over (2y)^\Delta} \ .
\ee
The coefficients $a_{\Delta}$ play a role in the bulk conformal block decomposition of the two-point function, as we will
see later.
One-point functions for operators with spin are however forbidden by conformal invariance.

To some extent, the planar boundary functions like a mirror.  In the context of two-point function calculations, in addition to the location $x = (y,{\bf x})$ and $x' = (y', {\bf x}')$ of the two operators, there are also mirror images at $(- y,{\bf x})$ and $(-y',{\bf x}')$.   With four different locations in play, one can construct cross ratios that are invariant under the action of the conformal subgroup.  Most of our results will be expressed in terms of the quantities 
\be
\xi &=& {(x-x')^2\over 4yy'}\ , \\
v^2 &=& {(x-x')^2 \over (x - x')^2 + 4yy'} = {\xi \over \xi + 1}\ .
\ee
Like four-point correlators in CFT without a boundary, the two-point correlators we consider can be characterized by a handful of functions of the cross ratios $\xi$ or equivalently $v$. 
In the physical region, one has $0\leq\xi\leq\infty$ and $0\leq v\leq1$.
It will be useful to introduce also the differences
\be
s \equiv x - x' \ , ~~ {\bf s} \equiv \bf x-\bf x' \ .
\ee 

Following ref.\ \cite{McAvity:1995zd}, we construct the two-point correlation functions out of weight zero tensors with nice 
bilocal transformation properties under $O(d,1)$.  
In addition to the metric $\delta_{\mu\nu}$, there are three:\footnote{In this section we follow the notation in   \cite{McAvity:1993ue,McAvity:1995zd} where the normal vector is inward-pointing. In sections 4, 5 and 6, we will adopt instead an outward-pointing normal vector.}
\be
I_{\mu\nu}(x)&=& \delta_{\mu\nu}-2 {x_\mu x_\nu\over x^2}\ , \\
X_\mu &=& y {v\over \xi}\del_\mu \xi= v \left(\frac{2y}{s^2} s_\mu -  n_\mu \right)
\ , \\
X'_{\mu} &=& y' {v\over \xi}\del'_{\mu} \xi=v \left(-\frac{2y'}{s^2} s_\mu -  n_\mu \right) \ .
\ee 
The transformation rules are $X \to R_g(x) \cdot X$,
${X'} \to R_g(x')\cdot  {X'} $, and the bilocal 
$I^{\mu\nu}(s) \to {R_g(x)^\mu}_\lambda {R_g(x')^\nu}_\sigma I^{\lambda \sigma}(s)$. One has $X_\mu'=I_{\mu\nu}(s) X^\nu$.
In enforcing the tracelessness of the stress tensor, it will be useful to note that
\be
X_\mu X^\mu = X'_{\mu} {X'}^{\mu} = 1\ .
\ee

\subsection*{Two-Point Functions}
We now tabulate the various two-point functions
\be
\langle O_1(x) O_2(x') \rangle &=& \frac{\xi^{-(\Delta_1+\Delta_2)/2}}{(2 y)^{\Delta_1} (2 y')^{\Delta_2}} G_{O_1 O_2} (v)
\label{OOtwo}
 \ , \\
\langle J_\mu(x) O(x') \rangle &=& \frac{ \xi^{1-d} }{(2y)^{d-1} (2y')^\Delta}X_\mu f_{JO}(v)
\label{JO}
 \ , \\
 \langle T_{\mu\nu}(x) O(x') \rangle &=& \frac{\xi^{-d} }{(2y)^d (2y')^\Delta} \alpha_{\mu\nu} f_{TO}(v) 
\label{TO}
\ , \\
\langle J_\mu(x) J_\nu(x') \rangle &=& \frac{ \xi^{1-d} }{(2y)^{d-1} (2y')^{d-1}} \Big( I_{\mu\nu}(s) P(v) + X_\mu X'_\nu Q (v) \Big) 
\label{JJ}
\ , \\
\langle T_{\mu\nu}(x) V_\lambda(x') \rangle &=& \frac{ \xi^{-d}}{(2 y)^d (2 y')^{\Delta}}
\Bigl[ \left( I_{\mu\lambda}(s) X_\nu + I_{\nu \lambda}(s) X_\mu - \frac{2}{d} g_{\mu\nu} X'_\lambda \right) \, f_{TV}(v) \nonumber \\
&&
~~~~~~~~~~~~~~~~+\alpha_{\mu\nu}  X'_\lambda \, g_{TV} (v) \Bigr]
 \label{TJ}
 \ , \\
\langle T_{\mu\nu}(x) T_{\lambda \sigma}(x') \rangle &=& {\xi^{-d} \over (2y)^d (2y')^{d}}  \Big[ \alpha_{\mu\nu} \alpha'_{\si\rho}  A(v) +  \beta_{\mu\nu,\sigma\rho} B(v)+ I_{\mu\nu,\si\rho}(s) C(v)\Big] \ ,
\label{TT}
\ee 
where $\Delta_1/\Delta_2$ is the scaling dimension of $O_1/O_2$ and 
\be
 \alpha_{\mu\nu}&=& \Big (
X_\mu X_\nu - {1\over d} \delta_{\mu \nu} \Big) \ , \; \; \;
 \alpha'_{\mu\nu}= \Big(
X'_\mu X'_\nu - {1\over d} \delta_{\mu \nu} \Big) \ , \\
\beta_{\mu\nu,\sigma\rho}&=& 
\Big(X_\mu X'_{ \si} I_{\nu\rho}(s)+X_\nu X'_{ \si} I_{\mu\rho}(s)+X_\mu X'_{ \rho} I_{\nu\si}(s)+ X_\nu X'_{ \rho} I_{\mu\si}(s) \nn\\
&&- {4\over d} \delta_{\si\rho}
X_\mu X_\nu  - {4\over d}\delta_{\mu\nu} X'_{\si} X'_{\rho} + {4\over d^2} \delta_{\mu\nu} \delta_{\si\rho} \Big)\ , \\
I_{\mu \nu,\si \rho} (s) &=& {1\over 2}\Big( I_{\mu \si}(s) I_{\nu \rho}(s) + I_{\mu \rho}(s) I_{\nu \si} (s) \Big)- {1\over d} \delta_{\mu\nu} \delta_{\si \rho}  \  .
\ee
In writing the tensor structures on the right hand side, we have enforced tracelessness $T^\mu_\mu = 0$.  However, we have not yet made use of the conservation conditions $\partial_\mu J^\mu = 0$ and $\partial_\mu T^{\mu\nu} = 0$.\footnote{%
While these conservation conditions may be altered by boundary terms involving displacement operators, away from the boundary they are strictly satisfied.
}  
The conservation conditions 
fix (\ref{JO}) and (\ref{TO}) up to constants $c_{JO}$ and $c_{TO}$:
\be
f_{JO} = c_{JO} v^{d-1} \ , \; \; \;
f_{TO} = c_{TO} v^d \ .
\ee
The mixed correlator $\langle T^{\mu\nu}(x) V^\lambda(x') \rangle$ is fixed up to two constants, $c_{TV}^\pm$:
\be
f_{TV} &=&  c_{TV}^+ v^{d+1} + c_{TV}^- v^{d-1}\ , \\
g_{TV} &=& -(d+2) c_{TV}^+ v^{d+1} + (d-2) c_{TV}^- v^{d-1} \ . 
\ee
If we further insist that the vector $V^\mu = J^\mu$ is a conserved current, such that $\Delta = d-1$, then the correlator is fixed up to one undetermined number, 
$c_{TV}^\pm = c_{TJ}$.

The $\langle J^\mu(x) J^\nu(x') \rangle$ and $\langle T^{\mu\nu}(x) T^{\lambda \sigma}(x') \rangle$ correlation functions on the other hand are fixed up to a single function by conservation. The differential equations are 
\be
v \partial_v (P + Q) &=& (d-1) Q \ , 
\label{Jcons}
\\ 
(v \partial_v - d)(C + 2 B) &=& - \frac{2}{d} (A + 4 B) - d C \ , 
\label{Tcons1}
\\
(v \partial_v - d)((d-1)A + 2 (d-2)B) &=& 2 A - 2(d^2 - 4) B \ .
\label{Tcons2}
\ee 
This indeterminancy stands in contrast to two (and three) point functions without a boundary, where conformal invariance uniquely fixes their form up to constants.

In the coincidental or bulk limit $v \to 0$, the operators are much closer together than they are to the boundary, and we expect to recover the usual conformal field theory results in the absence of a boundary.  We thus apply the boundary conditions $A(0) = B(0) = Q(0) = 0$.
The asymptotic values $C(0)$ and $P(0)$ are then fixed by the corresponding stress tensor and current two-point functions in the absence of a boundary; we adopt the standard notation, $C(0)=C_T$ and $P(0)=C_J$. The observables $C_T$ and $C_J$ play important roles when analyzing CFTs. In particular, for a free $d=4$ conformal field theory of $N_s$
scalars, $N_f$ Dirac fermions, and $N_v$ vectors, one has \cite{Osborn:1993cr} 
\be
C_T=  {1\over 4\pi^4} \left({4\over 3} N_s+ 8 N_f +16 N_{v} \right) \ . 
\ee 
By unitarity (or reflection positivity), $C_T>0$.\footnote{See \cite{Osborn:2016bev} for the discussion of $C_T$ in   non-unitary CFTs with four-and six-derivative kinetic terms.} A trivial theory has $C_T=0$. 
Similarly, we require that $G_{O_1 O_2} (0) = \K \delta_{\Delta_1 \Delta_2}$ for some constant $\K>0$.  

The decomposition of the two-point functions into $A(v)$, $B(v)$, $C(v)$, $P(v)$, and $Q(v)$ was governed largely by
a sense of naturalness with respect to the choice of tensors $X^\mu$ and $I^{\mu\nu}$ rather than by some guiding physical principal.  Indeed, an alternate decomposition was already suggested in the earlier paper ref.\ \cite{McAvity:1993ue}.  While uglier from the point 
of view of the tensors $X_\mu$ and $I_{\mu\nu}$, it is nevertheless in many senses a much nicer basis.
This alternate decomposition, discussed below, is more natural from the point of view of reflection positivity.  It also diagonalizes 
the contribution of the displacement operators in the boundary conformal block decomposition.    

This basis adopts the following linear combinations:
\be
\label{ax}
\alpha(v) &=& \frac{d-1}{d^2} \left[ (d-1) (A + 4 B) + d C \right] \ , \\
\label{gx}
\gamma(v) &=& - B - \frac{1}{2} C \ , \\
\label{ex}
\epsilon(v) &=& \frac{1}{2} C \ .
\ee
Ref.\ \cite{McAvity:1993ue} motivated these combinations by restricting 
$x$ and $x'$ to lie on a line perpendicular to the boundary, taking ${\bf x}={\bf x}'=0$:
\be
\label{alter}
\lim_{{\bf x}= {\bf x}'\to 0} \la T_{\mu\nu} (x) T_{\si\rho}(x') \ra ={A_{\mu\nu \sigma \rho}\over s^{2d}}\ .
\ee 
In this case, one finds
\be
A_{nnnn} &=& \alpha(v) \ ,   \\  \; \; \; 
A_{ABnn} &=& A_{nnAB} = - \frac{1}{d-1} \alpha(v)  \delta_{AB} \ , \\  \; \; \; 
\label{gammadef}
A_{AnBn} &=& \gamma(v)  \delta_{AB} \ , \\
A_{ABCD} &=& \epsilon(v)  (\delta_{AC} \delta_{BD} + \delta_{AD}\delta_{BC}) - \frac{1}{d-1}\left(2 \epsilon(v)  - \frac{\alpha(v) }{d-1} \right) \delta_{AB} \delta_{CD} \ .
\ee 
Recall that the coincidental limit corresponds to $v=0$ and the boundary limit to $v=1$, where
in this perpendicular geometry $v={|y-y'|\over y+y'}$.
Relating these new linear combinations to $C(0)$, for a non-trival unitary conformal field theory, we have 
\be
\label{unitary1}
\alpha(0)&=& {d-1\over d} C_T > 0 \ , \; \; \;
\gamma(0) =-\epsilon(0)=-{1\over 2} C_T <0 \ .
\ee  

One can play exactly the same game with the current:
\be
\lim_{{\bf x} = {\bf x'} \to 0} \langle J_\mu (x) J_\nu(x') \rangle = \frac{A_{\mu\nu}}{s^{2(d-1)}} \ ,
\ee
where 
\be
\label{pidef}
A_{nn} &=& \pi(v) = P(v) +Q(v)  \ , \\ 
A_{AB} &=& \rho(v) \delta_{AB} = P(v)  \delta_{AB}\ .
\label{rhodef}
\ee

\subsection*{Comments on Order of Limits}

There are subtleties when considering various limits of the objects $v$, $X_\mu$ and $X'_\mu$. 
We define the coincidental (or bulk) limit to be $s \to 0$ with $y$, $y' \neq 0$.  In this limit, $v \to 0$ and 
\be
\label{coincidentXX}
\lim_{s \to 0} X_\mu  = 0 = \lim_{s \to 0} X'_\mu\ .
\ee
We define the boundary limit to be $y \to 0$ and $y' \to 0$ with $s \neq 0$.  In this limit, we find instead that $v \to 1$ and
\be
\label{bryXX}
\lim_{y,y' \to 0} X_\mu = -n_\mu  = \lim_{y, y' \to 0} X'_\nu\ .
\ee 
We see that if one imposes the coincidental limit after the boundary limit has been imposed, the result is different from \eqref{coincidentXX}. 

%There is yet a third special case, that was used in \cite{McAvity:1993ue} in calculating the stress tensor two-point function. %where both $x$ and $x'$ lie on a perpendicular to the boundary, so that ${\bf s} = 0$.  In this case, depending on the sign of $y-y'$, one finds
In the special case where both $x$ and $x'$ lie on 
a perpendicular to the boundary, depending on the sign of $y-y'$, one instead finds
\be
\lim_{{\bf s}=0}X_\mu&=&-\lim_{{\bf s}=0}X'_\mu= n_\mu~~~ (y>y'\neq 0)\ , \\
\lim_{{\bf s}=0}X_\mu&=&-\lim_{{\bf s}=0}X'_\mu= -n_\mu~~~ (y< y'\neq 0) \ .
\ee  
The following quantity is then independent of the relative magnitudes of $y$ and $y'$,
\be
\label{bulkXX}
\lim_{{\bf s}=0} X_\mu X'_\nu=- n_\mu n_\nu \ .
\ee
A confusing aspect about this third case is that having taken this collinear limit, if we then
further take a boundary $y \to 0$ or a coincident $y \to y'$ limit, the answer does not
agree with either (\ref{coincidentXX}) or (\ref{bryXX}). 
In the near boundary limit, one finds that $A_{nAnB} = -\gamma \delta_{AB}$ while
restricting the insertions to a line perpendicular to the boundary, one finds instead $A_{nAnB} = \gamma \delta_{AB}$. 
In general, when comparing physical quantities, one will have to fix an order of limits to avoid the sign ambiguity.
In this case, however, due to our previous arguments in the introduction, we expect that $\gamma(1) = 0$ generically 
under conformal boundary conditions.

\subsection{Reflection Positivity and Bounds}

Unitarity in Lorentzian quantum field theory is equivalent to
the reflection positivity in quantum field theory with Euclidean signature.
To apply reflection positivity, let us consider the case where the coordinates 
\be
\label{Plane}
x = (y, z, {\bf 0}) \ ,  \; \;  x'=(y, -z, {\bf 0}) \ ,  \; \;  s_\mu=(0, 2z, {\bf 0})  \ ,
\ee 
lie in a plane located at a non-zero $y$, parallel to the boundary.  Denoting this plane as ${\cal P}$, we introduce a reflection operator $\Theta_{\cal P}$ such that the reflection with respect to ${\cal P}$ gives $\Theta_{\cal P}(x) = x'$.  The square of  $\Theta_{\cal P}$ is the identity operator.  
Acting on a tensor field,  $\Theta_{\cal P}(F_{\mu_1 \cdots \mu_n}(x))$, $\Theta_{\cal P}$ will flip the overall sign if there are an odd number of 2 ($z$-direction) indices.  The statement of reflection positivity for a tensor operator is that
\be
\langle F_{\mu_1 \cdots \mu_n}(x) \Theta_{\cal P}(F_{\nu_1 \cdots \nu_n}(x)) \rangle  \ , 
\ee
treated as a $d^n \times d^n$ matrix, has non-negative eigenvalues. (Note this reflection operator acts on just one of the points; when it acts on the difference it gives $\Theta_{\cal P}(s)=0$.)    
In our particular choice of frame \eqref{Plane}, $\Theta_{\cal P} (I_{\mu\nu}(s)) = \delta_{\mu\nu}$ 
and $\Theta_{\cal P}(X'_\mu) = X_\mu$.  Making these substitutions in the current and stress tensor correlators (\ref{JJ}) and (\ref{TT}), we can deduce eigenvectors and corresponding eigenvalues.  

For the current two-point function, $X_\mu$ is an eigenvector with eigenvalue proportional to $\pi$ while $\delta_{\mu 3}$
is an eigenvector with eigenvalue proportional to $\rho$, with positive coefficients of proportionality.  
(Instead of 3, we could have chosen any index not corresponding to the $y$ and $z$ directions.) 
Thus we conclude that 
\be
\label{currentreflection}
\pi(v) \geq 0 \ , \; \; \; \rho(v) \geq 0 \ ,
\ee 
for all values of $v$, $0 \leq v\leq 1$.  
For the stress tensor, $\alpha_{\mu\nu}$, $X_{(\mu} \delta_{\nu)3}$, and $\delta_{3(\mu} \delta_{\nu) 4}$ are eigenvectors with eigenvalues proportional to $\alpha$, $-\gamma$, and $\epsilon$, demonstrating the positivity that\footnote{%
For instance, the eigen-equation for $\alpha_{\mu\nu}$ is 
\be
\la T_{\mu\nu}(x) \Theta_P(T_{\lambda \sigma}(x))\ra \alpha^{\lambda \sigma}=  \frac{d}{d-1} \frac{\alpha(v)}{s^{2d}} \alpha_{\mu\nu}  \ .
\ee 
%Reflection positivity implies the eigenvalue is non-negative, which in turn implies the function $\alpha(v)$ is non-negative.
%Our reflection is parallel, not perpendicular to the boundary. 
}
\be
\label{stresstensorreflection}
\alpha(v) \geq 0 \ , \; \; \; -\gamma(v) \geq 0 \ , \; \; \; \epsilon(v) \geq 0 \ .
\ee 
With these positivity constraints in hand, one can deduce a couple of monotonicity properties from the conservation relations, re-expressed in terms of $\pi$, $\rho$, $\alpha$, $\gamma$, and $\epsilon$:
\be
\label{floweqpi}
(v \partial_v - (d-1)) \pi &=& -(d-1) \rho \leq 0 \ , \\
\label{floweqalpha}
(v \partial_v -d) \alpha &=& 2 (d-1) \gamma  \leq 0 \ ,  \\
(v \partial_v -d) \gamma &=& \frac{d}{(d-1)^2} \alpha + \frac{(d-2)(d+1)}{d-1} \epsilon \geq 0 \ .
\label{floweqgamma}
\ee The last two inequalities further imply  $(v \partial_v -d)^2 \alpha \geq 0$. 
While these inequalities provide some interesting bounds for all values of $v$, they unfortunately do not lead to a strong 
constraint on the relative magnitudes of the two end points of $\alpha$, $\alpha(1)$ and $\alpha(0)$, a constraint, as we will see, that could be interesting in relating the boundary charge $b_2$  in \eqref{4dtrace} 
to the usual central charge $c$ in $d=4$ CFTs. 

Using current conservation and our new basis of cross-ratio functions, we can write the stress tensor and current two-point functions in yet a third way, eliminating $\rho$, $\gamma$, and $\epsilon$ in favor of derivatives of $\pi$ and $\alpha$. This third way will be useful when we demonstrate the relationship between $\alpha(1)$ and the boundary central charge $b_2$.  We write
\be
\langle J_{\mu} (x) J_{\nu} (x') \rangle &=&
\frac{1}{s^{2d-2}} \left( \pi(v) I_{\mu\nu}(s) - \frac{v \partial_v \pi}{d-1} \hat I_{\mu\nu}(s) \right) \ , \\
\label{magicTT}
\langle T_{\mu\nu}(x)T_{\rho \sigma}(x') \rangle &=&
\frac{1}{s^{2d}} \Bigl[ 
 \alpha \frac{d}{d-1} I_{\mu\nu, \rho \sigma}(s) +  
v^2 \partial_v^2 \alpha \frac{ \hat I_{\mu\nu, \rho \sigma}  }{(d-2)(d+1)} 
\nn\\
&&
~~~~~ - v \partial_v \alpha
\left( \frac{ \hat \beta_{\mu\nu, \rho \sigma}}{2(d-1)}  + \frac{(2d-1) \hat I_{\mu\nu, \rho \sigma}}{(d-2)(d+1)}\right)
\Bigr] \ , 
\ee
where we have defined some new tensorial objects in terms of the old ones:
\be
\hat I_{\mu\nu}(s) &\equiv& I_{\mu\nu}(s) - X_\mu X'_\nu \ , \\
\hat I_{\mu\nu, \rho \sigma}(s) &\equiv& I_{\mu\nu, \rho \sigma}(s)  -  \frac{d}{d-1} \alpha_{\mu\nu} \alpha'_{\rho \sigma} 
- \frac{1}{2} \hat \beta_{\mu\nu, \rho \sigma} \ , \\
\hat \beta_{\mu\nu, \rho \sigma} &\equiv& \beta_{\mu\nu, \rho \sigma} -4 \alpha_{\mu\nu} \alpha'_{\rho \sigma} \ .
\ee
One nice feature of the hatted tensors is their orthogonality to the $X_\mu$ and $X_\rho'$ tensors.  In particular
\be
X_\mu \hat I^{\mu\rho} &=& 0 = \hat I^{\mu\rho} X'_\rho \ , \\
X_\mu \hat I^{\mu\nu, \rho \sigma} &=& 0 = \hat I^{\mu\nu, \rho \sigma} X'_\rho\ , \\
X_\mu X_\nu \hat \beta^{\mu\nu, \rho \sigma} &=& 0 = \hat \beta^{\mu\nu, \rho \sigma} X'_\rho X'_\sigma \ .
\ee 
In the near boundary limit, $v\to 1$, since $X_\mu$, $X_\mu' \to - n_\mu$, only the tangential components $\hat I_{AB}$ and $\hat I_{AB,CD}$ of $\hat I_{\mu\nu}$ and $\hat I_{\mu\nu,\rho\sigma}$ are nonzero.  
In fact, in this limit, these tensors may be thought of as the $d-1$ dimensional versions of the original tensors $I_{\mu\nu}$ and $I_{\mu\nu, \rho \sigma}$.  For $\hat \beta_{\mu\nu, \rho \sigma}$, only the mixed components $\hat \beta_{(n A), (n B)}$ survive in a near boundary limit.

\subsection{Conformal Block Decomposition}
\label{sec:conformalblockdecomp}

Like four point functions in CFT without a boundary, the two-point functions $\langle O_1(x) O_2(x') \rangle$, $\langle J^\mu(x) J^\nu(x') \rangle$, and $\langle T^{\mu\nu}(x) T^{\lambda \sigma}(x') \rangle$ admit conformal block decompositions.  We distinguish two such decompositions: the bulk decomposition in which the two operators get close to each other and the boundary decomposition in which the two operators get close to the boundary (or equivalently their images).  Our next task is to study the structure of these decompositions.  For simplicity, in what follows, we will restrict to the case that the dimensions of $O_1$ and $O_2$ are equal and take $\Delta_1=\Delta_2=\eta$. 

\subsubsection*{Bulk Decomposition}

Recall that in the presence of a boundary the one-point functions for operators with spin violate conformal symmetry.
As a result, the bulk conformal block decomposition will involve only a sum over scalar operators with coefficients proportional
to the $a_\Delta$.  

Allowing for an arbitrary normalization $\K$ of the two-point function, the bulk OPE for two identical scalar operators can be written as
\be
\label{OO}
O_\eta (x) O_\eta (x')= {\K\over s^{2\eta}}+\sum_{\Delta \neq 0}  \lambda_\Delta  B(x-x',\partial_{x'})O_\Delta (x')~~,~~\lambda_\Delta \in \mathbb{R}\ ,
\ee 
where the sum is over primary fields. 
The bulk differential operator $B(x-x',\partial_{x'})$ is fixed by bulk conformal invariance and produces the sum over descendants. 
As the OPE \eqref{OO} reflects the local nature of the CFT, this OPE is unchanged when a boundary is present.
The bulk channel conformal block decomposition is given by taking the expectation value of \eqref{OO} using \eqref{O} and then matching the result with \eqref{OOtwo}.
We write 
\be
G_{O_\eta O_\eta}(v) &=& \K  + \sum_{\Delta \neq 0} a_\Delta \lambda_\Delta G_{\rm bulk}(\Delta, v) \ ,
\ee
where we have pulled out the leading bulk identity block contribution.\footnote{%
For two-point functions of scalar operators of different dimension, $\Delta_1 \neq \Delta_2$, $G_{\rm bulk}$ will depend on $\Delta_1$ and $\Delta_2$.  We
 refer to the literature \cite{McAvity:1995zd, Liendo:2012hy} for the more general expression, but suppress it here as we are interested in 
 the simpler case.
}
  There are analogous expressions for the functions $P(v)$, $Q(v)$, $A(v)$, $B(v)$, and $C(v)$ out of which we constructed $\langle J^\mu(x) J^\nu(x') \rangle$ and $\langle T^{\mu\nu}(x) T^{\lambda\sigma}(x') \rangle$.  
  We can write for example
\be
%P(\xi) &=& P(0) + \sum_{\Delta \neq 0} a_\Delta \lambda_{\Delta } P_{\rm bulk}(\Delta, \xi) \ , \\
%\pi(\xi) &=&  \sum_{\Delta \neq 0} a_\Delta \lambda_{\Delta} \pi_{\rm bulk}(\Delta, \xi) \ , \\
Q(v) &=&  \sum_{\Delta \neq 0} a_\Delta \lambda_{\Delta} Q_{\rm bulk}(\Delta, v) \ , \\
A(v) &=& \sum_{\Delta \neq 0} a_\Delta \lambda_{\Delta} A_{\rm bulk}(\Delta, v) \ , 
%B(\xi) &=& \sum_{\Delta \neq 0} a_\Delta \lambda_{\Delta} B_{\rm bulk}(\Delta, \xi) \ , \\
%C(\xi) &=&C(0) + \sum_{\Delta \neq 0} a_\Delta \lambda_{\Delta } C_{\rm bulk}(\Delta, \xi) \ ,
\ee
where $G_{\rm bulk}(\Delta, v)$, $Q_{\rm bulk}(\Delta, v)$, and $A_{\rm bulk}(\Delta, v)$ have a very similar form: 
\be
G_{\rm bulk}(\Delta, v) &=& 
\xi^{\frac{\Delta}{2}}  {}_2 F_1 \left( \frac{\Delta }{2}, \frac{\Delta }{2}, 1 - \frac{d}{2} + \Delta; - \xi \right)
\ , \\
%\pi_{\rm bulk}(\Delta, \xi) &=& \xi^{\frac{\Delta}{2}}  \, {}_2 F_1 \left( \frac{\Delta}{2}, 1 + \frac{\Delta}{2}, 1 - \frac{d}{2} + \Delta; - \xi \right)
%\ , \\
%\alpha_{\rm bulk}(\Delta, \xi) &=& \xi^{\frac{\Delta}{2}}\,  {}_2 F_1 \left( \frac{\Delta}{2}, 2 + \frac{\Delta}{2}, 1- \frac{d}{2} + \Delta; - \xi \right)
% \ .
\label{Q2formula}
Q_{\rm bulk}(\Delta, v) &=& \xi^{\frac{\Delta}{2}}  \, {}_2 F_1 \left( 1 + \frac{\Delta}{2}, 1 + \frac{\Delta}{2}, 1 - \frac{d}{2} + \Delta; - \xi \right)
(1+\xi)
\ , \\
A_{\rm bulk}(\Delta, v) &=& \xi^{\frac{\Delta}{2}}\,  {}_2 F_1 \left( 2 + \frac{\Delta}{2}, 2 + \frac{\Delta}{2}, 1- \frac{d}{2} + \Delta; - \xi \right)
 (1+\xi)^2 \ .
\ee
Indeed, one is tempted to define a general form for which each of these functions is a special case:
\be
%G^{(s)}_{\rm bulk}(\Delta, \xi) &=& 
%\xi^{\frac{\Delta}{2}}  {}_2 F_1 \left( \frac{\Delta }{2}, s+\frac{\Delta }{2}, 1 - \frac{d}{2} + \Delta, - \xi \right)
%\ . 
G^{(s)}_{\rm bulk}(\Delta, v) &=& 
\xi^{\frac{\Delta}{2}}  {}_2 F_1 \left( s+\frac{\Delta }{2}, s+\frac{\Delta }{2}, 1 - \frac{d}{2} + \Delta, - \xi \right) ( 1 + \xi)^s
\ . 
\ee
The remaining functions $B(v)$, $C(v)$, and $P(v)$
 can be straightforwardly constructed from the conservation equations
\eqref{Jcons}-\eqref{Tcons2}, and can be represented as sums of hypergeometric functions.  Note that the bulk identity block does not contribute to $Q(v)$, $A(v)$, and $B(v)$, but it does to $C(v)$ and $P(v)$.  
We review the derivation of these conformal block decompositions using the null cone formalism in Appendix \ref{sec:nullcone}.

\subsubsection*{Boundary Decomposition}

In the presence of a boundary, a bulk scalar operator $O_\eta$ of dimension $\eta$ can be expressed as a sum over boundary operators denoted as $\oO_\Delta({\bf x})$. We write
\be
\label{bO}
O_\eta(x)= {a_O \over (2y)^\eta}+\sum_{\Delta \neq 0} \tilde \mu_\Delta \oB(y,\op) \oO_\Delta({\bf x}) ~~,~~\tilde\mu_\Delta \in \mathbb{R} \ ,
\ee 
where the sum is over boundary primary fields. Boundary conformal invariance fixes the operator $\oB(y,\op)$. 
The two-point function of two identical boundary operators is normalized to be
\be
\label{oOoO}
\la \oO_\Delta({\bf x})\oO_\Delta({\bf x})\ra = \frac{\K_{d-1}}{{\bf s}^{2 \Delta}}\ ,
\ee where $\K_{d-1}$ is a constant. The one-point function of the boundary operator vanishes.
Reflection positivity guarantees the positivity of these boundary two-point functions for unitary theories.
 The boundary channel conformal block decomposition is given by squaring \eqref{bO}, taking the expectation value using \eqref{oOoO}, and then matching the result with \eqref{OOtwo}. We write 
\be
 G_{O O}(v) =  \xi^\eta \left[ a_O^2 +  \sum_{\Delta \neq 0} \mu^2_\Delta G_{\rm bry} (\Delta, v) \right] \ ,
 \label{bryscalardecomp}
\ee
where \cite{Dolan:2003hv} 
\be
G_{\rm bry} (\Delta, v) &=& \xi^{-\Delta} {}_2 F_1 \left(\Delta, 1-  \frac{d}{2} +\Delta,2 - d+ 2 \Delta, - \frac{1}{\xi} \right)  \ .
\ee 
To remove the $\eta$ dependence from the conformal block, it is useful to include an explicit factor of $\xi^\eta$ in the decomposition (\ref{bryscalardecomp}).  We have made a redefinition $\mu_\Delta^2 = \tilde \mu_\Delta^2 \K_{d-1}$ to allow for more generally normalized two-point functions.  Reflection positivity applied to the boundary two-point functions along with the fact that $\tilde \mu_\Delta \in {\mathbb R}$ guarantees the coefficients $\mu_\Delta^2$ in the boundary expansion are non-negative.  There was no such constraint on the bulk conformal block decomposition.

For a field of spin $s$, there is an extra subtlety that the sum, by angular momentum conservation, can involve boundary fields of spin $s'$ up to and including $s$.  For a conserved current, we need to consider boundary fields of spin $s'=0$ and 1, while for the stress tensor we will need $s'=0$, 1, and 2 boundary fields.  
Fortunately, because of the restricted form of the $\langle J_\mu(x) O(x') \rangle$, $\langle T_{\mu\nu}(x) O(x') \rangle$, and $\langle T_{\mu\nu}(x) V_\lambda(x') \rangle$ correlation functions, the sum over fields with spin strictly less than $s$ is restricted, and the situation simplifies somewhat.  Consider first $\langle J_\mu(x) O(x') \rangle$ in the boundary limit, 
which vanishes for $\Delta < d-1$ and blows up for $\Delta > d-1$.  We interpret the divergence to mean that the corresponding coefficient $c_{JO}$ must vanish when $\Delta > d-1$.  
It follows that in the boundary conformal block expansion of $\langle J_\mu(x) J_\nu(x') \rangle$, the only scalar field that contributes will have $\Delta = d-1$.  An analogous argument in the stress tensor case implies that only scalar fields and vectors of dimension $\Delta = d$ can contribute in the boundary conformal block expansion.

These restrictions on the boundary conformal block expansion are reflected in the possible near boundary behaviors of the functions 
$\pi$, $\rho$, $\alpha$, $\gamma$, and $\epsilon$ allowed by the current conservation equations (\ref{floweqpi})--(\ref{floweqgamma}).  
From the definitions of $\pi$ (\ref{pidef}) and $\rho$ (\ref{rhodef}),
 $\rho$ corresponds to vector exchange on the boundary and $\pi$ to scalar exchange.
If we exchange a boundary vector $\hat V^A$ of dimension $d-2 + \delta_V$, where $\delta_V$ is an anomalous dimension,
the near boundary behavior for $\rho$ is $\rho \sim (1-v)^{-1+\delta_V}$, which can be deduced from the boundary conformal block expressions given in this section and the current conservation equations. 
The unitary bound implies $\delta_V > 0$, and there is a descendant scalar operator $\partial_A \hat V^A$ of dimension $d-1+\delta_V$.  
(A boundary vector operator at the unitarity bound $d-2$ would be conserved, $\partial_A \hat J^A = 0$.)  
Correspondingly, eq.~(\ref{floweqpi}) enforces the near boundary behavior $\pi \sim  (1-v)^{\delta_V}$.  
The exception to this rule is when $\delta_V = 1$.  Then the conservation equations allow $\rho$ and $\pi$ to have independent order one contributions near the boundary, corresponding to the possibility of having both vector and scalar primaries of dimension $d-1$.

The story is similar for the stress tensor with $\epsilon$ representing spin two exchange, $\gamma$ spin one exchange, and $\alpha$ scalar exchange.  
The generic case is boundary exchange of a spin two operator $\hat S_{AB}$ of dimension $d-1+\delta_S$ with $\delta_S > 0$.
The near boundary behaviors of the stress tensor correlation function are then $\epsilon \sim (1-v)^{-1 + \delta_S}$, $\gamma \sim (1-v)^{\delta_S}$, and
$\alpha \sim (1-v)^{1+\delta_S}$ where the scaling of $\gamma$ and $\alpha$ is consonant with the existence of descendants of the form 
$\partial_A \hat S^{AB}$ and $\partial_A \partial_B \hat S^{AB}$.  Again, there is one exception to this story, when $\delta_S = 1$.  In this case, the conservation equations allow $\alpha$, $\gamma$, and $\epsilon$ to have independent order one contributions near the boundary, corresponding to scalar, vector, and spin two exchange of scaling dimension $d$.

The scalar of $\Delta = d$ 
plays a special role in bCFT.  It is often called the displacement operator.  
The presence of a boundary affects the conservation of the stress tensor, 
$\partial_\mu T^{\mu n}(x) = D^n(x) \delta(y)$,
%\be
%\partial_\mu T^{\mu n}(x) = D^n(x) \delta(y) \ ,
% \\
%\label{JD}
%\partial_\mu J^\mu(x) = D_J(x) \delta(y) \ ,
%\ee
where $D^n$ is a scalar operator of $\Delta = d$.
% and $D_J$ is a scalar operator of $\Delta = d-1$. 
The scalar displacement operator $D^n$ is generally present in boundary and defect CFTs.

For interesting reasons, discussed in what follows, a vector of dimension $\Delta = d$ and scalar of dimension $\Delta = d-1$ are generically absent from the conformal block decompositions of these two-point functions.
 In the case of the current two-point function, a natural candidate for a scalar of dimension $\Delta = d-1$ is the boundary limit of $J^n$.
If there are no degrees of freedom on the boundary, then $J^n$ must vanish as a boundary condition or the corresponding charge is not conserved.  
If there are charged degrees of freedom on the boundary characterized by a boundary current $\hat J^A$, then current conservation 
 implies $J^n|_{\rm bry} = - \partial_A \hat J^A$ and the total charge is conserved by an inflow effect.  From the point of view of the 
 conformal field theory living on the boundary, the current $\hat J^A$ is no longer conserved, and $J^n|_{\rm bry}$ becomes a descendant of $\hat J^A$.  Because conservation on the boundary is lost, the scaling dimension of $\hat J^A$ must shift upward from $d-2$ by a positive amount $\delta_J$.  Correspondingly, the scaling dimension of $J^n$ shifts upwards by $\delta_J$ from $d-1$, and it will appear in the conformal block decomposition
 not as a primary but as a descendant of $\hat J^A$.  
We thus expect generically that a scalar primary of $\Delta = d-1$ is absent from the boundary conformal block expansion of the current-current two-point function.
%Therefore, we must have $J^n|_{\rm bry}=0$, as an operator statement. 
%On the other hand, $J^A|_{\rm bry}$ can be non-zero.
 %The two-point function of $J^n$ with itself will be padded with some extra factors of $y^\delta y'^\delta$ and will vanish in the boundary limit.
 
  The story for a vector of dimension $\Delta = d$ is similar.  
  A natural candidate for such an operator is the boundary limit of $T^{nA}$.  In the free models we consider, the boundary conditions force
 this quantity to vanish.  The interacting models we introduce in section \ref{sec:interaction}
  have extra degrees of freedom that propagate on the boundary and an associated 
  boundary stress tensor $\hat T^{AB}$.  
By conservation of the full stress tensor, the boundary limit of $T^{nA}$ is equal to the descendant operator $\partial_A \hat T^{AB}$, neither of which will necessarily vanish classically.  The scaling dimension of $\hat T^{AB}$ must shift upward from $d-1$ by a positive amount $\delta_T$.
%Despite not vanishing,  
The boundary operator corresponding to $T^{nA}|_{\rm bry}$ now enters the boundary conformal block decomposition not as a vector primary but  as a descendant of the spin two field $\hat T^{AB}$.  We expect generically that a vector of $\Delta = d$ is absent from the boundary conformal block expansion of $\langle T_{\mu\nu}(x) T_{\lambda \sigma}(x') \rangle$.
%However, as an operator statement, we 
%must have $T^{nA}|_{\rm{bry}}=0=\partial_A \hat T^{AB}$, since once $\hat T^{AB}$ is not conserved,
%its scaling dimension will be greater than the unitary bound $\Delta = d-1$ 
%for a spin two field in dimension $d-1$, and correspondingly the dimension of $T^{nA}|_{\rm{bry}}$ will be greater than $d$.
 
We will nevertheless keep these vectors and scalars in our boundary conformal block decomposition.  
The reason is that for these interacting models,
 we only perform leading order perturbative calculations.  At this leading order, we cannot see the shift in dimension of $T^{nA}$ and $J^n$, and it is useful to continue to treat them as primary fields.
 
The boundary block expansions for $\langle J^\mu(x) J^\nu(x') \rangle$ and $\langle T^{\mu\nu}(x) T^{\lambda \sigma}(x') \rangle$
have the forms
\be
\pi(v) &=& \xi^{d-1} \left( \mu_{(0)}^2 \pi^{(0)}_{\rm bry}(v) + \sum_{\Delta \geq d-2} \mu_\Delta^2 \pi^{(1)}_{\rm bry}(\Delta, v)  \right)
\label{Qexp}
\ , \\
\alpha(v) &=& \xi^{d} \left( \mu_{(0)}^2 \alpha^{(0)}_{\rm bry}(v) +\mu_{(1)}^2 \alpha^{(1)}_{\rm bry}(v) + \sum_{\Delta \geq d-1} \mu_\Delta^2 \alpha^{(2)}_{\rm bry}(\Delta, v) \right)
\label{Aexp}
\ ,
\ee
where the indices (0), (1) and (2) denote the spins. One has similar expressions for the other functions $\rho(v)$, $\gamma(v)$, and $\epsilon(v)$.  
%The boundary block expansions for $\langle J^\mu(x) J^\nu(x') \rangle$ and $\langle T^{\mu\nu}(x) T^{\lambda \sigma}(x') \rangle$
%have the forms
%\be
%Q(v) &=& \xi^{d-1} \left( \mu_{(0)}^2 Q^{(0)}_{\rm bry}(v) + \sum_{\Delta} \mu_\Delta^2 Q^{(1)}_{\rm bry}(\Delta, v)  \right)
%\label{Qexp}
%\ , \\
%A(v) &=& \xi^{d} \left( \mu_{(0)}^2 A^{(0)}_{\rm bry}(v) +\mu_{(1)}^2 A^{(1)}_{\rm bry}(v) + \sum_{\Delta \neq 0} \mu_\Delta^2 A^{(2)}_{\rm bry}(\Delta, v) \right)
%\label{Aexp}
%\ ,
%\ee
%where the indices (0), (1) and (2) denote the spins. One has similar expressions for the other functions $P(v)$, $B(v)$, and $C(v)$.  
%
%In the boundary limit $v \to 1$, 
%only spin zero and one operators of $\Delta = d-1$ can contribute (non-divergently) to $\langle J^\mu(x) J^\nu(x') \rangle$.  
%Similarly, only the spin zero, one, and two operators of $\Delta =d$ can contribute to $\langle T^{\mu\nu}(x) T^{\lambda \sigma}(x') \rangle$.
%A more natural basis is one which diagonalizes the contribution from $D^n$, $T^{nA}|_{\rm bry}$, and $J^n|_{\rm bry}$. 
%The basis which performs this diagonalization is the same combinations, given in (\ref{ax}-\ref{ex}), that exhibit reflection positivity -- $\alpha$, $\gamma$, $\epsilon$, $\pi$, and $\rho$ -- that we described previously.

In this basis, we find the following blocks\footnote{The results in the basis of $A(v),B(v),C(v)$ are given in \cite{Liendo:2012hy}.}
\be
\label{alphazero}
 \alpha_{\rm bry}^{(0)}(v) &=& \frac{1}{4(d-1)} (v^{-1} - v)^d (d (v^{-1} + v)^2 - 4) \ , \\
\gamma_{\rm bry}^{(0)}(v) &=& - \frac{d}{4(d-1)^2} (v^{-1} - v)^d (v^{-2} - v^2) \ , \\
 \epsilon_{\rm bry}^{(0)}(v) &=& \frac{d}{4(d-1)^2(d+1)} (v^{-1} - v)^d (v^{-2}-v^2)^2 \ .
\ee
In the boundary limit $\xi \to \infty$, the combinations $\xi^d \gamma_{\rm bry}^{(0)}$ and $\xi^d \epsilon_{\rm bry}^{(0)}$ vanish while  $\xi^{d} \alpha_{\rm bry}^{(0)} \to 1$.  
In this basis, the contribution of the displacement operator $D^n$ to the boundary block expansion is encoded purely by $\alpha_{\rm bry}^{(0)}$.  

Similarly, for the spin one exchange, we find
\be
 \alpha_{\rm bry}^{(1)}(v) &=& \frac{d-1}{d} (v^{-1} - v)^d (v^{-2} - v^2) \ , \\
\label{gammaone}
\gamma_{\rm bry}^{(1)}(v) &=& -\frac{1}{2} (v^{-1} - v)^d (v^{-2} + v^2) \ , \\
\epsilon_{\rm bry}^{(1)}(v) &=& \frac{1}{2(d+1)} (v^{-1} - v)^d (v^{-2} - v^2) \ , 
\ee
where now $\xi^d \gamma_{\rm bry}^{(1)} \to -1$ in the boundary limit while the other two vanish.  For spin two exchange with weight $\Delta = d$, we have
\be
 \alpha_{\rm bry}^{(2)}(d, v) &=&(v^{-1} - v)^d (v^{-1} - v)^2 \ , \\
\gamma_{\rm bry}^{(2)}(d, v) &=& -\frac{1}{d-1} (v^{-1} -v)^d (v^{-2} - v^2) \ , \\
 \epsilon_{\rm bry}^{(2)}(d, v) &=& \frac{1}{(d^2-1)(d-2)} (v^{-1} - v)^d(d (v^{-1} + v)^2 - 2 (v^{-2} + v^2) ) \ , 
\ee
where now $\xi^d \epsilon_{\rm bry}^{(2)} \to 4 / (d+1)(d-2)$ and the other two vanish.  We have shifted the normalization convention here relative to (\ref{alphazero}) and (\ref{gammaone})  so that we may write the higher dimensional blocks (\ref{tensortwoalpha}) for $\alpha_{\rm bry}^{(2)}(\Delta, v)$ in a simpler and uniform way.
 
Playing similar games with the current, we find
\be
%\xi^{-d} \gamma_{\rm bry}^{(2)}(\Delta, \xi) &=& 
%\xi^{-d} \epsilon_{\rm bry}^{(2)}(\Delta, \xi) &=&
%
\label{currentzeropi}
 \pi_{\rm bry}^{(0)}(v) &=& \frac{1}{2} (v^{-1} - v)^{d-1} (v^{-1} + v) \ , \\
\rho_{\rm bry}^{(0)}(v) &=& \frac{1}{2(d-1)} (v^{-1} - v)^{d}\ , 
\label{currentzerorho}
\ee
and
\be
\label{currentonefirstpi}
 \pi_{\rm bry}^{(1)}(d-1,v) &=&(v^{-1} - v)^{d} \ , \\
 \rho_{\rm bry}^{(1)}(d-1,v) &=& \frac{1}{d-1} (v^{-1} - v)^{d-1} (v^{-1} + v) \ .
 \label{currentonefirstrho}
\ee
For higher dimension operators, we have
\be
\label{tensortwoalpha}
 \alpha_{\rm bry}^{(2)}(\Delta, v) &=& \xi^{-\Delta-2} \, {}_2 F_1 \left( 2 + \Delta, 1 - \frac{d}{2} + \Delta, 2 - d + 2 \Delta; - \frac{1}{\xi} \right) \ , \\
 \pi_{\rm bry}^{(1)}(\Delta, v) &=& \xi^{-\Delta-1} \, {}_2 F_1 \left( 1 + \Delta, 1- \frac{d}{2} + \Delta, 2-d + 2 \Delta; - \frac{1}{\xi} \right) 
 \label{currentonepi}
 \ .
% \\
%\xi^{-d+1} \rho_{\rm bry}^{(1)}(\Delta, \xi) &=&
\ee
The remaining functions $\gamma_{\rm bry}^{(2)}(\Delta, v)$, $\epsilon_{\rm bry}^{(2)}(\Delta, v)$ and $\rho_{\rm bry}^{(1)}(\Delta, v)$ have a more cumbersome form but can be straightforwardly derived from the conservation equations \eqref{Jcons}-\eqref{Tcons2}. 
Evidently, $G_{\rm bry}(\Delta, v)$, $\pi_{\rm bry}^{(1)}(\Delta, v)$, and $\alpha_{\rm bry}^{(2)}(\Delta, v)$ all are special cases of the general form
\be
\xi^{-\Delta - s}  \, {}_2 F_1 \left( s + \Delta, 1- \frac{d}{2} + \Delta, 2-d + 2 \Delta; - \frac{1}{\xi} \right)  \ .
\ee

We have written all of these blocks to make a symmetry under $v \to v^{-1}$ apparent.  The transformation $v \to v^{-1}$ or equivalently $\xi \to -1 - \xi$ corresponds to a reflection $y' \to - y'$ keeping $y$ fixed.  
Under such a partial reflection, the blocks are eigenvectors with eigenvalue $\pm 1$ for integer $\Delta$:
\be
\label{reflectionrule}
f_{\rm bry}^{(s)} (\Delta, {1\over v}) = (-1)^{\Delta + s + \sigma} f_{\rm bry}^{(s)} (\Delta, v) \ .
\ee
The shift $\sigma$ is one for $\rho_{\rm bry}^{(s)}$ and $\gamma_{\rm bry}^{(s)}$ and zero otherwise.
For the higher dimensional exchanged operators, this reflection property relies on a hypergeometric identity
\be
{}_2 F_1 (a,b, c ;z) = (1-z)^{-a} \, {}_2 F_1 \left( a, c-b, c; \frac{z}{z-1} \right) \ ,
\ee
in the special case where $c = 2b$.  

\subsection{Crossing Relations}

\begin{figure}
\begin{center}
\includegraphics[width=2 in]{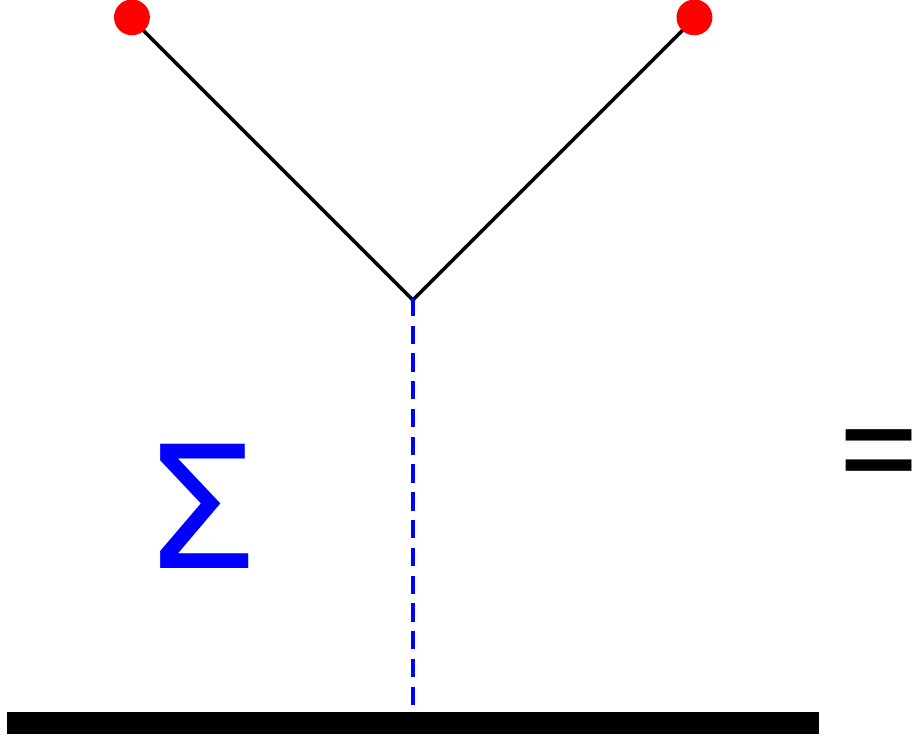}
\includegraphics[width=2 in]{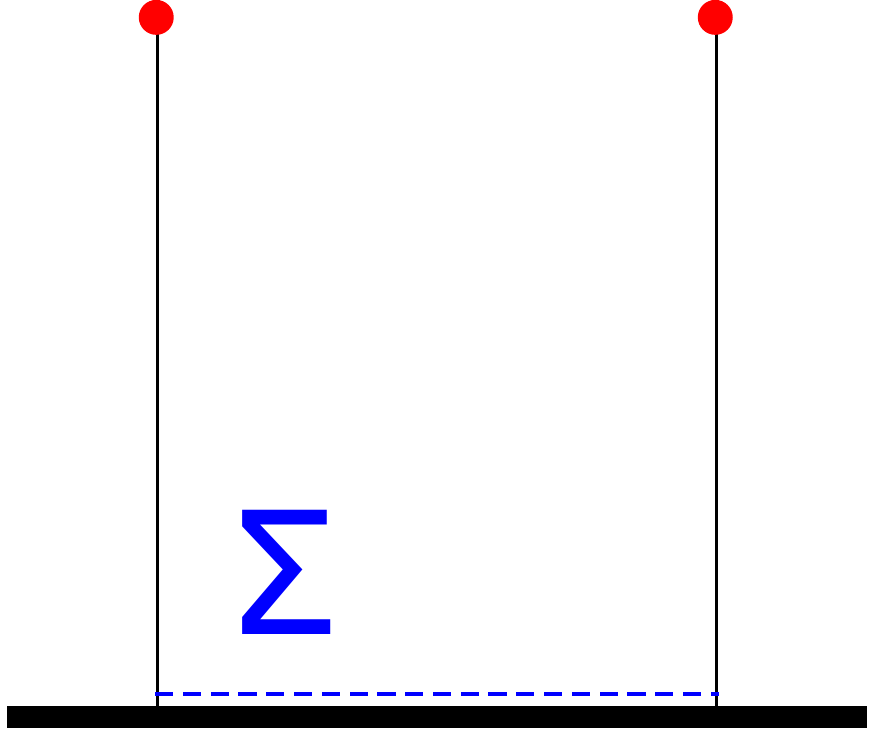}
\end{center}
\caption{Crossing symmetry for two-point functions in bCFTs. 
\label{fig:crossing} 
}
\end{figure}

A crossing relation for boundary conformal field theory is the statement that two-point functions can be expressed either
as a sum over boundary conformal blocks or as a sum over bulk conformal blocks. (See figure \ref{fig:crossing}. The left/right plot represents the bulk/boundary channel.)

The field theories we consider in this paper
are either free or have some weak interactions that are constrained to live on the boundary.  The solutions to crossing for the current and stress tensor correlation functions are remarkably universal for the family of theories we consider.  
Roughly speaking, they all involve a decomposition of a function of an invariant cross ratio of the form 
\be
\label{Gform}
G(v) = 1 + \chi v^\eta \ .
\ee    The parameter $\chi$ will depend on the boundary conditions.
Roughly, one can think of this expression in terms of the method of images, where  the 1 reproduces the answer in the coincident/bulk limit, in the absence of a boundary, and the $v^\eta$ represents the correlation between points and their images on the other side of the boundary.   
In the bulk channel, $1$ is the identity block and $v^\eta$ will generically involve a sum over a tower of fields.  In the boundary channel, we first decompose $G(v) = \frac{1}{2} (1 + \chi) (1+v^\eta) + \frac{1}{2} (1-\chi) (1-v^\eta)$ into eigenfunctions of the reflection operator $v \to 1/v$ and then find infinite sums of boundary blocks that reproduce $1 \pm v^\eta$.  The two-point function may not be precisely of the form $1 + \chi v^\eta$, but the discrepancy can always be accounted for by adjusting the coefficients of a few blocks of low, e.g.\ $\Delta = d-1$ or $d$, conformal dimension. 

A number of the blocks have a very simple form.  In the bulk, we find 
\be
G_{\rm bulk} (d-2, v) = v^{d-2} \ , \; \; \; Q_{\rm bulk}(d, v) = v^d \ , \; \; \; A_{\rm bulk}(d+2, v) = v^{d+2} \ .
\ee
In the boundary, we already saw that the blocks of dimension $d-1$ for $\langle J^\mu(x) J^\nu(x') \rangle$ and of
dimension $d$ for $\langle T^{\mu\nu}(x) T^{\lambda \sigma}(x') \rangle$ have a polynomial form.  However, we neglected
to point out that for the scalar two-point functions, the boundary blocks of dimension $\frac{d-2}{2} + n$ where $n$ is a non-negative integer also have a simple polynomial form.   The polynomial like expressions satisfy the recursion relation 
\be
\label{recrel}
G_{\rm bry} \Big( \frac{d-2}{2} + n , v \Big) &=&  \frac{4 (2n-1)}{(2n-d)} 
\Big[
(1 + 2 \xi)
G_{\rm bry} \left( \frac{d-2}{2} + n-1 , v \right)
\nn\\
&&~~~~~+
\frac{4 \xi (\xi+1)}{(d-4+2n)}
\partial_\xi G_{\rm bry} \left( \frac{d-2}{2} + n-1 , v \right)
\Big] \ .
\ee
The first two values are
\be
\xi^{\frac{d-2}{2}} G_{\rm bry}\left( \frac{d-2}{2} , v \right) &=& \frac{1}{2}(1+v^{d-2} )
\ , \\
\xi^{\frac{d-2}{2}} G_{\rm bry} \left( \frac{d}{2}, v \right) &=& \frac{2}{d-2} (1 - v^{d-2} ) 
\ .
\ee
These two particular cases are degenerate in fact:  they satisfy the same differential equation (see Appendix \ref{sec:nullcone}).   
We have imposed boundary conditions that are consistent with the recursion relation (\ref{recrel}) and the reflection symmetry (\ref{reflectionrule}). 

These simple expressions for the conformal blocks motivate the following
remarkably
simple relation: 
\be
\label{simplecrossing}
\xi^{\frac{d-2}{2}} \left[\frac{1+\chi}{2} G_{\rm bry} \left( \frac{d-2}{2}, v \right) + \frac{1-\chi}{2} \frac{d-2}{2} G_{\rm bry} \left(\frac{d}{2} , v\right) \right] = 1 + \chi G_{\rm bulk}(d-2, v) \ .
\ee (For $\chi=\pm 1$, this relation is pointed out in \cite{Liendo:2012hy}.)
In the next section, we will compute the two-point function for a free scalar field of dimension $\Delta = \frac{d-2}{2}$. 
We find a free scalar takes advantage of precisely such a crossing relation (\ref{simplecrossing}). Moreover,
the case $\chi = 1$ corresponds to Neumann boundary conditions, in which case the contribution from a boundary operator $\partial_n \phi$ of dimension $\Delta = \frac{d}{2}$ is absent.  
Correspondingly, the case $\chi = -1$ is Dirichlet boundary conditions, and the boundary operator $\phi$ itself is absent.  
An absent or trivial boundary is the case $\chi = 0$.  
The contribution from the bulk comes simply from the identity operator and the composite operator $\phi^2$.  
By adding an interaction on the boundary, we will be able to move perturbatively away from the limiting cases $\chi = \pm 1$.  However, positivity of the boundary decomposition (\ref{bryscalardecomp}) implies the bounds:
\be
\label{boundchi}
-1 \leq \chi \leq 1 \ .
\ee 
Given these bounds, one might interpret that $\chi = \pm 1$ correspond to ``corners'' in the bootstrap program.

More generally, for a function of the form $G_{OO}(v) = a_O^2 \xi^\Delta +  1\pm v^{2 \Delta}$, the boundary and bulk decompositions will involve a sum over infinite numbers of operators.  Here $\xi^\Delta$ corresponds to the boundary identity block and the 1 to the bulk identity block.  With a little bit of guess work, one can deduce a general form of these series expansions.  (For a more rigorous derivation, one can use the $\alpha$-space formalism \cite{Hogervorst:2017sfd, Hogervorst:2017kbj}.)  One has the boundary decompositions
\be
\label{evensum}
  \frac{\xi^{-\Delta}}{2} \left(1 + v^{2 \Delta} \right)=\sum_{n \in 2 {\mathbb Z}^*} \mu_n^2 G_{\rm bry}(\Delta+n, v)  \ , \\
\label{oddsum}
  \frac{\xi^{-\Delta}}{2} \left(1 - v^{2 \Delta} \right)=\sum_{n \in 2 {\mathbb Z}^* +1} \mu_n^2 G_{\rm bry}(\Delta+n, v)  \ ,
\ee
where ${\mathbb Z}^*$ denotes a non-negative integer and the coefficients are
\be
\mu_n^2 &=& \frac{2^{d-2 \Delta-2n} \sqrt{\pi} \Gamma(n + 2 \Delta - d + 1) \Gamma(n+\Delta)}
{ \Gamma(\Delta) \Gamma\left( n + \Delta - \frac{d-1}{2} \right)\Gamma(n+1) \Gamma \left( \Delta + 1 - \frac{d}{2} \right)} \ ,
\ee
where $\mu_0^2 = 1$.  
%\be
%\mu_{n}^2 &=& \frac{(\Delta)_n \left(\frac{d-n}{2} - \Delta\right)_{n/2}}{2^n n! \left(\frac{1+d}{2} - \Delta-n \right)_{n/2}} \; ; \; \; \; n \in 2 {\mathbb Z}^*\ , \\
%\mu_{n}^2 &=&  \frac{(\Delta)_n \left(\frac{d-n+1}{2} - \Delta\right)_{(n-1)/2}}{2^n n! \left(\frac{1+d}{2} - \Delta-n \right)_{(n-1)/2}} \; ;  \; \; \; n \in 2 {\mathbb Z}^*+1 \ .
%\ee
In contrast, for the bulk decomposition, the boundary identity block decomposes into bulk conformal blocks
\be
\label{bulkidblock}
\xi^\Delta &=& \sum_{n=0}^\infty \frac{[(\Delta)_n]^2}{n! \left( 2 \Delta - \frac{d}{2}+n \right)_n} G_{\rm bulk}(2 \Delta+2n, v) \ .
\ee
One also has the bulk decomposition
\be
\label{bulkmoregenblock}
v^{2 \Delta} =   \sum_{n=0}^\infty \frac{(-1)^n}{n!}\frac{(\Delta)_n \left( \Delta - \frac{d}{2}+1 \right)_n}{\left( 2 \Delta +n -\frac{d}{2}  \right)_n} G_{\rm bulk}(2 \Delta+2n, v) \ .
\ee

There are similar decompositions for the $\langle J^\mu(x) J^\nu(x') \rangle$ and $\langle T^{\mu\nu} (x) T^{\lambda \sigma}(x') \rangle$ correlation functions.  
For the current, we need to give the decomposition of 
\be
Q(v) = 2 \chi^2 v^{2d-2} \ , \; \; \;
\pi(v) = 1 - \chi^2 v^{2d-2} \ ,
\ee
and for the stress tensor, we need to give the decomposition of
\be
A(v) = \frac{4d}{d-1} \chi^2 v^{2d} \ , \; \; \;
\alpha(v) = 1 + \chi^2 v^{2d} \ .
\ee  (For free theories, $\chi^2=1$.)  
Using the relations
\be
\label{pidecompodd}
\frac{1}{2}(1 + v^{2d-2} )&=&\xi^{d-1} \left(  \pi_{\rm bry}^{(0)}(v) + \sum_{n \in 2 {\mathbb Z}^* + 1} \mu_n^2 \pi_{\rm bry}^{(1)}(d-1+n, v) \right)\ , \\
\label{pidecompeven}
\frac{1}{2}(1 - v^{2d-2}) &=& \xi^{d-1} \sum_{n \in 2 {\mathbb Z}^*} \mu_n^2 \pi_{\rm bry}^{(1)}(d-1+n, v) \ , 
\ee
where
\be
\mu_n^2 =  \frac{2^{1-d-2n} \sqrt{\pi} \Gamma(d+n-2) \Gamma(d+n)}{\Gamma(d-2) \Gamma \left( \frac{d}{2} \right) \Gamma(n+2) \Gamma \left( \frac{d-1}{2} + n \right)} \ ,
\ee
and $\mu_0^2 = (d-1)/2$, we can find a decomposition similar in spirit to the lhs of (\ref{simplecrossing}).
Similarly, for the stress tensor
\be
\label{alphaevendecomp}
\frac{1}{2} (1 + v^{2d}) &=&\xi^{d} \left( \alpha_{\rm bry}^{(0)}(v) + \sum_{n \in 2{\mathbb Z}^*} \mu_n^2 \alpha_{\rm bry}^{(2)}(d+n, v) \right)\ , \\
\label{alphaodddecomp}
\frac{1}{2}(1 - v^{2d}) &=& \xi^{d}\left(  \frac{d^2}{4(d-1)} \alpha_{\rm bry}^{(1)}(v) +  \sum_{n \in 2 {\mathbb Z}^*+1} \mu_n^2 \alpha_{\rm bry}^{(2)}(d+n, v) \right) \ ,
\ee
where
\be
\mu_n^2 =  \frac{2^{-d-2n} \sqrt{\pi} \Gamma(d+n-1) \Gamma(d+n+2)}{\Gamma(d) \Gamma \left( \frac{d}{2} -1 \right) \Gamma(n+3) \Gamma \left( \frac{d+1}{2} + n \right)} \ ,
\ee
where $\mu_0^2 = (d-2)d(d+1)/8(d-1)$.  
Finally, there are also corresponding bulk decompositions for which there is no obvious positivity constraint. 
We can write decompositions for the scalar, conserved current, and stress tensor two-point functions in a unified form:
\be
\label{bulkunified}
v^{2 \Delta} &=&  \sum_{n=0}^\infty \frac{(-1)^n}{n!} \frac{\left(\Delta+s \right)_n \left( \Delta+1 - \frac{d}{2}-s \right)_n}
{\left( 2 \Delta + n  - \frac{d}{2} \right)_n }
G^{(s)}_{\rm bulk}(2 \Delta + 2n, v) \ .
\ee
%\be
%v^{2 \Delta}&=&  \sum_{n=0}^\infty \frac{(-1)^n}{n!} \frac{\left(\Delta+1 \right)_n \left( \Delta - \frac{d}{2} \right)_n}
%{\left( 2 \Delta + n  - \frac{d}{2} \right)_n }
%Q_{\rm bulk}(2 \Delta + 2n, \xi) \ ,
%\\
%&=& \sum_{n=0}^\infty \frac{(-1)^n}{n!} \frac{\left( \Delta+2 \right)_n \left( \Delta - \frac{d}{2} - 1\right)_n}
%{\left( 2 \Delta + n - \frac{d}{2} \right)_n  }A_{\rm bulk}(2 \Delta + 2n, \xi) \ .
%\ee

Similar decompositions of $1 \pm v^{2 \Delta}$ were discussed in the appendices of ref.\ \cite{Liendo:2012hy}.
As a result, many of the formulae here are not entirely new.  We have made an attempt to present them in a
way that stresses their symmetry properties under $v \to 1/v$ and also stresses the important role played by
the decomposition of $1 \pm v^{2 \Delta}$ in free theories --  for scalar, vector, and tensor operators.

%% file: proof.tex
Consider $d=4$ CFTs in curved space with a smooth codimension-one boundary $\partial {\cal M}$. The conformal anomaly is given by
\bea
\label{4dtrace} 
&&\la T^\mu_\mu\ra=
{1\over 16 \pi^2} \Big( c W_{\mu\nu\lambda\rho}^2- a E_4\Big)\nn\\
&&~~~~~~~~~~~~+{\delta(y)\over 16 \pi^2}  \Big(a E^{\rm{(bry)}}_4-b_1 \tr\hat{K}^3-b_2
 h^{AB}\hat K^{CD} W_{ACBD} \Big) \ .
\eea   
We construct a projector onto the boundary metric $h_{\mu\nu}=g_{\mu\nu}-n_\mu n_\nu$ with $n_\mu$ being a unit, outward normal 
vector to $\partial {\cal M}$; $E_4$ is the $d=4$ Euler density, $ W_{\mu\nu\lambda\rho}$ is the Weyl tensor and $\hat K_{AB}= K_{AB}- {K\over 3 } h_{AB}$ is the traceless part of the extrinsic curvature.

The energy-momentum (stress) tensor in the Euclidean signature is defined by
\be
\la T_{\mu\nu} (x) \ra=-{2\over \sqrt{g}} {\delta W\over g^{\mu\nu}(x)} \ ,
\ee 
where $W$ is the generating functional for connected Green's functions.  
The two-point function in flat space is 
\be
\label{tt} 
\la T_{\mu\nu} (x) T_{\sigma\rho}(x') \ra &=&
\lim_{g_{\mu\nu}\to \delta_{\mu\nu}}\Big((-2)^2 {\delta^2 \over \delta
g^{\sigma\rho}(x')\delta g^{\mu\nu}(x) } W \Big) \ .
\ee  
We will denote $\widetilde W$ as the anomalous part of $W$. 
Note in general there can be Weyl invariant contributions to correlation functions. 
The theory is assumed to be regulated in a diffeomorphism-invariant way.  

We will adopt the dimensional regulation and will be interested in the mass scale, $\mu$, dependence in the correlation functions.
The $a$-anomaly is topological so it does not produce any $\mu$ dependence. 
The $b_1$-charge does not contribute to the two-point function in the flat limit, since $K^3 \sim {\cal O}(g_{\mu\nu})^3$.
One will be able to extract $b_1$ from a study of three-point functions in the presence of a boundary, but we leave
such a project for the future.
We here only consider the $c$ and $b_2$ anomalies.  The relevant pieces of the anomaly effective action are
\bea
{\widetilde W}^{(c)}= {c\over 16 \pi^2} {\mu^\epsilon \over \epsilon }\int_{\cal M} W^2_{\m\n\lambda\rho} \ , ~~ {\widetilde W}^{(b_2)}= {b_2\over 16 \pi^2}{\mu^\epsilon \over \epsilon } \int_{\partial \cal M}  K^{AB}W_{nAnB} \ . 
\eea 
These pieces should allow us to compute anomalous contributions to stress tensor correlation functions in the coincident limit.\footnote{We remark that the $\Box R$ anomaly in $d=4$ does not affect the scale dependent contribution to the two-point function, since the corresponding effective action, $R^2$, is finite.}  

We will perform the metric variation twice on the anomaly action to obtain anomalous contributions to
the two-point function of the stress tensor.
We work in Gaussian normal coordinates.  While we do not impose that $\delta g_{\mu\nu}=0$ on the boundary,  
we do keep $\delta g_{nA}=0$.  
In the flat boundary limit (see Appendix \ref{sec:Weyl}), 
\be
\lim_{g_{\mu\nu}\to \delta_{\mu\nu}}  \delta K_{AB}= {1\over 2} \partial_n \delta g_{AB} \ . 
\ee   
Note the $\delta g_{nn}$ contribution vanishes in the flat limit in the transformed extrinsic curvature. 
The transformed Weyl tensor can be written as
\be
\lim_{g_{\mu\nu}\to \delta_{\mu\nu}}  \delta W_{\mu\si\rho\nu} =- 2 P_{\mu\si\rho\nu,\alpha\gamma\delta\beta} \del^\gamma \del^\delta \delta
g^{\alpha\beta} \ ,
\ee
where  $P_{\mu\si\rho\nu,\alpha\gamma\delta\beta}$, defined in \eqref{P}, is a projector that shares the same symmetries as the Weyl tensor: 
\be
P_{\mu\sigma\rho\nu,\alpha\gamma\delta\beta}&=&P_{\alpha\gamma\delta\beta,\mu\sigma\rho\nu}\ , \\
P_{\mu\sigma\rho\nu,\alpha\gamma\delta\beta} P_{\mu\sigma\rho\nu, \eta\chi\epsilon\omega}&=&P_{\alpha\gamma\delta\beta, \eta\chi\epsilon\omega} \ .
\ee
It will be convenient to define the following fourth order differential operator using the projector:
\be
\label{rewriteforbulk}
P_{\mu\sigma\rho\nu,\alpha\gamma\delta\beta} \partial^\sigma \partial^\rho \partial^\gamma \partial^\delta&=& {(d-3)\over 4 (d-2)}\Delta^T_{\m\n\a\b} \ .
\ee
Some additional properties of this tensor along with its definition can be found in Appendix \ref{sec:Weyl}.

It is useful first to recall the story \cite{Osborn:1993cr} without a boundary.  The argument that gives a relation between $c$ and $\alpha(0)$ will also work with a boundary, provided
we arrange for the variation $\delta g_{\mu\nu}$ to vanish as we approach the boundary, eliminating any boundary terms that may arise through integration by parts.   
We then have, in the bulk limit, that 
\be
\label{bulkvary}
\lim_{g_{\mu\nu}\to \delta_{\mu\nu}}  \delta^2  \Big( \lim_{v\to 0}{\widetilde W}^{(c)} \Big) 
&=& {c_{} \over 4 \pi^2} {\mu^\epsilon \over \epsilon } \int_{\cal M} P_{\a\g\delta\b,\eta\chi\phi\psi} (\delta g^{\eta\psi}) (\partial^\phi \partial^\chi \partial^\gamma \partial^\delta \delta g^{\a\b}) \ .
\ee 
From the definition of the stress tensor as a variation with respect to the metric, one infers the scale dependent contribution:
\be
\label{c1bulk}
\mu {\partial \over \partial \mu} \la  T_{\m\n}(x') T_{\a\b}(x'') \ra^{(c)}&=& {c_{} \over 4\pi^2} \Delta^T_{\m\n\a\b} \delta^4 (x'-x'') \ .
\ee 
The general form of the two-point function without a boundary (or with a boundary but in the bulk limit) is given by
\be
\la T_{\mu\nu} (x) T_{\si\rho}(x') \ra &=& C_T {I_{\mu\nu,\si\rho} \over s^{8}} = \frac{C_T}{320} \Delta_{\mu\nu\sigma\rho}^T \frac{1}{s^4} \ ,
\ee 
where we have used \eqref{rewriteforbulk} in $d=4$.  
We next regularize the UV divergence in the two-point function in $d=4$ by taking \cite{dimreg}
\be
{\cal R} \frac{1}{x^4} = - \partial^2 \left( \frac{ \ln \mu^2 x^2}{4 x^2} \right) \ ,
\ee
from which we obtain
\be
\label{reg4d}
\mu {\partial \over \partial \mu} \Big({\cal R} {1\over x^{4}}\Big) = 2\pi^2 \delta^4 (x) \ ,
\ee  
and hence
\be
\label{c2}
\mu {\partial \over \partial \mu}  \la T_{\mu\nu} (x) T_{\si\rho}(0) \ra=C_T {\pi^2 \over 160} \Delta^T_{\m\n\si\rho}  \delta^4 (x) \ .
\ee 
Matching \eqref{c2} with \eqref{c1bulk}, one identifies 
\be
\label{bulkcon}
c_{} = {\pi^4\over 40} C_T\ ,
\ee  where $C_T=C(0) = \frac{4}{3} \alpha(0)$. 

Now let us consider the variation of the boundary term in the trace anomaly.
Given the variation rules, the $b_2$-anomaly action gives 
\be
\label{v2KW}
\lim_{g_{\mu\nu}\to \delta_{\mu\nu}}\delta^2 {\widetilde W}^{(b_2)}
&=&{b_2\over 16 \pi^2}{\mu^\epsilon \over \epsilon } \int_{\partial \cal{M}} (\partial_n \delta g^{AB})
(P_{AnBn ,\alpha\gamma\delta\beta} \del^\gamma \del^\delta \delta g^{\alpha\beta}) \ .
\ee 
Thus,
\be
\label{b22}
\mu {\partial \over \partial \mu} \la  T_{AB}(x') T_{\a\b}(x'') \ra^{(b_2)}
&=& {b_2\over 2 \pi^2}\partial_y \delta (y-y') P_{AnBn ,\alpha\gamma\delta\beta}\del^\gamma \del^\delta  \delta^4 (x' - x'') |_{y\to0}\ . 
\ee
However, it is peculiar that such a boundary term should be present at all.  By simple power counting, we do not expect a pure boundary, log divergent contribution to the stress tensor two-point function.  The corresponding momentum space correlator has odd mass dimension, $4+4-3 = 5$, which naively should not involve a logarithmic divergence.  
More convincing, perhaps, is the flip in sign of this term under reflection $y \to -y$.  As we saw in the boundary conformal block decomposition of the stress tensor, under reflection the $ABCD$ and $ABnn$ components of the two-point function restricted to the boundary should be even.       
Although these two arguments fall short of a rigorous proof, it seems natural for
such a pure boundary log divergence to cancel against something else.  

Consider whether this boundary term (\ref{b22}) may cancel against boundary terms we dropped in calculating
(\ref{bulkvary}).    There is an immediate subtlety associated with the noncommutativity of the boundary and coincident limits.  
The boundary term (\ref{b22}) exists in a strict boundary limit, while the calculation (\ref{bulkvary}), which reproduces the anomalous part of the $\frac{1}{s^8} I_{\mu\nu, \rho \sigma}$ tensor structure, was performed in the coincident limit.  As we can see from the two-point function (\ref{magicTT}), the coefficient $\alpha(v)$ of the $I_{\mu\nu, \rho \sigma}$ structure will vary as $v$ changes from the coincident limit 0 to the boundary limit 1.  

We posit the existence of an effective action which computes correlation functions of the stress tensor.  Almost everywhere, the scale dependent part of this action is  ${\widetilde W}^{(c)}$.  However, if we introduce a small distance $\epsilon$ to separate the stress tensor insertions, in a very thin layer of thickness less than $\epsilon$ along the boundary, we should replace
the constant $c$ in ${\widetilde W}^{(c)}$ with a generally different constant $c_{\rm bry}$.  
The idea is that $c_{\rm bry}$ will give us both the freedom to reproduce the scale dependence of the $\alpha(1) I_{\mu\nu, \rho \sigma}$ contribution to the two-point function (\ref{magicTT}) and to cancel the offensive boundary term (\ref{b22}).
In contrast, the terms in the expression \eqref{magicTT} proportional to $\partial_v \alpha$ and $\partial_v^2 \alpha$ give vanishing contribution to the $nnnn$ and $nnAB$ components of the two-point function.  
The term proportional to $\hat \beta_{\mu\nu\rho\sigma}$ in \eqref{magicTT} near the boundary only has $nAnB$ contributions.
Because of this index incompatibility, it seems unlikely to us that terms in an effective action that would produce
this index structure would also lead to a cancellation of the boundary term  (\ref{b22}).  Unfortunately, we cannot offer a rigorous proof.

Keeping the surface terms, by varying the metric such that $\delta g_{\mu\nu}$ is nonzero close to the boundary, the near-boundary limit of the $c$-anomaly action gives 
\be
\label{revisit}
\lim_{g_{\mu\nu}\to \delta_{\mu\nu}}  \delta^2  \Big( \lim_{v\to 1}{\widetilde W}^{(c_{\rm bry})} \Big) 
&=& {c_{\rm bry} \over 4 \pi^2} {\mu^\epsilon \over \epsilon } \int_{\cal M} P_{\a\g\delta\b,\eta\chi\phi\psi} (\delta g^{\eta\psi}) (\partial^\phi \partial^\chi \partial^\gamma \partial^\delta \delta g^{\a\b})  \nn\\
&&+{c_{\rm bry} \over 4 \pi^2} {\mu^\epsilon \over \epsilon } \int_{\partial {\cal M}} P_{\a\g\delta\b,\eta n \phi\psi} (\partial^\phi \delta g^{\eta\psi}) (\partial^\gamma \partial^\delta \delta g^{\a\b}) \nn\\
&&-{c_{\rm bry} \over 4 \pi^2} {\mu^\epsilon \over \epsilon } \int_{\partial {\cal M}} P_{\a\g\delta\b,\eta \chi n \psi} (\delta g^{\eta\psi})  (\partial^\chi \partial^\gamma \partial^\delta \delta g^{\a\b}) \ ,
\ee 
where we have performed integration by parts near the boundary. 
Consequently, we find for the scale dependence of the two-point function in the near boundary limit that\footnote{The two-point functions presented in this section generalize the results given in \cite{Huang:2016rol}, which has assumed a certain boundary condition on boundary geometry that removes normal derivatives acting on the metric variations \cite{Solodukhin:2015eca}.}
\be
\label{c1new}
\mu {\partial \over \partial \mu} \la T_{\m\n}(x') T_{\a\b}(x'') \ra^{(c)}
&=& {c_{\rm bry} \over 4 \pi^2} \Delta^T_{\m\n\a\b} \delta^4 (x'-x'') \nn\\
&&- {2c_{\rm bry} \over \pi^2} \partial_y \delta (y-y') P_{\m n \n n,\a\g\delta\b} \partial^\gamma \partial^\delta \delta^4 (x' - x'') |_{y\to0} \nn\\
&&- {2c_{\rm bry} \over \pi^2} \delta (y-y') P_{\m n A \n,\a\g\delta\b} \partial^\gamma \partial^\delta \partial^A \delta^4 (x' - x'') |_{y\to0} \nn\\
&&- {2c_{\rm bry} \over \pi^2} \delta (y-y') P_{\m \phi n \n,\a\g\delta\b} \partial^\gamma \partial^\delta \partial^\phi \delta^4 (x' - x'') |_{y\to0} \ .
\ee 
Next observe, through a direct computation, that
\be
\lim_{y\to 0} P_{\m n A \n,\a\g\delta\b} \partial^\gamma \partial^\delta \partial^A {1\over x^4}
= \lim_{y\to 0} P_{\m \phi n \n,\a\g\delta\b} \partial^\gamma \partial^\delta \partial^\phi {1\over x^4} = 0 \ .
\ee
This implies, after adopting the regularized expression \eqref{reg4d}, the last two lines of \eqref{c1new} 
do not contribute.\footnote{%
	If we also turn on $\delta g_{nA}$ in the Gaussian normal coordinates when varying the $b_2$ action, restoring the last term of	
	\eqref{varyKab} in the flat limit, we find the additional contributions to the two-point function do not 
	have a scale dependence.
}
The second line of \eqref{c1new} suggests to evaluate
\be
\lim_{y\to 0} P_{\m n \n n,\a\g\delta\b} \partial^\gamma \partial^\delta {1\over x^4} \ ,
\ee
which turns out to be non-zero. 
However, this second line has precisely the right form to cancel the earlier boundary contribution we
found from varying the $b_2$ anomaly \eqref{b22}.  As explained above, we will eliminate this problematic boundary term by requiring a cancellation between $b_2$ and $c$-contributions:
\be
b_2 = 4 c_{\rm bry} \ .
\ee
On the other hand, to reproduce the near boundary structure of the stress tensor two-point function, $\alpha(1) I_{\mu\nu,\rho\sigma}$, we must have 
that 
 $c_{\rm bry} = \pi^4 \alpha(1) / 30$.
Thus, we conclude that
\be
\label{mainresult}
b_2 = \frac{2\pi^4}{15} \alpha(1) \ .
\ee  

With the relation \eqref{mainresult},
we can achieve a better understanding of the previously conjectured equality \eqref{8c}  (i.e $b_2=8 c_{}$), and discuss how general it is.
Observe first that the relation \eqref{8c} is true only when $\alpha(1)=2\alpha(0)$. (Recall in general one has $c_{}=\frac{\pi^4}{30}\alpha(0)$.)
We will find that $\alpha(1)=2\alpha(0)$ indeed holds for a large class of free CFTs in the following sections.  However, in the 4d mixed dimensional QED theory which we discuss in section \ref{sec:interaction}, the boundary value
$\alpha(1)$ depends on the coupling, while the bulk theory is the standard Maxwell theory with an unchanged value of $c$ or $\alpha(0)$.
In other words, the mixed dimensional QED can 
provide a counterexample to the relation \eqref{8c}.  

%We have not analyzed the $b_1$ boundary charge yet.  It will presumably require analysis of the structure of the stress tensor three-point function with a boundary.

%% file: free.tex
In this section, we consider three families of free conformal field theories: 
a conformally coupled massless scalar in $d$ dimensions,  a massless fermion in $d$ dimensions and
an abelian $p$-form in $2p+2$ dimensions. 
We will see that the corresponding two-point functions take a remarkably universal form.  
They correspond to special cases of the crossing relations we found in section \ref{sec:twopoint} with the parameter $\chi = \pm 1$.  
The parameter $\chi$ can be promoted to a matrix, with $\chi^2=\id$, an identity.  
To construct CFTs with more general eigenvalues of $\chi^2$ away from unity, we will include boundary interactions in the next section. 

\subsection{Free Scalar}

We start with the classical Minkowski action for a conformally coupled scalar in $d$-dimensions with a possibly curved codimension-one boundary term:
\be
\label{Iphi}
I=-\int_{\cal M} {1\over 2} \Big((\partial \phi)^2+{(d-2)\over 4 (d-1)} R \phi^2\Big)- {(d-2)\over 4 (d-1)} \int_{\partial \cal M}  K \phi^2 \ ,
\ee  where $R$ is the Ricci scalar and $K$ is the trace of the extrinsic curvature. The surface term is required by Weyl invariance. 
Restricting to flat space with a planar boundary at $y=0$, the usual improved stress tensor is given by
\be
T_{\mu \nu} 
= \del_\mu \phi \del_\nu \phi - {1\over 4} \, {1\over d-1} \Big(
(d-2) \del_\mu \del_\nu + \delta_{\mu \nu} \del^2 \Big) \phi^2 - \frac{(d-2)}{4(d-1)} \delta(y) h_{\mu\nu} (\partial_n \phi^2) \ ,
\ee 
with $n_\mu$ an outward-pointing unit normal vector to the boundary.
While in the bulk, the stress tensor is traceless (on shell), the boundary term requires either Dirichlet $\phi = 0$ or Neumann $\partial_n \phi = 0$ boundary conditions to preserve the tracelessness. 

Let us consider a more general case with a vector of scalar fields, i.e $\phi \to \phi^a$. (We will suppress the index $a$ in what follows.)  Then, we can introduce two complementary projectors $\Pi_\pm$ such that $\Pi_+ + \Pi_- = \id$ and $\Pi_\pm^2 = \Pi_\pm$.  The generalized boundary conditions are then\footnote{%
 These boundary conditions are sometimes called mixed in the literature; for instance, see section 5.3 in \cite{Vassilevich:2003xt}.
  }
\be
\label{gbcs}
\partial_n  (\Pi_+ \phi)|_{y=0} = 0 \ , \; \; \; \Pi_- \phi|_{y=0} = 0 \ .
\label{pibry}
\ee For a single scalar, one can only have either $\Pi_+=1, \Pi_-=0$ or $\Pi_+=0, \Pi_-=1$.
For the scalar, the $nA$ component of the stress tensor is 
\be
T_{nA} = \frac{d}{2(d-1)} (\partial_n \phi) (\partial_A \phi) - \frac{(d-2)}{2(d-1)} \phi \, \partial_A \partial_n \phi \ .
\ee 
The boundary conditions (\ref{pibry}) force that $T_{nA}$ vanishes at $y=0$.
 
It is perhaps useful to discuss the case of a transparent boundary.  We have fields $\phi_R$ and $\phi_L$ on each side of the boundary.  Given the second order equation of motion, the boundary conditions are continuity of the field $\phi_R = \phi_L$ and its derivative $\partial_n \phi_R = \partial_n \phi_L$.  We can use the folding trick to convert this interface CFT into a bCFT by replacing the $\phi_R$ fields with their mirror images $\tilde \phi_R$ on the left hand side.  We still have continuity of the fields as a boundary condition $\tilde \phi_R = \phi_L$, but having reflected the normal direction, continuity of the derivative is replaced with $\partial_n \tilde \phi_R = -\partial_n \phi_L$.  In terms of the projectors \eqref{gbcs}, we have 
\be
\Pi_\pm = \frac{1}{2} \left( \begin{matrix} 1 & \pm 1 \\ \pm 1 & 1 \end{matrix} \right) \ ,~~  \phi= \left( \begin{matrix} \tilde \phi_R \\  \phi_L \end{matrix} \right) \ .
\ee 
As the fields $\tilde \phi_R$ and $\phi_L$ do not interact, it is straightforward to go back to the unfolded theory.  One slightly tricky point relates to composite operators like the stress tensor.  In the original theory, there is no reason 
for a classical $T_{nA}$ to vanish at the boundary.  
However, in the folded theory (or bCFT), by our previous argument, we saw the $T_{nA}$ does vanish classically.  
In this case there are really two, separately conserved stress tensors, one associated with $\tilde \phi_R$ and one associated with $\phi_L$.  
The statement that $T_{nA}$ vanishes classically in the bCFT is really the statement that $T_{nA}$ computed from the $\tilde \phi_R$ fields cancels $T_{nA}$ computed from the $\phi_L$ fields at the boundary.  
More generally, a nonzero classical $T_{nA}$ in a bCFT corresponds to a discontinuity in $T_{nA}$ for the interface theory.  From the pill box argument mentioned at the end of the introduction, this situation corresponds to non-conservation of the boundary stress tensor $\partial_B \hat T^{AB}$. 
(As mentioned  in the introduction, we expect quantum effects to restore the condition $T_{nA}=0$ on the boundary for general bCFTs.)

We note in passing that the component $T^{nn}$ of the scalar field will in general not vanish on the boundary. 
Indeed, as discussed in the introduction,
 it corresponds to the displacement operator which is generally present in bCFTs. 

The two-point function for the elementary fields $\phi$ can be constructed using the method of images: 
%It is useful to consider the perpendicular geometry where ${\bf x}={\bf x'}=0$ and write the two-point function as  
\be
\langle  \phi(y) \phi (y')\rangle = \frac{\K}{s^{d-2}}
(\id + \chi v^{d-2} )\ , 
\label{scalarimage}
\ee where we denote
\be
\K= {1\over (d-2) {\rm{Vol}}(S_{d-1})}\ , ~~~ {\rm Vol}(S_{d-1})= {2 \pi^{d\over 2}\over \Gamma{({d\over 2})}} \ .
\ee
Applying the boundary conditions (\ref{pibry}), one finds that 
\be
\chi = \Pi_+ - \Pi_- \ .
\ee 
 From the properties of the projectors, $\chi^2 = \id$. The eigenvalues of $\chi$ must be $\pm 1$, $+1$ for Neumann boundary conditions and $-1$ for Dirichlet.  
The relevant cross-ratio function \eqref{OOtwo} is then $G_{\phi\phi}(v) = \id + \chi v^{d-2}$. 
In section \ref{sec:twopoint}, we saw that this particular $G_{\phi\phi}(v)$ admitted the decomposition (\ref{simplecrossing}) into a pair of bulk and a pair of boundary blocks.    In fact, because of the restriction on the eigenvalues of $\chi$, we only require a single boundary block, of dimension $\frac{d-2}{2}$ for Neumann boundary conditions or dimension $\frac{d}{2}$ for Dirichlet.  
We will see in the next section how to move away from eigenvalues $\pm 1$ perturbatively by adding a boundary interaction.  

Next we consider $\langle \phi^2(x) \phi^2(x') \rangle$.  There is a new element here because $\phi^2$ has a nontrivial one-point function
\be
\langle \phi^2 (y) \rangle = \frac{\K \tr(\chi)}{(2y)^{d-2}} \ .
\ee
For $N$ scalars, one finds the following cross-ratio function for the two-point correlator:
\be
\label{phi2phi2}
G_{\phi^2 \phi^2}(v) = 2 \K^2 \tr( \id + \chi v^{d-2})^2 +  \K^2 \tr(\chi)^2 \xi^{d-2}\ .
\ee
This function $G_{\phi^2 \phi^2}(v)$ is straightforward to decompose into boundary and bulk blocks, using the results of section \ref{sec:twopoint}.  For the boundary decomposition, the last term on the rhs of (\ref{phi2phi2}), proportional to $\xi^{d-2}$, is the boundary identity block. We may decompose $1 + v^{2(d-2)}$ using the infinite sum (\ref{evensum}). The piece proportional to $2 \tr(\chi) v^{d-2}$ can be expressed using $v^{d-2}=\xi^{d-2} G_{\rm bry}(d-2, v)$.   One may worry that this term comes with a negative coefficient when $\tr(\chi) < 0$, violating reflection positivity.  In fact, in the infinite sum (\ref{evensum}), the block $G_{\rm bry}(d-2,v)$ has coefficient one, which, in the case of Dirichlet boundary conditions, precisely cancels the $G_{\rm bry}(d-2, v)$ reproduced from 
$-v^{d-2}$.  Indeed, for Dirichlet boundary conditions, the boundary $\phi^2$ operator is absent. There is no issue for Neumann boundary conditions since all the coefficients are manifestly positive.
The bulk decomposition is similarly straightforward.  The ``one'' in (\ref{phi2phi2}) is the bulk identity block.  The term proportional to $v^{d-2}$ can be expressed again as a single block, this time in the bulk, $G_{\rm bulk}(d-2, v) = v^{d-2}$.  The pieces proportional to $\xi^{d-2}$ and $v^{2(d-2)}$ decompose into bulk blocks using (\ref{bulkidblock}) and (\ref{bulkmoregenblock}).  

For the stress tensor two-point function, using Wick's theorem one obtains 
\be
\alpha(v) &=& (d-2)^2 \K^2 \left( \tr(\id) + \tr(\chi^2) v^{2d} + \tr(\chi) \frac{d(d-2)(d+1)}{4(d-1)} v^{d-2}(1-v^2)^2  \right) \ , \\
A(v) &=& \frac{d(d-2)^2 \K^2}{4(d-1)^2} \Big( \tr(\chi) v^{d} \left(- 2d (d^2-4)  + d(d-2)^2 v^{-2} + (d^2-4)(d+4) v^{2}  \right)\nonumber \\
&& 
\hspace{2in}+ 16 (d-1) \tr(\chi^2) v^{2d} \Big)  \ .
\ee  Setting $\chi=\pm 1$ we recover the results computed in \cite{McAvity:1993ue, McAvity:1995zd} for a single scalar under Dirichlet or Neumann boundary condition.  
In the boundary decomposition, looking at $\alpha(v)$, we recognize the $v^{d-2}(1 - v^2)^2$ piece as a contribution from $\alpha_{\rm bry}^{(2)}(d, v)$, with a sign depending on the boundary conditions.
Then, decomposing $1 + v^{2d}$ using (\ref{alphaevendecomp}), we see that the coefficient of the $\alpha_{\rm bry}^{(2)}(d,v)$ is precisely of the right magnitude to cancel out the possibly negative contribution from $v^{d-2}(1 - v^2)^2$, consistent with the absence of a $(\partial_A \phi) (\partial_B \phi)$ type boundary operator for Dirichlet boundary conditions.
Regarding the bulk decomposition, we can write $\alpha_{\rm bry}^{(2)}(d, v)$ as a linear combination of
$\alpha_{\rm bulk}(d-2, v)$, $\alpha_{\rm bulk}(d, v)$, and $\alpha_{\rm bulk}(d+2, v)$, all of which are polynomials in $v^{d \pm 2}$ and $v^d$,  giving a trivial solution of the crossing equations.

Let us also consider a complexified scalar $\phi = \phi_1 + i \phi_2$, or equivalently a pair of real scalars to define a conserved current. 
We have
\be
J_\mu = \frac{i}{2} \left[ \phi^* (\partial_\mu \phi) - (\partial_\mu \phi^*) \phi \right]= -\phi_1 \partial_\mu \phi_2 + \phi_2 \partial_\mu \phi_1 \ .
\ee
We introduce real projectors, $\Pi_\pm^\dagger = \Pi_\pm$, acting on the complexified combinations, $ \partial_n(\Pi_+\phi) = 0$ and $\Pi_- \phi = 0$. With these boundary conditions, the current is conserved at the boundary, $J_n = 0$.
Changing the $\phi(x)$ to $\phi^*(x)$ in (\ref{scalarimage}) and using Wick's Theorem, one finds
\be
Q(v) &=& \frac{(d-2) \K^2}{2} \left( \tr(\chi) v^{d-2} ((d-2) - d v^2) - 2 \tr(\chi^2) v^{2d-2} \right) \ , \\
\pi(v) &=&  \frac{(d-2) \K^2}{2} \left( \tr(\id) + (d-1)  \tr(\chi)  v^{d-2}( 1 - v^2) - \tr(\chi^2) v^{2d-2} \right) \ . 
\ee
Looking at $\pi(v)$, we recognize $(d-1) v^{d-2}  (1 - v^2)$ as a contribution from $\pi^{(1)}_{\rm bry}(d-1, v)$.
 The 
$1 -  v^{2d-2}$ dependence of $\pi(v)$ decomposes into boundary blocks according to (\ref{pidecompeven}).  Similar to the 
$\langle \phi^2(x) \phi^2(x') \rangle$ case we analyzed above, one might again be worried that the contribution from $\pi^{(1)}_{\rm bry}(d-1, v)$  is negative, violating reflection positivity.  However, for Dirichlet boundary conditions, the contributions from $1 -  v^{2d-2}$ and $(d-1) v^{d-2}  (1 - v^2)$ precisely cancel, consistent with the absence of a $\phi \partial_A \phi$ type boundary operator.  
It turns out that $\pi^{(1)}_{\rm bry}(d-1, v)$ and $\pi_{\rm bulk}(d-2, v)$ are proportional, giving a trivial solution of the crossing equations.  
Indeed, looking at $Q(v)$ we recognize $v^{d-2} ((d-2) - d v^2)$ as a contribution from $Q_{\rm bulk}(d-2, v)$.
Similar to what we found for the $\langle \phi^2(x) \phi^2(x') \rangle$ correlation function, looking now at the $1 -  v^{2d-2}$ dependence of $\pi(v)$, we recognize the one as the bulk identity block and decompose the $v^{2d-2}$ using (\ref{bulkunified}).

\subsection{Free Fermion}

The Minkowski action for Dirac fermions in curved space is 
\be
I= {i\over 2} \int_{\cal M}  \Big( \bar \psi \gamma_\mu  {D^\mu} \psi-({D^\mu}\bar \psi) \gamma_\mu  \psi\Big) \ ,
\ee 
where, as usual, the covariant derivative contains the spin connection and the bar is defined by $\bar \psi = \psi^\dagger \gamma^0$.  
The scaling dimension of the fermion $\psi$ is $\Delta={1\over 2} (d-1)$. The action is conformally invariant without any boundary term needed. 
Using a Minkowski tensor with mostly plus signature 
the Clifford algebra is given by $\{ \gamma_\mu, \gamma_\nu \} = - 2 \eta_{\mu\nu}$. 
In the flat space, the current and stress tensor in terms of the spinor field $\psi$ are
\be
J_\mu &=& \bar \psi \gamma_\mu \psi \ , \\
T_{\mu\nu} &=& \frac{i}{2} \left( (\partial_{(\mu} \bar \psi) \gamma_{\nu)} \psi - \bar \psi \gamma_{(\mu} \partial_{\nu)} \psi \right) \ .
\ee
We symmetrize the indices with strength one, such that
\be
T_{nn} = \frac{i}{2} \left( (\partial_n \bar \psi) \gamma_n \psi - \bar \psi \gamma_n \partial_n \psi \right)\ .
\ee
Following \cite{Mixed, McAvity:1993ue},
we define the following hermitian projectors $\Pi_+$ and $\Pi_-$: 
\be
\Pi_{\pm}={1\over 2} (\id \pm \chi)  \ ,
\ee 
with the parameter $\chi=\Pi_+ - \Pi_-$ for the fermion theory acting on the Clifford algebra such that 
\be
\chi \gamma_n = - \gamma_n \bar \chi \ , \; \; \; \chi \gamma_A = \gamma_A \bar \chi \ , \; \; \; \chi^2 = \bar \chi^2 = \id \ ,
\ee where $\bar \chi= \gamma^0 \chi^\dagger \gamma^0$. 
Since the action only has first-order derivatives we only need boundary
conditions imposed on half of the spinor components.
We consider boundary conditions $\Pi_-\psi=0$ and its conjugate $\bar \psi \Pi_-=0$. 
In terms of $\chi$, they become
\be
(\id - \chi ) \psi |_{\partial {\cal M}} = 0  \ , \; \; \; \bar \psi(\id- \bar \chi)|_{\partial {\cal M}} = 0 \ .
\ee 
As a consequence, from the equation of motion one can deduce a related but not independent Neumann boundary condition
$ \partial_n (\Pi_+ \psi)=0$.
A physical interpretation of these boundary conditions is that they make $J_n$ and $T_{nA}$ vanish on the boundary.
The two-point function of the spinor field is then 
\be
\label{psi2}
\langle \psi(x) \bar \psi(x') \rangle = \K_f \left( \frac{i \gamma \cdot (x - x')}{|x-x'|^d} + \chi \frac{i \gamma \cdot (\bar x - x')}{|\bar x - x'|^d} \right) \ ,
\ee
where $\bar x = (-x_1, {\bf x})\equiv(-y,{\bf x})$.   The parameter $\chi$ enters naturally in the fermion theory with a boundary. We consider a typical choice of normalization of the two-point function $\K_f = (d-2) \K = 1 / \Vol(S^{d-1})$. 

A straightforward application of Wick's theorem then allows us to calculate the $\langle J_\mu(x) J_\nu(x') \rangle$ and 
$\langle T_{\mu\nu}(x) T_{\lambda \sigma}(x') \rangle$ correlators.  In fact, as we have seen, it is enough to work out just 
the components with all normal indices.  The remaining components can then be calculated using the conservation relations.
One finds
\be
\pi(v) &=& \K_f^2 \tr_\gamma (\id) \left(1 - \tr(\chi^2) v^{d-1}\right) \ , \\
\alpha(v) &=& \frac{1}{2} (d-1) \K_f^2 \tr_\gamma (\id)  \left(1 + \tr(\chi^2) v^{2d}\right) \ ,
\ee 
where the value of $\tr_\gamma(\id)$ depends on the particular Clifford algebra we choose.
Essentially the same result for $\alpha(v)$ can be found in ref.\ \cite{McAvity:1993ue};  
for Dirac fermions, it is common in the literature to take $\tr_\gamma(\id) = 2^{\lfloor d/2 \rfloor}$.
%
%In even dimensions, for Dirac fermions, $\tr_\gamma(\id) = 2^{d/2}$.   
%While it is common in the literature to take $\tr_\gamma(\id) = 2^{\lfloor d/2 \rfloor}$  
%for an odd dimensional Dirac fermion, here we leave our the expressions general. 

The same conformal block decompositions that we worked out for the scalar apply to the free fermions as well.
Observe that,
$(d-2) \tr_\gamma (\id)$ scalars, half of which have Dirichlet and half of which have Neumann boundary conditions, produce the same $\langle J_\mu(x) J_\nu(x') \rangle$ two-point function as the spinor. 
Similarly, $\frac{d-1}{2} \tr_\gamma (\id)$ scalars, again split evenly between Neumann and Dirichlet boundary conditions, produce the same stress tensor two-point function as our spinor field. 

%With these expressions, one can consider a similar boundary
% and bulk conformal block decomposition using the results in section \ref{sec:twopoint}.  

\subsection{Free $p$-Form Gauge Fields}

Now we consider an abelian $p$-form in $d$ dimensions in the presence of a planar, codimension-one boundary. The Minkowski action is 
\be
I = - \frac{1}{2(p+1)!}
\int_{{\cal M}} \d^d x\, H_{\mu_1 \cdots \mu_{p+1}} H^{\mu_1 \cdots \mu_{p+1}} \ ,
\ee
where $H_{\mu_1 \cdots \mu_{p+1}} = D_{\mu_1} B_{\mu_2 \cdots \mu_{p+1}} \pm$ cylic permutations; $D_\mu$ is the standard covariant derivative. The action in $d = 2(p+1)$ is conformally invariant without any boundary term neeeded. Important special cases are a Maxwell field in four dimensions and a $2$-form in six dimensions.
We will again work in a flat half-space with coordinate system $x_\mu = (y, {\bf x})$ with a boundary at $y=0$. 
In ref.\ \cite{Bastianelli:1999ab}, the authors computed two- and three-point functions of the stress tensor in the absence of a boundary.  Here we will generalize their two-point calculations to include a planar boundary. 
The stress tensor
in flat space is given by
\be
T_{\mu\nu} = \frac{1}{p!} H_{\mu \mu_1 \cdots \mu_p} {H_\nu}^{\mu_1 \cdots \mu_p} - \frac{1}{2(p+1)!} \delta_{\mu \nu} H_{\mu_1 \cdots \mu_{p+1}} H^{\mu_1 \cdots \mu_{p+1}} \ .
\ee
This stress tensor is traceless only when $d=2p+2$.

We fix a generalization of Feynman gauge by adding $\frac{1}{2 (p-1)!} (\partial_\mu B^{\mu \nu_1 \cdots \nu_{p-1}})^2$ to the 
action.\footnote{%
We remark that there are additional subtleties in $p$-form theories that are worthy of further consideration. 
First, the gauge fixing process breaks conformal invariance.  An ameliorating factor is that the ghost and gauge fixing sectors to a large extent decouple from the rest of the theory.  
For example, the two-point function of $\partial \cdot B$ and $H = d B$ vanishes in general. 
Second, the ghosts required in the gauge fixing process require further ghost degrees of freedom, so-called ``ghosts for ghosts"  (see e.g.\ \cite{Cappelli:2000fe, HT}).
}
The two-point function of the $B$-field is then
\be
\label{BBp}
\langle B_{\mu_1 \cdots \mu_p}(x) B^{\nu_1 \cdots \nu_p}(x') \rangle = \K \delta^{\nu_1 \cdots \nu_p}_{\mu_1 \cdots \mu_p}\left( \frac{1}{(x-x')^{d-2}} + \chi \frac{1}{(({\bf x}-{\bf x}')^2 + (y+y')^2)^{(d-2)/2}} \right) \ .
\ee
The choice of $\chi$ is based on the presence or absence of a normal index.\footnote{The parameter $\chi$ is a c-number for gauge fields, not a matrix.}  There are two possible choices of boundary conditions, generalizing the ``absolute'' and ``relative'' boundary conditions of the Maxwell field $F_{\mu\nu}$ \cite{Vassilevich:2003xt}.  The Neumann-like or ``absolute'' choice corresponds to setting the normal component of the field strength to zero $H_{n A_1 \cdots A_p} = 0$ and leads to the two conditions $\partial_n B_{A_1 \cdots A_p} = 0$
and $B_{n A_2 \cdots A_p} = 0$.  The Dirichlet-like or ``relative'' choice means $B_{A_1 \cdots A_p} = 0$ which, along with the gauge fixing condition $\partial_\mu B^{\mu \mu_2 \cdots \mu_p} = 0$, leads to the additional constraint $\partial_n B^{n A_2 \cdots A_p} = 0$.  
To keep things general, we set $\chi = \chi_\perp$ when one of the indices of $B$ is the normal index and $\chi = \chi_\parallel$ otherwise. 

Conformal covariance suggests that the two-point function of $H$ with itself can be written in the form
\be
\label{abstract}
\langle H_{\mu_1 \cdots \mu_{p+1}}(x) H_{\nu_1 \cdots \nu_{p+1}}(x') \rangle &=&
\frac{1}{s^{d}} \sum_{g,h \in \Sigma^{p+1}} (-1)^{g+h} \Bigl( a(v) \prod_{i=1}^{p+1} I_{g(\mu_i) \, h(\nu_i)}(s) \nonumber \\
&& + b(v) X_{g(\mu_{p+1})} X'_{h(\nu_{p+1})} \prod_{i=1}^p I_{g(\mu_i) \, h(\nu_i)}(s)  \Bigr) \ ,
\label{twoptH}
\ee
where $\Sigma^p$ is the permutation group of $p$ elements.
The objects $I_{\mu \nu}$, $X_{\mu}$ and $X'_{\nu}$ were defined in section \ref{sec:twopoint}. 

To fix $a(v)$ and $b(v)$ in \eqref{abstract}, we don't need to calculate all components of the two-point function.  Let us focus on the diagonal components.
In fact, we can further restrict to the perpendicular geometry where ${\bf s}=0$. From \eqref{BBp}, we find
\be
\langle H_{2 \cdots p+2}(x) H^{2 \cdots p+2}(x') \rangle &=& \frac{\K(d-2)}{s^d}(p+1) (1 + \chi_\parallel v^d) \ , \\
\langle H_{1 \cdots p+1}(x) H^{1 \cdots p+1}(x') \rangle &=& \frac{\K(d-2)}{s^d}\Big(p+1-d  +(p \chi_\perp+(d-1) \chi_\parallel)  v^d\Big) \ .
\ee
We then compare these expressions with (\ref{twoptH}) in the same limit,
\be
\langle H_{2 \cdots p+2}(x) H^{2 \cdots p+2}(x') \rangle &=& \frac{(p+1)!}{s^{d}} a \ , \\
\langle H_{1 \cdots p+1}(x) H^{1 \cdots p+1}(x') \rangle &=& -\frac{p!}{s^{d}} \Big( (p+1) a + b \Big)\ . 
\ee
Solving for $a(v) $ and $b(v) $ yields
\be
a(v) &=& \frac{(d-2) \K}{p!} (1 + \chi_{\parallel} v^d) \ , \\
b(v)  &=& \frac{(d-2) \K}{p!} \Big(d -2(p+1) - (\chi_\parallel (d+p) + \chi_\perp p) v^d\Big) \nonumber \\
&=&  -\frac{(d-2)\K}{p!} \left(\chi_\parallel (d+p) + \chi_\perp p\right) v^d \ ,
\ee where we have set $d =2(p+1)$ to have a traceless stress tensor. 
In the absolute and relative cases where $\chi_\parallel = - \chi_\perp = \pm 1$, we find the simpler
\be
a(v)  &=& \frac{(d-2)\K}{p!} (1 \pm v^d) \ , \\
b(v)  &=& \frac{(d-2)\K}{p!} \Big(d - 2(p+1) \mp d v^d\Big)  = \mp d \frac{(d-2)\K}{p!}  v^d \ .
\ee

To pin down the form of the stress tensor, we need the following three two-point functions: 
\be
\langle T_{nn}(x) T_{nn}(x') \rangle &=& \frac{(p!)^2}{2 s^{2d}} \left( {d-1 \choose p} ((p+1)a+b)^2 
+ {d-1 \choose p+1} (p+1)^2 a^2 \right)
\ , \\
\langle T_{n2}(x) T_{n2}(x') \rangle &=& - \frac{(p!)^2}{s^{2d}} {d-2 \choose p} ((p+1) a + b)(p+1) a
\ , \\
\langle T_{23}(x) T_{23}(x') \rangle &=& \frac{(p!)^2}{s^{2d}} \left({d-3 \choose p-1} ((p+1) a + b)^2 +
{d-3 \choose p} (p+1)^2 a^2 \right)
\ .
\ee 
Away from $d = 2p+2$,  the calculation becomes inconsistent because the stress tensor is no longer traceless and there should be additional structures that need to be matched to fix the complete form of the stress tensor two-point function. 
For $d=2p+2$, we find
\be
A(v) &=& 2 (2p)! b^2 \nn\\
&=& \frac{2(d-2)^2 \K^2 (2p)!}{(p!)^2}\left(\chi_\parallel(d+p) + \chi_\perp p\right)^2  v^{2d}  \ , \\
B(v) &=& - \frac{1}{2} (2p)! b^2\nn\\
&=& - \frac{ (d-2)^2 \K^2 (2p)!}{2(p!)^2} \left(\chi_\parallel(d+p) + \chi_\perp p\right)^2v^{2d} \ , \\
C(v) &=& (2p)! \left( 2 (p+1)^2 a^2 + 2 a b (p+1) + b^2\right) \nonumber \\
&=& \frac{(d-2)^2 \K^2 (2p)!}{2 (p!)^2} \big[ d^2 
- (d-2) d v^{d} ( \chi_\parallel + \chi_\perp) + \nonumber \\
&& +\frac{1}{2} \Big( (4 + d(5d-8)) \chi_\parallel^2 + 4(d-2)(d-1) \chi_\parallel \chi_\perp
+ (d-2)^2 \chi_\perp^2 \Big) v^{2d} \big] \ .
\ee 
%In general, the expressions for $A$, $B$, and $C$ appear to be 
%\be
%A &=& \frac{2^{2k+1} (d-2)^2 K^2 \Gamma(k+1/2)}{\sqrt{\pi} k!} ( \pm (d-2k-2) + d v^d)^2 \ , \\
%B &=& -\frac{2^{2k-1} (d-2)^2 K^2 \Gamma(k+1/2)}{\sqrt{\pi} k!}  ( \pm (d-2k-2) + d v^d)^2 \ , \\
%C &=&  \frac{2^{2k} (d-2)^2 K^2 \Gamma(k+1/2)}{\sqrt{\pi} k!} ( d(d-2k-2)(1\pm v^d)^2  + 2(1+k)^2 (1+v^{2d}) )  \ .
%\ee
%I believe this result is inconsistent with the current conservation Ward identities when $d \neq 2k+2$.  
Note that in the bulk limit $v \to 0$, this result agrees with \cite{Bastianelli:1999ab}, as it should.
Restricting to the absolute and relative boundary conditions where $\chi_\parallel = - \chi_\perp$, we find that
\be 
\label{alphap}
\alpha(v) &=& \frac{d-1}{d} C(v) \nn\\
&=& \frac{d(d-1)(d-2)^2 \K^2 (2p)!}{2(p!)^2} ( 1 + \chi^2 v^{2d} ) \ .
\ee 
Observe that, $\frac{(2p+2)!}{2(p!)^2}$ scalars, split evenly between Neumann and Dirichlet boundary conditions, reproduce the same stress tensor as this $p$-form with either absolute or relative boundary conditions.  
This equivalence means that the conformal block decomposition for the $p$-form is the same as that for the scalar.
% one can again consider a similar boundary and bulk conformal block decomposition using the results in section \ref{sec:twopoint}. 

From \eqref{alphap}, the 4d U(1) gauge field has the following values:
\be
\alpha(0)= {3\over \pi^4} \ , ~~~ \alpha(1)=  {6\over \pi^4} \ .
\ee 
From the bulk relations \eqref{bulkcon} and \eqref{unitary1}, we indeed recover the bulk $c$-charge given in \eqref{cs1}. 
From the relation \eqref{mainresult}, we get $b_2= {4\over 5}$, which is consistent with the heat kernel computation of the gauge field \cite{Fursaev:2015wpa}. 
Indeed, the free theories considered in this section all have the relation $\alpha(1)=2 \alpha(0)$, which implies that $b_2=8c$ as we mentioned earlier. 
In the next section, we will see how the story changes when interactions are introduced on the boundary.

%% file: interaction.tex
The free theories we studied generically have a current two-point function 
characterized by a $\pi(v) \sim 1 - v^{2d-2}$ and 
stress tensor two-point function characterized by an
$\alpha(v) \sim 1 + v^{2d}$.\footnote{\label{susynote} The story was slightly more complicated for a vector of
free scalars, $\phi^a$, where additional pieces proportional to $\tr (\chi)$ appear. 
While we keep our discussion general, we remark that by having 
an equal number of Dirichlet and Neumann boundary conditions, we obtain 
$\tr(\chi) = 0$.  
In supersymmetric theories, 
an equal number of Neumann and Dirichlet boundary conditions appears to correlate with preserving a maximal amount of supersymmetry.  In ${\mathcal N}=4$ Super-Yang Mills theory in 3+1 dimensions, a $3+3$ splitting of the scalars preserves a $SO(3) \times SO(3) \subset SO(6)$ subgroup of the R-symmetry and a $OSp(4|4)$ subgroup of the $PSU(4|4)$ superalgebra \cite{DHoker:2006qeo,Gaiotto:2008sa}.  Similarly for ABJM theory, a $4+4$ splitting of the scalars preserves a $SO(4) \times SO(4) \subset SO(8)$ subgroup of the R-symmetry \cite{DHoker:2009lky}.
}
Since we saw generally that $\chi^2 =\id$, there was as a result no way to 
modify the coefficients of  $v^{2d-2}$ and $v^{2d}$ 
in $\pi(v)$ and $\alpha(v)$ (respectively) relative 
to the bulk identity block contribution.
On the other hand, we saw in the boundary conformal block decomposition 
that it should be straightforward to realize a bCFT with $\pi(v) \sim 1 - \chi^2 v^{2d-2}$ 
and $\alpha(v) \sim 1 + \chi^2 v^{2d}$, $\chi^2 < 1$, simply 
by taking advantage of the sums over blocks (\ref{pidecompodd}) 
and (\ref{alphaodddecomp}) with the opposite parity under $v \to 1/v$.
An obvious question poses itself.  
Is it possible to realize physically interesting bCFTs with $\chi^2 \neq \id$?
In this section we provide several examples below where we can move 
perturbatively away from the case where all the eigenvalues of $\chi$ are $\pm 1$. 
Moreover, we will see that a model with perturbative corrections to $\chi^2 = \id$  provides 
a counter-example to the $b_2=8c$ relation in 4d.
%A definitive conclusion, unfortunately, is hindered by an order of limits issue.

The idea is to couple a free field in the bulk to a free field 
in the boundary with a classically marginal interaction that lives
purely on the boundary.  For simplicity, we will restrict the 
bulk fields to a scalar field and Maxwell field in four dimensions.  
For boundary fields, we will allow only scalars and fermions.  
The fermions require less fine tuning as their larger 
engineering dimension allows for fewer relevant interactions.
We again consider a planar boundary located at $y=0$ 
while the bulk fields live in $y>0$. 
Here is our cast of characters:  
\begin{enumerate}
\item
A mixed dimensional Yukawa theory,
\be
I=-\frac{1}{2} \int_{\cal M} \d^4 x (\partial^\mu \phi)(\partial_\mu \phi) 
+  \int_{\partial {\cal M}} \d^3 x  \left( i\bar \psi \slashed{\partial} \psi - g\phi  \bar \psi \psi \right) \ ,
\ee
with the modified Neumann boundary condition $\partial_n \phi = -g \bar \psi \psi$.  
In our conventions, the unit normal $n^\mu$ points in the negative $y$-direction.

\item
A mixed dimensional QED, 
\be
I=-\frac{1}{4} \int_{\cal M}  \d^4 x F^{\mu\nu} F_{\mu\nu} 
+  \int_{\partial {\cal M}} \d^3 x  \left( i\bar \psi \slashed{D} \psi  \right) \ ,
\label{mixedQED}
\ee
where $D_\mu = \partial_\mu - i g A_\mu$.  The boundary conditions 
are a modification of the absolute boundary conditions discussed before, with 
$A_n = 0$, and $F_{nA} = \partial_n A_A = g \bar \psi \gamma_A \psi$. 

\item
A $d=4$ mixed dimensional scalar theory,
\be
I=-\frac{1}{2} \int_{\cal M} \d^4 x (\partial^\mu \phi)(\partial_\mu \phi) 
-   \int_{\partial {\cal M}} \d^3 x  \Big(\frac{1}{2} (\partial_A \eta)(\partial^A \eta) 
+  (\partial_n \phi)(-\phi + g \eta^2) \Big) \ ,
\ee
with the modified Dirichlet boundary condition $\phi = g \eta^2$. 
Another scalar field $\eta$ is introduced on the boundary.

\end{enumerate}

The boundary conditions are determined by having a well-posed 
variational principle for these classical actions.
The coupling $g$ is dimensionless.  
The limit $g\to 0$ results in two decoupled free theories, one living 
in the bulk space and another propagating on the boundary.
We should perhaps emphasize that 
in each of these models, there is an alternate trivial choice of boundary conditions
-- Dirichlet, relative, and Neumann  respectively -- which leaves the boundary
and bulk theories decoupled.
In this case, only the free bulk theory contributes to central charges, since the free boundary theory can be defined independent of the embedding 
space, without ``knowing"  about extrinsic curvature or bulk curvature.

One can generalize these models to curved space with actions that are 
explicitly Weyl invariant. 
Here we have again focused on flat space. 
The improved stress tensors of these models are traceless on shell. 
This list is not meant to be exhaustive.  
In general, one can add additional classical marginal 
interactions on the boundary, but these toy models are sufficient 
to illustrate several interesting features of this class of interacting theories. 

As was discussed in the introduction, among several other remarkable properties, 
the mixed QED theory is likely to be exactly conformal. 
For the other theories, using dimensional regularization and suitably tuning to 
eliminate relevant operators, we will find fixed points in the 
$\epsilon$ expansion using dimensional regularization.

Apart from the mixed dimensional QED, to our knowledge 
none of these theories has been studied in the literature. 
The canonical example of an interacting bCFT appears to be 
scalar $\phi^4$ theory in the bulk with no extra propagating 
degrees of freedom living on the boundary 
\cite{McAvity:1993ue,McAvity:1995zd,Liendo:2012hy,Gliozzi:2015qsa,Diehl:1996kd,Diehl2}.
  
The classically marginal interaction serves to alter slightly the 
boundary conditions on the bulk field away from Dirichlet or Neumann cases.  
One may think of these interactions as a coupling between an 
operator of dimension $\frac{d-2}{2}$ and an operator of dimension $\frac{d}{2}$.  
In the Neumann case, the operator of dimension $\frac{d-2}{2}$ is the boundary limit of the bulk field $\phi$ or $A_A$.  
In the Dirichlet case, the operator of dimension $\frac{d}{2}$ is the boundary limit of $\partial_n \phi$.  

Recall in the discussion of crossing relations, we found the simple relation (\ref{simplecrossing}).  
The free fields we discussed in the previous section take advantage of this 
relation only in the limiting Dirichlet or Neumann cases $\chi \to \pm 1$ 
(or more generally when the eigenvalues of $\chi$ are $\pm 1$).  
In these cases, the two-point function decomposes either into a single 
boundary block of dimension $\frac{d}{2}$ in the Dirichlet case or a 
single boundary block of dimension $\frac{d-2}{2}$ in the Neumann case.  
Indeed, the operator of the other dimension is missing because of the 
boundary conditions.  Now we see, at least perturbatively, how the story will generalize.  
The boundary interaction adds back a little bit of the missing block, and the 
two-point function for the bulk free field will be characterized instead by a $\chi = \pm (1 - {\cal O}(g^2))$.  
(The story with the bulk Maxwell field is complicated by the lack of 
gauge invariance of $\langle A_\mu(x) A_\nu(x') \rangle$, but morally the story is the same.)
Through Feynman diagram calculations below, we will confirm this over-arching picture.

With the modified two-point function of the bulk fields in hand, it will be 
straightforward to modify the corresponding two-point functions of the 
current and stress tensor, using Wick's theorem, to leading order in the interaction $g$.  
We just need to keep a general 
value of $\chi$, instead of setting $\chi = \pm 1$.  For the stress tensor, 
one finds the structure $\alpha(v) = 1 + \chi^2 v^{2d}$ instead of $\alpha(v) = 1 + v^{2d}$, and 
similarly for the current two-point function.
%At higher order in $g$,  additional Feynman diagrams will contribute to these 
%two-point functions,
%but as our interest was a proof of 
%principle that $\chi$ is not always $\pm 1$, we will stop at leading order in the expansion.

In the special case of mixed QED, where the theory is purported to be conformal in 
$d=4$ dimensions, we have an example of a conformal field theory where $\alpha(1) < 2 \alpha(0)$ 
and $b_2$ cannot be directly related to  the central charge $c$ in the 
bulk trace anomaly.
In fact, the situation is more subtle.  In order to evaluate $\alpha(v)$ at $v=1$, we take a near boundary limit.
It is in fact not necessarily true that the $v \to 1$ limit commutes with the perturbative $g \to 0$ limit in these
theories.  

For the related function $\gamma(v)$, a similar perturbative 
computation indicates that $\gamma(1) = {\cal O}(g^2)$ where the nonzero contribution comes from $T^{nA}$ exchange
in the boundary conformal block decomposition.  
%However, because of the relation $T^{nA}|_{\rm bry} = - \partial_B \hat T^{AB}$, 
%at higher order in perturbation theory one should see $T^{nA}|_{\rm bry}$ develop an anomalous dimension $\delta_T > 0$ and be absorbed into the spin 2 conformal block associated with the boundary stress tensor $\hat T^{AB}$.  
However, as mentioned in the introduction, we must 
have $T^{nA}|_{\rm bry}=0$ as an operator statement since the dimension of  $T^{nA}$ is protected.
Mathematically, one expects $\gamma(v) \sim g^2 (1-v)^{ \delta_T}$ where $\delta_T \sim {\cal O}(g^2)$, leading to noncommuting
small $g$ and $v\to 1$ limits and allowing $\gamma(1)$ to remain zero.\footnote{%
 We would like to thank D.~Gaiotto for pointing out an error in an earlier version of the manuscript where we claimed $\gamma(1) \neq 0$.
}

From the conservation relations, one could worry there is a similar issue with $\alpha(1)$.  
But, looking more carefully, the behavior $\gamma(v) \sim g^2 ( 1-v)^{\delta_T}$ leads to $\alpha(v) \sim g^2 (1-v)^{1 +  \delta_T}$ which vanishes at $v=1$ independent of the order of limits, and $\epsilon(v) \sim g^2 \delta_T \, (1-v)^{-1 + \delta_T}$ whose associated divergence will only show up at the next order in perturbation theory.
We therefore claim the ${\cal O}(g^2)$ contribution to $\alpha(1)$ we find is independent of the order of limits and comes from an alteration in the contribution
of the displacement operator conformal block to the two-point function.  
Indeed, if we were to find a behavior of the form $\alpha(v) \sim g^2 (1-v)^{\delta_T}$, which has the order of limits issue, that behavior through stress tensor conservation corresponds to 
an $\epsilon(v) \sim g^2 \delta_T \, (1-v)^{-2+\delta_T}$ or equivalently exchange of a boundary spin two operator of dimension $d-2 + \delta_T$ which is below the unitarity bound of $d-1$ for small $\delta_T$.  
%Thus, we believe we can trust the leading order answer we get for $\alpha(1)$ in these boundary interacting theories.
To check these arguments that $\alpha(1) \neq 2 \alpha(0)$, 
ideally we should go to higher loop order in perturbation theory.
We leave such calculations for the future.
%
%
%However, as we have seen in section \ref{sec:twopoint}, a nonzero value of $\gamma(1)$ corresponds to a contribution 
%from a vector primary of dimension $\Delta = d$ to the boundary conformal block decomposition of the stress tensor 
%two-point function.  
%
%

It would be interesting furthermore to see if one can bound $\alpha(1)$ and 
correspondingly the boundary trace anomaly $b_2$.  
It is tempting to conjecture that free theories saturate an upper bound 
$\alpha(1) \leq 2 \alpha(0)$ in four dimensions.\footnote{
Away from $d=4$, there are already counterexamples.  
For $\phi^4$ theory and Neumann (special) boundary conditions, $\alpha(1) > 2 \alpha(0)$ 
both in the large $N$ expansion in the range $5/2 < d < 4$ and also at leading order in the 
$\epsilon$ expansion for any $N$.  See (7.31) and (7.23) of ref.\ \cite{McAvity:1995zd}.  
In $d=4$, the theory becomes free and one has $\alpha(1) = 2\alpha(0)$ or $b_2=8c$.}
The phenomenon that $\alpha(1) = 2 \alpha(0)$ at this point appears to be a special feature of free bCFTs.

\subsection{Mixed Yukawa Theory}

Let us begin with a one loop analysis of the Yukawa-like theory,
\be
I=-\frac{1}{2} \int_{\cal M} \d^4 x (\partial^\mu \phi)(\partial_\mu \phi) 
+  \int_{\partial {\cal M}} \d^3 x  \left( i\bar \psi \slashed{\partial} \psi - g\phi  \bar \psi \psi \right) \ ,
\ee
with the modified Neumann boundary condition $\partial_n \phi = -g \bar \psi \psi$.  
Again, the normal coordinate will be denoted by $y$ and the coordinates tangential to the boundary by ${\bf x}$: $x = ({\bf x}, y)$.  

Our first task will be to calculate a $\beta$-function
for the interaction $\phi \bar \psi \psi$ to see if we can find a conformal fixed point.
We should comment briefly on the space of relevant operators and the amount of fine tuning we need to achieve our goal.
The engineering dimension of the $\psi$ field is one, and thus a $(\bar \psi \psi)^2$ term should be perturbatively irrelevant. 
One could in principle generate relevant $\phi$ and $\phi^2$ and a classically marginal $\phi^3$ interactions on the boundary through loop effects.  We will assume that we can tune these terms away.\footnote{%
We have checked explicitly that $\phi^3$ is not generated at one loop in this theory.}

As we use dimensional regularization, we need the propagators for the scalar and spinor fields in arbitrary dimension.
The Euclidean propagators are 
\be
G_\phi(x; x') &=& C_S \left( \frac{1}{\big(({\bf x}-{\bf x'})^2 + (y-y')^2\big)^{d-2\over 2}} + \frac{1}{\big(({\bf x}-{\bf x'})^2 + (y+y')^2\big)^{d-2\over 2}} \right) \ , \\
G_\psi({\bf x} ) &=& C_F \frac{\gamma_A x^A}{{\bf x}^{d-1}}  = -\frac{C_F}{d-3} \gamma^A \partial_A \left(\frac{1}{{\bf x}^{d-3}}\right)\ .
\ee
A canonical normalization is $C_S =\K= 1 / (d-2) \Vol(S^{d-1})$ for the scalar
and $C_F = 1/\Vol(S^{d-2})$ for the boundary fermion, where $\Vol(S^{d-1}) = 2 \pi^{d/2} / \Gamma(d/2)$. 
Note that, unlike what we did in section 5.1, here we have started with a propagator with $\chi=1$, fixed by the required Neumann boundary condition (when $g=0$) on a single scalar in this toy model. 

For our Feynman diagram calculations, we need the Fourier transforms along the boundary directions: 
\be
\tilde G_\phi(p) &\equiv& \int_{\partial {\cal M}} \d^{d-1} {\bf x} ~  e^{-i p \cdot {\bf x}} G_\phi(y,{\bf x};0,0)= \frac{e^{-py}}{p}
\label{scalarprop}
 \ , \\
\tilde G_\psi(p) &\equiv& \int_{\partial {\cal M}} \d^{d-1}{\bf x}  ~e^{-i p \cdot {\bf x}} G_\psi({\bf x})= -i \frac{\gamma \cdot p}{p^2} \ . 
\ee
While $\tilde G_\psi(p)$ takes its canonical, textbook form, the scaling
of $\tilde G_\phi(p)$ is $1/p$ instead of the usual $1/p^2$.  This shift leads to many of the physical effects
we now consider.
We will perform our Feynman diagram expansion in Lorentzian signature.  Analytically continuing, we find the usual $-i / \slashed{p}$ rule for an internal spinor line and a $-i  / |p|$ for an internal scalar line.  As the beginning and end point of the scalar line must lie on the $y=0$ plane, we can remove the $e^{-p y}$ factor from the momentum space propagator. 

\begin{figure}
\begin{center}
(a)
\includegraphics[width=1.3in]{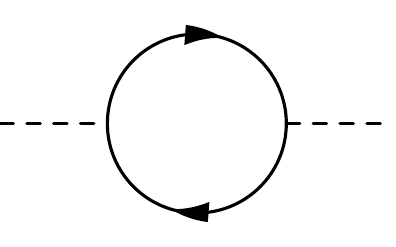}
(b)
\includegraphics[width=1.3in]{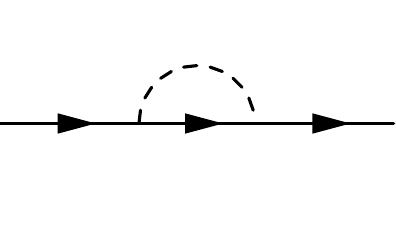}
(c)
\includegraphics[width=1.3in]{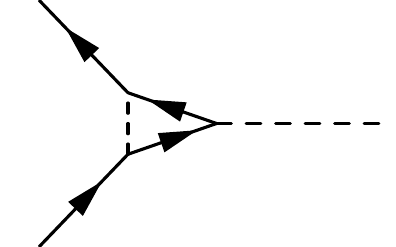}
\end{center}
\caption{For the mixed dimensional Yukawa theory: (a) scalar one loop propagator correction; (b) fermion one loop propagator correction; (c) one loop vertex correction.
\label{fig:Yukawa}
}
\end{figure}

We now calculate the one loop corrections shown in figure \ref{fig:Yukawa}.
We begin with the scalar propagator.  The diagram has a linear UV divergence which is invisible in dimensional regularization: 
\be
i\tilde \Pi_\phi(q) &=& (-1) (-ig)^2 \int \frac{\d^{d-1} p}{(2 \pi)^{d-1}}\frac{\tr [ i \slashed{p} \, i (\slashed{p} + \slashed{q})]}{p^2 (p+q)^2} 
\\
&=& -ig^2 \frac{2^{5-2d} \pi^{2 - \frac{d}{2}}}{\cos \left( \frac{\pi d}{2} \right) \Gamma \left( \frac{d}{2}-1 \right)} q^{d-3} \ ,
\ee where we have used $\tr (\gamma_A \gamma_B)= - 2 \eta_{AB}$ and  $\tr [ \slashed{p} (\slashed{p} + \slashed{q})]=-2 (p^2+ p\cdot q)$.\footnote{In 
this section we take $\tr \id=2$ for the three dimensional Clifford space.}
In $d=4$, the self-energy reduces to 
\be
\label{scalaroneloopyukawa}
\tilde \Pi_\phi = -\frac{q}{8} g^2  \ .
\ee
This result is in contrast to the usual self-energy correction for the 4d Yukawa theory, which has a logarithmic divergence.  
As the fermion momentum space propagators are the same in 3d and 4d, the difference comes from integrating over
three rather than four momentum space dimensions.  

The correction to the fermion propagator, in contrast, has a logarithmic divergence:
\be
i\tilde \Pi_\psi(q) &=& (-ig)^2  \int \frac{\d^{d-1} p}{(2\pi)^{d-1}} \frac{( i \slashed{p}) (-i) }{p^2 |p-q|} \\
%&=&- g^2  \int \frac{\d^{d-1} p}{(2\pi)^{d-1}} \frac{ \gamma \cdot p}{p^2 |p+q|} \\
%&=& -\frac{g^2}{2} \int_0^1 \d u \int \frac{\d^{d-1} p}{(2\pi)^{d-1}} \frac{ \gamma \cdot( p-u q) u^{-1/2}}{(p^2 + u(1-u) q^2)^{3/2}} \\
&=&- i g^2 \frac{4^{2-d} \pi^{\frac{1-d}{2}} \Gamma\left( 2 - \frac{d}{2} \right) \Gamma(d-2)}{\Gamma \left(d - \frac{3}{2} \right)}
\frac{ \gamma \cdot q}{q^{4-d}} \ .
\ee
In $d=4-\epsilon$, the result becomes
\be
\tilde \Pi_\psi(q) &=&  -\slashed{q} g^2 \left[ \frac{1}{6 \pi^2 \epsilon} + \frac{1}{36 \pi^2} (10 - 3 \gamma - 3 \log(q^2/\pi)) \right] + {\cal O}(\epsilon) \ .
\ee
The logarithmic divergence is evidenced by the $1/\epsilon$ in the dimensionally regulated expression, or we could have seen it explicitly by performing the original integral in $d=4$ dimensions with a hard UV cut-off.
%Or we can do the integral directly with a large $p_{\rm max}$ cut-off:
%\be
%\tilde \Pi_\psi(q) = \slashed{q} \frac{g^2}{18 \pi^2} \left[  -3\log \frac{q}{p_{\rm max}} + 2 \right] \ .
%\ee
%

Third, we look at the one loop correction to the vertex:
\be
-i g \tilde \Gamma(q_1, q_2) &=& 
(-ig)^3 \int \frac{\d^{d-1} p}{(2\pi)^{d-1}}  \frac{ i(\slashed{p} + \slashed{q}_1) i(\slashed{p} + \slashed{q}_2)(-i)}{(p+q_1)^2 (p+q_2)^2 |p| } \ .
% \\
%&=&
%g^3 \int \frac{\d^{d-1} p}{(2\pi)^{d-1}}  \frac{ (\slashed{p} + \slashed{q}_1) (\slashed{p} + \slashed{q}_2)}{(p+q_1)^2 (p+q_2)^2 |p| } \\
%&=& g^3 \frac{3}{4} \int_\triangle \d^3 u \int  \frac{\d^{d-1} p}{(2\pi)^{d-1}}
%\frac{(\slashed{p} + \slashed{q}_1)(\slashed{p} + \slashed{q}_2)u_3^{-1/2} \delta (1 - \sum_j u_j) }{(u_1 (p+q_1)^2 + u_2 (p+q_2)^2 + u_3 p^2)^{5/2}} \\
%&=&  g^3 \frac{3}{4} \int_\triangle \d^2 u \int  \frac{\d^{d-1} p}{(2\pi)^{d-1}}
%\frac{(\slashed{p} - u_i \slashed{q}_i + \slashed{q}_1) (\slashed{p} -u_i \slashed{q}_i + \slashed{q}_2) (1-u_1-u_2)^{-1/2}}{(p^2  - 2 u_1 u_2 q_1 \cdot q_2+ \sum_{i=1}^2 u_i (1-u_i) q_i^2 )^{5/2}} \\
%&=&i  g^3 \frac{\Gamma \left( 3 - \frac{d}{2} \right)}{2^{d-1} \pi^\frac{d}{2}} \int_\triangle \d^2 u  \, M^{d-6} 
%\left( \frac{d-1}{d-4} M^2 + (u_i \slashed{q}_i - \slashed{q}_1)(u_i \slashed{q}_i - \slashed{q}_2) \right)(1-u_1-u_2)^{-1/2}
\ee
%where
%\be
%M^2 \equiv  - 2 u_1 u_2 q_1 \cdot q_2+ \sum_{i=1}^2 u_i (1-u_i) q_i^2
%\ee
Using Feynman parameters, we can extract the most singular term.  In $d = 4-\epsilon$ dimensions, we find that
\be
g \tilde \Gamma(q_1, q_2) &=& -g^3 \frac{1}{2 \pi^2 \epsilon} + {\rm finite} \ .
\ee

To compute the $\beta$-function for $g$, we introduce the wave-function renormalization factors $Z_\phi$ and $Z_\psi$ for the scalar and fermion kinetic terms as well as a vertex renormalization factor $Z_g$.  The $\beta$-function follows from the relation
\be
g_0 Z_\phi^{1/2} Z_\psi = g \mu^{\epsilon/2} Z_g \ ,
\ee
where we can extract the $Z$-factors from our one loop computations: 
\be
Z_\psi &=& 1 + g^2 \left( -\frac{1}{6 \pi^2 \epsilon} + {\rm finite} \right) \ , \\
\label{ZphiYukawa}
Z_\phi &=& 1 + g^2 ({\rm finite}) \ , \\
Z_g &=& 1 + g^2 \left( \frac{1}{2 \pi^2 \epsilon} + {\rm finite} \right) \ ,
\ee 
and $g_0$ denotes the bare coupling which is $\mu$-independent. It follows that the $\beta$-function, $\beta(g(\mu))=\mu {\partial \over \partial \mu}g(\mu)$, is given by 
\be
\beta
= - \frac{\epsilon}{2} g + \frac{2}{3 \pi^2} g^3 + {\cal O}(g^4) \ .
\ee 
For $d\geq 4$, the function remains positive which indicates that the coupling flows to zero at large distance. 
For $d< 4$, the coupling increases or decreases with the distance depending on the strengh of $g$. 
Given our fine tuning of relevant operators, we obtain an IR stable fixed point:
\be
g_*^2 = {3 \pi^2\over 4} \epsilon \ ,
\ee
in $d<4$ dimensions. 
Note that $Z_\phi$ has no divergent contribution.  Indeed, 
a general feature of our collection of theories is that the bulk field will not be renormalized at one loop.  In the case of the mixed dimensional QED theory, we can in fact make a stronger argument.

We claimed above that one effect of the classically marginal interaction was to shift slightly the form of the scalar-scalar two-point function.  Let us see how that works by Fourier transforming the result (\ref{scalaroneloopyukawa}) back to position space\footnote{For the loop computation, we use a propagator from one point on the boundary to another where we set $y=0$. When Fourier transforming back to real space, we are sewing on external propagators, taking us from points in the bulk (with non-zero $y_1$ and $y_2$) to points on the boundary.}: 
\be
 \Pi_\phi(x_1; x_2) &=& \int \frac{\d^{d-1} p}{(2\pi)^{d-1}} \tilde \Pi_\phi(p) \frac{e^{-p(y_1 + y_2)}}{p^2} e^{i p \cdot \delta {\bf x}} \\
 &=& -\frac{g^2}{16 \pi^{2}} \frac{1}{(y_1 + y_2)^2 + \delta {\bf x}^2} \ .
 \label{Yukawachichange}
\ee
As we started with a single component scalar with Neumann boundary conditions $\chi=1$, this Fourier transform implies
that we have ended up with a two-point function with a slightly shifted $\chi$:
\be
\chi \to \chi = 1 - {\cal O}(g^2) \ . 
\ee
The corrections to the current and stress tensor two-point functions will be controlled by the shift in the scalar two-point function, at this leading order ${\cal O}(g^2)$.  Thus, we can read off the corresponding current and stress tensor two-point functions merely by inserting the modified value of $\chi$ in the formulae we found for the free scalar. 
Note this mixed Yukawa model becomes free in $d=4$ where $\chi=1$ is recovered. 
Our next example will be an interacting CFT in $d=4$ where the parameter $\chi$ can be different from one.

\subsection{Mixed Quantum Electrodynamics}

\begin{figure}
\begin{center}
(a)
\includegraphics[width=1.3in]{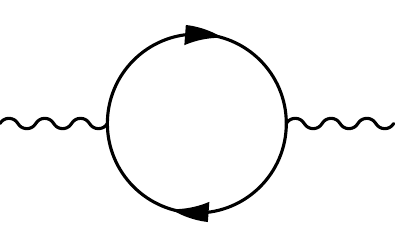}
(b)
\includegraphics[width=1.3in]{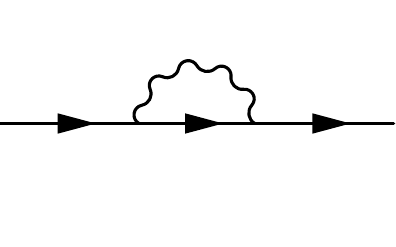}
(c)
\includegraphics[width=1.3in]{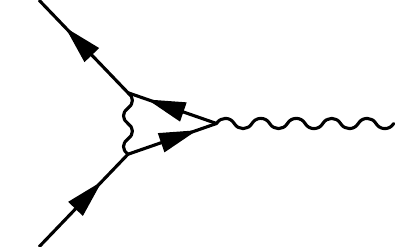}
\end{center}
\caption{For the mixed dimensional QED: 
(a) photon one loop propagator correction; 
(b) fermion one loop propagator correction; 
(c) one loop vertex correction.
\label{fig:QED}
}
\end{figure}

The action for the mixed dimensional QED is\footnote{%
 There is a slight variant of the mixed QED theory 
where the boundary becomes transparent in the limit $g \to 0$ instead of satisfying absolute
 boundary conditions.  Through the folding trick, this theory can be mapped to the one under consideration 
 plus an extra decoupled Maxwell field, provided we make the redefinition $g \to g / \sqrt{2}$.
} 
\be
I=-\frac{1}{4} \int_{\cal M}  \d^4 x F^{\mu\nu} F_{\mu\nu} 
+  \int_{\partial {\cal M}} \d^3 x  \left( i\bar \psi \slashed{D} \psi  \right) \ ,
\ee
where $D_\mu = \partial_\mu - i g A_\mu$.
Note there is a potential generalization to include a Chern-Simons term on the boundary for this mixed QED model.
We will work with a four component fermion to avoid generating a parity anomaly \cite{Redlich1,Redlich2}, and proceed with a standard
evaluation of the one loop corrections (see figure \ref{fig:QED}) using the following Feynman rules:
photon propagator, $-i \frac{e^{-p y}}{p} \eta^{AB}$; fermion propagator, $\frac{i \slashed{p}}{p^2}$;
interaction vertex, $i g \gamma^A$.  The ghosts are decoupled in this abelian theory so below we do not need to consider them. 
A more general version of this calculation can be found in ref.\ \cite{Teber:2012de}.

The photon self-energy can be evaluated in a completely standard way: 
\be
i\tilde \Pi_\gamma^{AB} (q) &=& (-1) (i g)^2 \int \frac{\d^{d-1} p}{(2\pi)^{d-1}} \frac{\tr[\gamma^A\, i\slashed{p} \gamma^B \, i(\slashed{p}+\slashed{q})]}{p^2 (p+q)^2} \\ 
%
%&=& - e^2  \int \frac{\d^{d-1} p}{(2\pi)^{d-1}} \frac{\tr[\gamma^\mu\, \slashed{p} \gamma^\nu \, (\slashed{p}+\slashed{q})]}{p^2 (p+q)^2} \\ 
%&=&-e^2  \int_0^1 \d u \int \frac{\d^{d-1}p}{(2\pi)^{d-1}} \frac{\tr[ \gamma^\mu \, (\slashed{p}- u \slashed{q}) \gamma^\nu \, (\slashed{p}+(1-u)\slashed{q})]}
%{ (p^2 + u(1-u) q^2)^2} \\
%&=& - 2e^2 \int_0^1 \d u \int \frac{\d^{d-1}p}{(2\pi)^{d-1}} \frac{ 2 p^\mu p^\nu - 2 u(1-u) q^\mu q^\nu - \eta^{\mu\nu}(p^2 -u(1-u)q^2)}
%{(p^2 + u(1-u)q^2)^2} \\
%&=&  - 2e^2 \int_0^1 \d u \int \frac{\d^{d-1}p}{(2\pi)^{d-1}} \frac{ - 2 u(1-u) q^\mu q^\nu +\eta^{\mu\nu}\left( \left(\frac{2}{d-1} - 1 \right) p^2 +u(1-u)q^2\right)}
%{(p^2 + u(1-u)q^2)^2} \\
%&=& - 4 e^2 \int_0^1 \d u \, u(1-u) \int \frac{\d^{d-1}p}{(2\pi)^{d-1}} \frac{q^2 \eta^{\mu\nu} - q^\mu q^\nu}{(p^2 + u(1-u) q^2)^2} \\
&=& - 2ig^2 (q^2 \eta^{AB} - q^A q^B) \frac{(d-3) \pi^{2 - \frac{d}{2}}}{4^{d-2} \cos \left( \frac{\pi d}{2} \right) \Gamma \left( \frac{d}{2} \right)} \frac{1}{q^{5-d}} \ .
\ee
In $d=4$, one gets the finite answer in dimensional regularization 
\be
\label{photonselfenergy}
\tilde \Pi_\gamma^{AB}(q) = - \frac{g^2}{8 q} (q^2 \eta^{AB} - q^A q^B) \ .
\ee

There is in fact never a logarithmic divergence at any order in the loop expansion for $\tilde \Pi_{\gamma}^{AB}(q)$, and the wave-function renormalization for the photon $Z_\gamma$ will be finite in dimensional regularization.
The usual topological argument shows that the photon self-energy diagrams have a linear superficial degree of divergence.
Consider a general $n$-loop correction to the scalar propagator with $\ell$ internal propagators and $v$ vertices.  Momentum conservation tells us that $n-\ell + v = 1$.  We can divide up $\ell$ into photon lines $\ell_\gamma$ and fermion lines $\ell_\psi$.  As each vertex involves two fermion lines and one photon, it must be that $\ell_\psi = v$ and (recalling that two photon lines are external) $\ell_\gamma = (v-2)/2$.  Therefore $n=v/2$.  
The superficial degree of divergence of the photon self-energy diagrams is thus
\be
n(d-1) - \ell_e - \ell_\gamma = n(d-1)-\frac{3v}{2} +1 = n(d-4) +1 \ , 
\ee
which in $d=4$ dimensions is equal to one. 
Gauge invariance implies that we can strip off a $q^A q^B - \eta^{AB} q^2$ factor from the self-energy.  As a result, it is conventionally argued that the degree of divergence is reduced by 2.  Thus the photon self-energy is finite in this mixed dimensional context.  (In QED, the superficial degree of divergence is 2, and the gauge invariance argument changes the divergence to a log.  There is then a corresponding renormalization of the photon wave-function.) 

Let us again Fourier transform back to position space.
There is a subtle issue associated with gauge invariance.  Our Feynman gauge breaks conformal symmetry, and if we proceed naively, we will
not be able to write the correlator $\langle A_\mu (x) A_\nu(x') \rangle$ as a function of the cross-ratio $v$, making it difficult to make use of the results from section \ref{sec:free}.  To fix things up, we have the freedom to perform a small gauge transformation that changes the bare propagator by a term of ${\cal O}(g^2)$.  In fact, we claim we can tune this transformation such that there is a ${\cal O}(g^2)$ term in the bare propagator that cancels the $q^A q^B$ dependence of (\ref{photonselfenergy}).  The details are in appendix \ref{sec:gaugefixing}.  In our slightly deformed gauge, the corrections to the position space
correlation function become 
\be
\label{pigammaf}
\Pi_\gamma^{AB}(x; x') &=& -c \int \frac{\d^{d-1} p}{(2\pi)^{d-1}} \frac{e^{-p(y_1+y_2) + i p \cdot \delta {\bf x}}}{p^{5-d}} \eta^{AB} \ ,
\ee
where
\be
c =(d-3) \frac{2g^2 \pi^{2 - \frac{d}{2}}}{4^{d-2} \cos \left(\frac{\pi d}{2} \right) \Gamma \left(\frac{d}{2}\right)} \ .
\ee
In four dimensions, we obtain 
\be
\Pi_\gamma^{AB}(x; x') &=& - \frac{g^2}{16 \pi^2 (\delta {\bf x}^2 + (y_1 + y_2)^2)} \eta^{AB} \ . 
\label{QEDchichange}
\ee 
Analogous to the Yukawa theory, we can interpret this shift as a shift in the $\chi_\parallel$ parameter of the $\langle A_A(x) A_B(x') \rangle$ two-point function.   
 The corresponding current and stress tensor two-point functions can then be deduced at leading order ${\cal O}(g^2)$ by making the appropriate substitutions for $\chi_\parallel$ in the Maxwell theory results obtained in section \ref{sec:free}.

As in the Yukawa theory case, 
the corrections to the fermion propagator are modified slightly by the reduced  dimensionality of the theory.  The calculation is almost identical: 
\be
i \tilde \Pi_\psi(q) &=& (i g)^2 \int \frac{\d^{d-1}p}{(2\pi)^{d-1}} \frac{ \gamma^A \, i \slashed{p}  \gamma^B (-i) \eta_{AB}}{p^2 |p-q|}
\\
&=& (ig)^2 (d-3) \int \frac{\d^{d-1}p}{(2\pi)^{d-1}} \frac{i \slashed{p}(-i)  }{p^2 |p-q|} \\ 
&=& - \slashed{q} g^2 \frac{1}{6 \pi^2 \epsilon} + {\rm finite} \ .
\ee
The result is precisely the result for the fermion self-energy in the Yukawa theory.

Finally, we calculate the singular contributions to the one loop vertex correction:
\be
i g \tilde \Gamma^A (q_1, q_2) &=& (i g)^3 \int \frac{\d^{d-1} p}{(2 \pi)^{d-1}} \frac{\gamma^C i (\slashed{p} +\slashed{q}_1)\gamma^A  i (\slashed{p} + \slashed{q}_2) \gamma^B (-i) \eta_{CB}}{(p+q_1)^2 (p+q_2)^2 |p|} \ .
\ee
Evaluating this integral in $d = 4 - \epsilon$ dimensions yields
\be
\tilde \Gamma^A(q_1, q_2) &=& g^2 \gamma^A \frac{1}{6 \pi^2 \epsilon} \ .
\ee
There is a relative factor of -1/3 compared to the Yukawa theory.  
In fact, there is a well known and relevant Ward identity argument (see e.g.\ \cite{Srednicki}) that can be employed here.
Current conservation applied to the correlation function $\langle J^\mu(z) \bar \psi(x) \psi(y) \rangle$ implies that $Z_g/Z_\psi$ is finite in perturbation theory.  In the minimal subtraction scheme where all corrections to $Z_g$ and $Z_\psi$ are divergent, we conclude that $Z_g = Z_\psi$.  

At one loop, we have all the information we need to compute the $\beta$-function:
\be
g_0 Z_\gamma^{1/2} Z_\psi = g \mu^{\epsilon/2} Z_g \ ,
\ee
where 
\be
Z_\psi &=& 1 - g^2 \left( \frac{1}{6 \pi^2 \epsilon} + {\rm finite} \right) \ , \\
\label{Zgamma}
Z_\gamma &=& 1 + g^2 ({\rm finite}) \ , \\
Z_g &=& 1 - g^2 \left( \frac{1}{6 \pi^2 \epsilon} + {\rm finite} \right) \ . 
\ee
Hence the beta function is 
\be
\beta= - \frac{\epsilon}{2} g +{\cal O}(g^4) \ .
\ee
In other words, the $\beta$-function vanishes in 4d at one loop.  In fact, as we have sketched, the Ward identity argument $Z_\psi = Z_g$ and the non-renormalization $Z_\gamma = 1 + g^2({\rm finite})$ are expected to hold order by order in perturbation theory, and so we can tentatively conclude 
that this mixed dimensional QED is $exactly$ conformal in four dimensions, making this theory rather special.

From the relation between $b_2$ and $\alpha(1)$ \eqref{mainresult}, the Fourier transformed propagator \eqref{QEDchichange} and the two-point function of U(1) gauge fields in $d=4$ \eqref{alphap}, we obtain the boundary charge $b_2$ for the mixed conformal QED as
\be
\label{b2mqed}
b_{2(\rm{Mixed~QED})}&=& {2 \over 5} \Big(2-{g^2\over 2} + \ldots\Big) < {4\over 5}= 8 c_{(\rm{Mixed~QED})} \ ,
\ee 
where ${4\over 5}=b_{2(\rm{EM})}$ is the boundary charge for the standard bulk U(1) theory. 
This weakly interacting conformal model therefore provides an example of $b_2\neq 8c$ in 4d bCFTs.

In addition to $\alpha(v)$, consider the behavior of $\gamma(v)$, defined in \eqref{gammadef}, and
representing the correlation function of the boundary limit of $T^{nA}$.  While for free theories, it vanishes
universally, $\gamma(1) = 0$, in this mixed conformal QED we find instead that, from the one loop computation 
given here, $\gamma(1)=- {3 g^2\over 2 \pi^4}$.
But, as mentioned earlier, we must have a vanishing $T^{nA}$ in the boundary limit as an operator statement. We expect 
\be
\gamma(v) \sim - {3 g^2\over 2 \pi^4} (1-v)^{ \delta_T} \ , 
\ee
where $\delta_T \sim {\cal O}(g^2)$ is the anomalous dimension. 
In this case, the small $g$ and $v \to 1$ limits do not commute.  
While perturbatively, we might be fooled
into thinking that $\gamma(1) \neq 0$, in point of fact $\gamma(1)$ should vanish.

%\footnote{In free theories we observe that the boundary limit of $\gamma$, defined in \eqref{gammadef}, 
%representing the correlation function of the spin-1 displacement operator, vanishes universally, i.e. $\gamma(1)=0$. 
%In this mixed conformal QED, we find instead that, from the one loop computation 
%given here, $\gamma(1)=- {3 g^2\over 2 \pi^4}$, as an example of $\gamma(1)\neq 0$ in 4d bCFTs.}

While we do not do so here, there are two further calculations of great interest.  
The first is to look at the next loop order in the stress tensor two-point function.
The stress tensor conservation equations suggest that the order of limits will not be an issue 
for evaluating $\alpha(1)$. It would be nevertheless nice to verify this claim by actually computing more 
Feynman diagrams.
While we have no expectation that the value of $\alpha(1)$ is somehow protected in interacting theories,
it would be fascinating if it were.
The second project is to calculate the trace anomaly of this theory
directly in curved space with a boundary to verify the relation between $\alpha(1)$ and $b_2$.  
We leave such projects for the future.

\subsection{Mixed Scalar}

In the two examples we considered so far, the boundary interaction modified a Neumann boundary condition.  In this third
example, the boundary interaction modifies a Dirichlet condition.  There will be a corresponding all important change in sign in the 
correction to $\chi=-1$.  
The theory is 
\be
I=-\frac{1}{2} \int_{\cal M} \d^4 x (\partial^\mu \phi)(\partial_\mu \phi) 
-   \int_{\partial {\cal M}} \d^3 x  \Big( \frac{1}{2} (\partial_A \eta)(\partial^A \eta) 
+  (\partial_n \phi)(-\phi + g \eta^2) \Big) \ .
\ee
This theory has many possible relevant interactions on the boundary that can be generated by loop effects, e.g.\ $\phi^2$, $\eta^2$, $\eta^4$, etc.  We will assume we can fine tune all of these relevant terms away.  We will also ignore additional classically marginal interactions such as $\phi^2 \eta^2$ and $\eta^6$.  

\begin{figure}
\begin{center}
(a)
\includegraphics[width=1.3in]{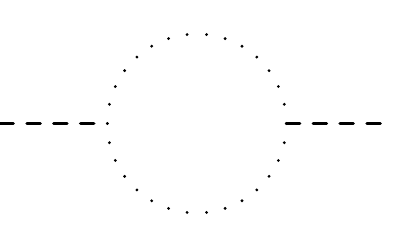}
(b)
\includegraphics[width=1.3in]{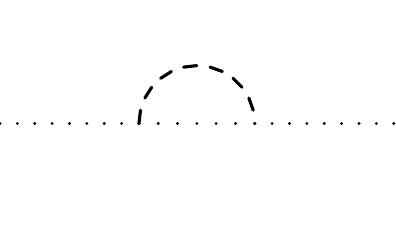}
(c)
\includegraphics[width=1.3in]{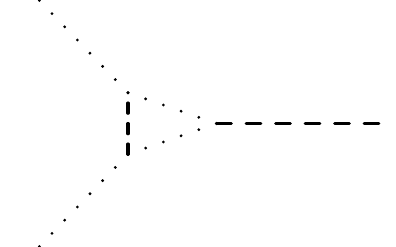}
\end{center}
\caption{For the mixed dimensional scalar theory: 
(a) a 4d bulk scalar one loop self energy correction; 
(b) a 3d boundary scalar one loop self energy correction; 
(c) one loop vertex correction.
\label{fig:scalar}
}
\end{figure}

We proceed to a calculation of the three Feynman diagrams in figure \ref{fig:scalar}.
The propagator correction for the bulk scalar is
\be
i \tilde \Pi_\phi &=& 2 (i g)^2 \int \frac{\d^{d-1} p}{(2 \pi)^{d-1}} \frac{(-i)^2}{p^2 (p+q)^2} \\
%&=& 2 g^2 \int_0^1 \d u \int \frac{\d^{d-1}p}{(2\pi)^{d-1}} \frac{1}{(p^2 + u(1-u) q^2)^2} \\
%&=&ig^2\frac{4^{3-d} \pi^{2 - \frac{d}{2}}}{\cos \left( \frac{\pi d}{2} \right) \Gamma \left( \frac{d}{2}-1 \right)} q^{d-5} \\
&=& i\frac{g^2}{4 q} \ .
\ee
We can Fourier transform this result back to position space to see how the two-point function will be modified: 
\be
\Pi_\phi &=& \int \frac{\d^{d-1} p}{(2 \pi)^{d-1}} \tilde \Pi_\phi(p) e^{-p(y_1 + y_2)} e^{i p \cdot \delta {\bf{x}}} \\
 &=&  \frac{ g^2}{8 \pi^2 (\delta {\bf{x}}^2 + (y_1+y_2)^2)} \ ,
\ee
where in the last line, we set $d=4$.
Crucially, the sign here is different from (\ref{Yukawachichange}) and (\ref{QEDchichange}), corresponding to a 
shift in the two-point function for the scalar away from Dirichlet conditions $\chi = -1 + {\cal O}(g^2)$ instead of away 
from Neumann conditions $\chi = 1 - {\cal O}(g^2)$.  
Note these results are consistent with the bounds on $\chi$ \eqref{boundchi}.
At leading order ${\cal O}(g^2)$, we can compute the
corrected current and stress tensor two-point functions as well, merely by making the appropriate replacement for
$\chi$ in the free scalar results.

The correction to the boundary scalar propagator is
\be
i \tilde \Pi_\eta &=& 4(-i g)^2 \int \frac{\d^{d-1} p}{(2 \pi)^{d-1}} \frac{(-i)^2(-1)|p|}{(p+q)^2} 
\\
&=&- i \frac{2 g^2 q^2}{3 \pi^2 \epsilon} + {\rm finite} \ .
\ee
Finally, we give the divergent contribution to the one loop vertex correction:
\be
-i g \tilde \Gamma(q_1, q_2) &=& 8(-ig)^3 \int \frac{\d^{d-1} p}{(2\pi)^{d-1}} \frac{(-i)^3(-1) |p|}{(p+q_1)^2 (p+q_2)^2} \ .
\ee
In $d= 4 - \epsilon$ dimensions, this reduces to
\be
 g \tilde \Gamma(q_1, q_2) &=& -  g^3 \frac{4}{\pi^2 \epsilon} + {\rm finite} \ .
\ee

We compute the $\beta$-function for $g$ using
$
g_0 Z_\phi^{1/2} Z_\eta = g \mu^{\epsilon/2} Z_g
$
and\footnote{%
We note in passing that bulk fields are not renormalized in our one loop computations.
$Z_\phi$ in (\ref{ZphiYukawa}) and (\ref{Zphimixed}) and $Z_\gamma$ in (\ref{Zgamma}) are
finite.  There should be an argument based on locality, that boundary interactions
can never renormalize the bulk fields.  We are not sure how to make precise the relationship
between locality and the actual Feynman diagram computations, however. 
We thank D.~Gaiotto for discussions on this point.
}
\be
Z_\eta &=& 1 - g^2 \left(\frac{2}{3 \pi^2 \epsilon} +  {\rm finite} \right) \ , \\
\label{Zphimixed}
Z_\phi &=& 1 + g^2 ({\rm finite}) \ , \\
Z_g &=& 1 + g^2 \left( \frac{4}{\pi^2 \epsilon} + {\rm finite} \right) \ . 
\ee
The result is that 
\be
\beta= - \frac{\epsilon}{2} g + \frac{14}{ 3\pi^2} g^3 + {\cal O}(g^4) \ .
\ee
There is an IR stable fixed point at 
\be
g_*^2 = \frac{3 \pi^2}{28} \epsilon \ , 
\ee in $d<4$ dimensions. In the $d=4$ limit, the theory becomes free and one has $\alpha(1)=2\alpha(0)$ and $b_2=8c$ relations.

%% file: nullcone.tex
The null cone formalism is a useful tool for linearizing the action of the conformal group $O(1,d+1)$ \cite{Dirac}.  The linearization in turn makes a derivation of the conformal blocks straightforward \cite{Dolan:2003hv,Costa:2011mg,Costa:2011dw,Weinberg:2010fx}, especially for higher spin operators, as we now review, drawing heavily on \cite{Liendo:2012hy}.

Points in physical space $x^\mu \in {\mathbb R}^d$ are in one-to-one correspondence with null rays
in ${\mathbb R}^{1,d+1}$. Given a point written in light cone coordinates, 
\be
P^A = (P^+, P^-, P^1, \ldots, P^d) \in {\mathbb R^{1,d+1}} \ , 
\ee 
a null ray
corresponds to the equivalence class $P^A \sim \lambda P^A$ such that $P^A P_A = 0$.  
A point in physical space can then be 
recovered via
\be
x^\mu = \frac{P^\mu}{P^+} \ .
\ee
A linear $O(1,d + 1)$ transformation of ${ \mathbb R}^{1,d+1}$ which maps null rays into null rays corresponds to a conformal transformation on the physical space.

We are further interested in correlation functions of symmetric traceless tensor fields $F_{\mu_1 \cdots \mu_n}$.  For a tensor field lifted to embedding space $F_{A_1 \cdots A_n}(P)$ and inserted at $P$, 
\be
F_{A_1 \cdots A_n} (\lambda P) = \lambda^{-\Delta} F_{A_1 \cdots A_n}(P) \ ,
\ee
we reduce this problem to that of correlation functions of scalar operators by contracting the open indices with a vector $Z$:
\be 
F(P, Z) = Z^{A_1} \cdots Z^{A_n} F_{A_1 \cdots A_n} \ .
\ee
Tracelessness means that we can take $Z^2 = 0$.  
In the embedding space, the tensor must be transverse $P^{A_1} F_{A_1 \cdots A_n} = 0$, which implies that
$P \cdot \partial_Z F(P,Z) = 0$.  Given the redundancy in the embedding space, we can also choose $Z \cdot P = 0$ without harm.

In the presence of a boundary, we have an extra unit normal vector $V = (0, \ldots, 0,1)$ which breaks the
symmetry $O(1,d+1)$ down to $O(1,d)$.  For two-point functions with operators inserted at $P$ and $P'$, we can form the following scalar quantities invariant under $O(1,d)$:
\be
P \cdot P' \ , \; \; \; V \cdot P \ , \; \; \; V \cdot P' \ ,  \; \; \; Z \cdot P'  \ , \; \; \; Z' \cdot P  \ , \; \; \; V \cdot Z \ , 
\; \; \; V \cdot Z' \ .
\ee
Note the cross ratio $\xi$ can be written as
\be
\xi = - \frac{P \cdot P'}{2 (V \cdot P) (V \cdot P')} \ ,
\ee
in this formalism.
The game is then to write down functions of these invariants which correspond to a correlation function with the correct scaling weights and index structure.  For the operator $F(P_i, Z_i)$ of weight $\Delta_i$, we need one $Z_i$ field for each index of the original $F_{\mu_1 \cdots \mu_n}$.  Also, the expression should be homogeneous in $P_i$ with degree $-\Delta_i$.  Furthermore,
we will need to make sure that the expressions satisfy transversality.

The one-point function of a scalar operator is 
\be
\langle O(P) \rangle &=& {a_{\Delta}\over (2 V \cdot P)^\Delta} \ .
\ee 
Note the one-point function of an operator with spin $l$ would introduce a factor $(V\cdot Z)^l$, which violates the  transversality condition. Indeed, only the one-point function of a scalar is allowed in the presence of a boundary.

The scalar two-point function is 
\be
\label{nullOO}
\langle O_1(P) O_2(P') \rangle &=& \frac{ 1}{(2 V \cdot P)^{\Delta_1} (2 V \cdot P')^{\Delta_2}} f(\xi) \ ,
\ee 
where
\be
f(\xi) = \xi^{-{(\Delta_1 + \Delta_2)\over 2}} G(\xi)  \ .
\ee
And, for current and stress tensor, we have
\be
 \langle Z \cdot J_1(P) Z' \cdot J_2(P') \rangle &=& \frac{P(\xi) S_1 + v^2 Q(\xi) S_2}{\xi^{d-1} (V \cdot P)^{\Delta_1} (V \cdot P')^{\Delta_2} } \ , \\
\langle Z \cdot T_1 (P) \cdot Z \; Z' \cdot T_2(P') \cdot Z' \rangle 
 &=& \frac{C(\xi) S_1^2 +4 v^2 B(\xi) S_1 S_2 + v^4 A(\xi) S_2^2}{(4 \xi)^{d} (V \cdot P)^{\Delta_1} (V \cdot P_2)^{\Delta_2}} \ ,
\ee
where
\be
S_1 &=& \frac{ (Z \cdot Z') (P \cdot P') - (Z \cdot P') (Z' \cdot P) }{P \cdot P'} \ , \\
S_2 &=& \left( \frac{ (V \cdot P) (Z \cdot P')}{P \cdot P'} - V \cdot Z \right) \left( \frac{(V \cdot P') (Z' \cdot P)}{P \cdot P'} - V \cdot Z' \right) \ .
\ee

The conservation conditions can be expressed in terms of the Todorov differential operator
\be
D^{(d)}_A = \left( \frac{d}{2} -1 + Z \cdot \frac{\partial}{\partial Z} \right) \frac{\partial}{\partial Z^A} - \frac{1}{2} Z_A \frac{\partial^2}{\partial Z \cdot \partial Z} \ . 
\ee
Conservation for an operator $F(P, Z)$ means that $(\partial_P \cdot D^{(d)}) F = 0$.  The conservation conditions will enforce that 
$\Delta_i = d-1$ for the current and $\Delta_i =d$ for the stress tensor, but we leave them arbitrary for now.

The Todorov differential is also useful for writing the action of an element $L_{AB}$ of the Lie algebra ${\mathfrak o}(1,d+1)$ on a symmetric traceless tensor:
\be
L_{AB} F(P,Z) = \left( P_A \frac{\partial}{\partial P^B} - P_B \frac{\partial}{\partial P^A} + \frac{1}{\frac{d}{2}+s-2} (Z_A D_B^{(d)} - Z_B D_A^{(d)} ) \right) F(P,Z) \ . \nn\\
\ee
The conformal Casimir equation  is then
\be
\frac{1}{2} L_{AB} L^{AB} F(P,Z) = - C_{\Delta, l} F(P,Z) \ ,
\ee
where $C_{\Delta, l} = \Delta (\Delta-d) + l(l+d-2)$.  The conformal blocks in the bulk expansion are then determined by an equation of the form
\be
\frac{1}{2}(L_{AB} + L_{AB}') (L^{AB} + {L'}^{AB}) G(P,Z,P',Z') = - C_{\Delta, 0} G(P,Z,P',Z') \ ,
\ee
acting on the two-point function $G(P,Z,P',Z')$ expressed in the null-cone formalism.

In the boundary conformal block expansion, we need to consider instead the generators of $O(1,d)$, $a,b = \pm, 1, \ldots, d-1$:
\be
L_{ab} = P_a \frac{\partial}{\partial P^b} - P_b \frac{\partial}{\partial P^a} + \frac{1}{\frac{d-1}{2}+s-2} (Z_a D_b^{(d-1)} - Z_b D_a^{(d-1)} ) \ .
\ee
In this case, the conformal blocks in the boundary expansion are determined by an equation of the form
\be
\frac{1}{2} L_{ab} L^{ab} G(P,Z, P',Z') = - \tilde C_{\Delta, l} G(P,Z,P',Z') \ ,
\ee
where the Casimir operator acts on just the pair $P$ and $Z$ and $\tilde C_{\Delta, l} = \Delta (\Delta -d +1) + l(l+d-3)$.

We give some details of the derivation for the conserved current, which is new. (For conformal blocks of stress tensor two-point function, we refer the reader to \cite{Liendo:2012hy} for details.)  In this case, because of the linearity of the two-point function in $Z$ and $Z'$, the Todorov differentials can be replaced by  ordinary partial differentials with respect to $Z$:
\be
\frac{1}{\frac{d}{2}-1}D_A^{(d)} \rightarrow \frac{\partial}{\partial Z^A} \ , \; \; \;
\frac{1}{\frac{d-1}{2} - 1} D_a^{(d-1)} \rightarrow \frac{\partial}{\partial Z^a} \ .
\ee

For what follows, we define the functions
\be
\tilde f \equiv P \ , \; \; \; \tilde g \equiv v^2 Q \ .
\ee
In the bulk conformal block decomposition, exchanging a scalar of dimension $\Delta$ with the boundary leads to the following pair of differential equations:
\be
&F:&~4 \xi^2 (1+\xi) \tilde f'' +2 \xi(2 \xi+2-d)\tilde f' \nn\\
&&~~~~~~~~~~~ +[(d-\Delta) \Delta - (\Delta_1-\Delta_2)^2] \tilde f - 2 \tilde g = 0 \ , \\
&G:&~4 \xi^2 (1+\xi) \tilde g'' + 2\xi (2 \xi -  2-d) \tilde g' \nn\\
&&~~~~~~~~~~~ +[(2+d-\Delta)(2+\Delta) -(\Delta_1-\Delta_2)^2] \tilde g = 0 
\ .
\ee
The tensor structure $S_1$ gives rise to the differential equation $F$ while the structure $S_2$ gives the equation $G$.
This system is compatible with the conservation relation. 
Restricting to $\Delta_i = d-1$, current conservation gives
\be
J:~(d+1)\tilde g-2\xi \tilde g' -2 \xi^2 (\tilde f'+\tilde g')=0 \ .
\ee
One can construct a linear relation of the form $c_1 F' + c_2 G' + c_3 F + c_4 G + J'' + c_5 J' + c_6 J$, indicating that either of the second-order differential equations for $\tilde f$ and $\tilde g$ can be swapped for current conservation.

The differential equation $G$ may be solved straightforwardly: 
\be
\label{gbulkjj}
\tilde g_{\rm bulk}(\Delta, \xi) = \xi^{1 + {\Delta\over2}} {}_2 F_1 \left( 1 + \frac{\Delta+\Delta_1-\Delta_2}{2},  1 + \frac{\Delta-\Delta_1+\Delta_2}{2} , 1 - \frac{d}{2} + \Delta, - \xi \right) \ ,
\ee where another solution with the behaviour $\sim \xi^{1-{\Delta\over 2}}$ is dropped.  Note $\tilde g_{\rm bulk}(\Delta, 0)=0$. 
We introduce un-tilde'd functions that will simplify the equations for the boundary blocks:
\be
\tilde f(\xi) &=& \xi^{(\Delta_1 + \Delta_2)/2 -d+1} f(\xi) \ , \\
\tilde g(\xi) &=& \xi^{(\Delta_1 + \Delta_2)/2 -d+1} g(\xi) \ . 
\ee
Note the distinction disappears for conserved currents.  
Plugging the soluton \eqref{gbulkjj} into the conservation equation $J$ one obtains 
%\be
%f_{\rm bulk}(\Delta, \xi) &=& \xi^{\Delta/2} \Bigl[ \frac{(2 + \Delta) \xi(1+\xi)}{2 - d + 2 \Delta} \,  {}_2 F_1\left( 1 + \frac{\Delta}{2}, 2 + \frac{\Delta}{2} ; 2 - \frac{d}{2} + \Delta; - \xi \right) \nonumber \\
%&& - \frac{1 - d + \Delta(1 + \xi)}{\Delta} {}_2 F_1 \left( \frac{\Delta}{2}, 1 + \frac{\Delta}{2}; 1 - \frac{d}{2}+ \Delta; - \xi \right) \Bigr] \ .
%\ee
\be
f_{\rm bulk}(\Delta, \xi) + v^{-2} g_{\rm bulk}(\Delta, \xi) &=& \frac{d-1}{\Delta} \xi^{\Delta/2} 
{}_2 F_1 \left( \frac{\Delta}{2}, 1+\frac{\Delta}{2}, 1- \frac{d}{2}+\Delta, -\xi \right) \ .
\ee

In the boundary block decomposition, we find the following differential equations for $f_{\rm bry}$ and $g_{\rm bry}$:
\be
&&\xi (1 + \xi) g'' + \left( 2\xi - \frac{d}{2}(3 + 2 \xi) \right) g' \nn\\
&&~~~~~~~~~+ \left( \frac{2 + d + d^2}{2 \xi} - C_{\Delta, \ell} \right) g = (d-2) f \ , \\
&&\xi (1+\xi) f'' + \left( \xi(2-d) +2 - \frac{3d}{2} \right) f' \nn\\
&&~~~~~~~~~+ \left( \frac{(d-2)(1+d+2\xi)}{2 \xi} - C_{\Delta, \ell} \right) f = \frac{1+2 \xi}{2 \xi^2} g \ ,
\ee
where
\be
C_{\Delta, \ell} = \ell(\ell + d-3) + \Delta(\Delta-d +1) \ .
\ee
As in the bulk case, these differential equations are compatible with the conservation condition, as can be verified by constructing a similar linear dependence between the equations.

We need to solve these equation for ($\ell = 0$ and $\Delta = d-1$) and also for ($\ell = 1$ and all $\Delta$).  
In the first case
\be
f_{\rm bry}^{0}(d-1, \xi) &=& 
%\xi^{-1+d/2} {}_2 F_1 \left( h, 1 -h, 1 - h, - \xi \right) \nonumber \\
%&=& 
\frac{1}{\xi} \left( \frac{\xi}{1+\xi} \right)^{h} = v^{d-2} (1 - v^2) \ , \\
g_{\rm bry}^0(d-1, \xi) &=& \xi^{h} (1 + \xi)^{-1-h} ( d-2 + 2(d-1) \xi) = v^d ( d-2 + d v^2)\ .
\ee
There are similarly simple expressions for $\ell = 1$ and $\Delta = d-1$:
\be
f_{\rm bry}^1(d-1, \xi) &=& \frac{1}{2} \xi^{h-1} (1+\xi)^{-h} (1 + 2 \xi)  = \frac{1}{2} v^{d-2} (1+v^2)\ , \\
g_{\rm bry}^1(d-1, \xi) &=& \frac{1}{2} \xi^{h} (1+\xi)^{-h-1} (d-2 - 2 \xi) = \frac{1}{2} v^d (d-2 - d v^2) \ .
\ee
%We also mention in passing that $g_{\rm bry}^1(d-3, \xi) = \xi^2$.  
In general, the spin one exchange is given by 
\be
g_{\rm bry}^{1}(\Delta, \xi) &=& -\xi^{d-1-\Delta} {}_3 F_2 \left( 
\begin{array}{c}
1 + \Delta, 3 - d + \Delta, 1 - \frac{d}{2} + \Delta \\
 2 - d + \Delta, 2 - d + 2 \Delta
 \end{array}
 ; - \frac{1}{\xi} \right) \ , \\
f_{\rm bry}^{1}(\Delta, \xi) &=& \frac{\xi ^{d-\Delta -2} }{2 (\Delta +2-d)}
\Big[ 2\xi ( \Delta+1-d) \,{}_2 F_1 \left( \Delta, - \frac{d}{2} + \Delta + 1; -d + 2 \Delta + 2; - \frac{1}{\xi} \right) \nn \\
&& 
 + (2 \xi +1) \, _2F_1\left(\Delta +1,-\frac{d}{2}+\Delta +1;-d+2 \Delta +2;-\frac{1}{\xi }\right)
 \Big] \ .
%\Big[2 (\Delta +1) (\xi +1) \, _2F_1\left(\Delta +2,-\frac{d}{2}+\Delta +1;-d+2 \Delta +2;-\frac{1}{\xi }\right)\nn\\
%&&-(d-1) (2 \xi +1) \, _2F_1\left(\Delta +1,-\frac{d}{2}+\Delta +1;-d+2 \Delta +2;-\frac{1}{\xi }\right)\Big] \ .
\ee